\documentclass[a4paper,11pt,draft]{article}

\usepackage{jheppub} 

\usepackage[T1]{fontenc} 
\usepackage{amsmath}
\usepackage{tikz}
\usepackage{mathdots}
\usepackage{yhmath}
\usepackage{cancel}
\usepackage{color}
\usepackage{array}
\usepackage{multirow}
\usepackage{amssymb}
\usepackage{gensymb}
\usepackage{makecell}
\usepackage{tabularx}
\usepackage{booktabs}
\usetikzlibrary{fadings}
\usetikzlibrary{patterns}
\usetikzlibrary{shadows.blur}
\usepackage{amsmath,amsthm}
\usepackage{amssymb,amsfonts}
\usepackage{graphicx}
\usepackage{dsfont}
\usepackage{relsize}
\usepackage{stmaryrd}
\usepackage{lmodern}
\usepackage{slantsc}
\usepackage{scalefnt}
\usepackage{subfigure}
\usepackage{xspace}
\usepackage{boxedminipage}
\usepackage{ifpdf}
\usepackage{multirow}
\usepackage{xifthen}
\usepackage{tensor}
\usepackage{array}
\usepackage{tabu}
\usepackage{diagbox}
\usetikzlibrary{arrows,matrix,calc,scopes,decorations.markings}
\allowdisplaybreaks[1]
\usepackage[mathcal]{euscript}

\newcommand{\be}{\begin{equation}}
\newcommand{\ee}{\end{equation}}
\newcommand{\bse}{\begin{subequations}}
	\newcommand{\ese}{\end{subequations}}
\newcommand{\ket}[1]{\left|{#1}\right\rangle}

\newcommand{\bket}[1]{\left\vert{#1}\right\rangle}

\newcommand{\bpm}{\begin{pmatrix}}
	\newcommand{\epm}{\end{pmatrix}}
\newcommand{\bmm}{\begin{matrix}}
	\newcommand{\emm}{\end{matrix}}

\allowdisplaybreaks[4]
\newtabulinestyle{dash=off 2pt}

{\null\,\vcenter\bgroup\scriptsize
	\arraycolsep=.13885em
	\hbox\bgroup$\array{@{}#1@{}}}
{\endarray$\egroup\egroup\,\null}

\newcolumntype{C}{>{\centering\arraybackslash} m{1.5em} }

\makeatletter
\newcommand*{\Relbarfill@}{\arrowfill@\Relbar\Relbar\Relbar}
\newcommand*{\xeq}[2][]{\ext@arrow 0055\Relbarfill@{#1}{#2}}
\makeatother

\usetikzlibrary{decorations.markings}
\usetikzlibrary{decorations.pathmorphing,positioning}
\tikzset{>=latex}
\tikzset{snake it/.style={decorate, decoration={snake,amplitude=0.15mm,segment length=1mm}}}
\tikzset{->-/.style={decoration={
			 markings,
			 mark=at position .55 with {\arrow{>}}},postaction={decorate}}}

\tikzset{-<-/.style={decoration={
			 markings,
			 mark=at position .55 with {\arrow{<}}},postaction={decorate}}}

\newcommand{\algebraQD}{
\begin{tikzpicture}[baseline={([yshift=-.8ex]current bounding box.center)},scale=1]
\draw [->-] (1.9,12.8) ..controls (1.879,12.521) and (1.857,12.243) .. (2.,12.1)
			 ..controls (2.143,11.957) and (2.45,11.95) .. (2.6,12.1)
			 ..controls (2.75,12.25) and (2.743,12.557) .. (2.6,12.7)
			 ..controls (2.457,12.843) and (2.179,12.821) .. (1.9,12.8);
\draw [fill,rotate around={0.:(1.9,12.8)}] (1.9,12.8) ++(0.:0.08 and 0.08) arc(0.:360.:0.08 and 0.08);
\node [rotate=0.] at (1.6,12.8) {\scriptsize $g_0$};
\node [rotate=0.] at (2.9,12.1) {\scriptsize $h_0$};
\end{tikzpicture}
}

\newcommand{\betaByTetrahedraAA}{
\begin{tikzpicture}[baseline={([yshift=-.8ex]current bounding box.center)},scale=1.6]
\draw [->-] (3.5,10.7) -- (2.5,10.7);
\draw [->-] (2.9,12.8) -- (1.9,12.8);
\draw [->-] (3.5,10.7) -- (2.9,12.8);
\draw [dashed] (2.5,10.7) -- (2.31,11.365);
\draw [dashed] (2.255,11.557) -- (2.242,11.604);
\draw [dashed] (2.187,11.796) -- (1.9,12.8);
\draw [dashed] (3.5,10.7) -- (2.799,11.62);
\draw [dashed] (2.677,11.78) -- (1.9,12.8);
\draw [->-] (2.5,10.7) -- (1.9,11.7);
\draw [->-] (1.9,11.7) -- (1.9,12.8);
\draw [->-] (3.5,10.7) -- (2.9,11.7);
\draw [->-] (2.9,11.7) -- (2.9,12.8);
\draw [->-] (2.9,11.7) -- (1.9,12.8);
\draw [->-] (2.9,11.7) -- (1.9,11.7);
\draw [->-] (3.5,10.7) -- (1.9,11.7);
\node [rotate=0.] at (1.6,12.2) {\scriptsize $g_1$};
\node [rotate=0.] at (1.9,11.1) {\scriptsize $g_2$};
\node [rotate=0.] at (2.6,11.6) {\scriptsize $h_1$};
\node [rotate=0.] at (3.1,10.5) {\scriptsize $h_2$};
\node [rotate=0.] at (2.7,12.4) {\scriptsize $g_1$};
\node [rotate=0.] at (3.,11.3) {\scriptsize $g_2$};
\end{tikzpicture}
}

\newcommand{\betaByTetrahedraAB}{
\begin{tikzpicture}[baseline={([yshift=-.8ex]current bounding box.center)},scale=1.6]
\draw [->-] (3.5,10.5) -- (2.5,10.5);
\draw [->-] (2.9,12.6) -- (1.9,12.6);
\draw [->-] (3.5,10.5) -- (2.9,12.6);
\draw [dashed] (2.5,10.5) -- (2.31,11.165);
\draw [dashed] (2.255,11.357) -- (2.242,11.404);
\draw [dashed] (2.187,11.596) -- (1.9,12.6);
\draw [dashed] (3.5,10.5) -- (2.799,11.42);
\draw [dashed] (2.677,11.58) -- (1.9,12.6);
\draw [->-] (2.5,10.5) -- (1.9,11.5);
\draw [->-] (1.9,11.5) -- (1.9,12.6);
\draw [->-] (3.5,10.5) -- (2.9,11.5);
\draw [->-] (2.9,11.5) -- (2.9,12.6);
\draw [->-] (2.9,11.5) -- (1.9,12.6);
\draw [->-] (2.9,11.5) -- (1.9,11.5);
\draw [->-] (3.5,10.5) -- (1.9,11.5);
\node [rotate=0.] at (1.6,12.) {\scriptsize $g'_1$};
\node [rotate=0.] at (1.9,10.9) {\scriptsize $g'_2$};
\node [rotate=0.] at (2.4,12.8) {\scriptsize $h'_1$};
\node [rotate=0.] at (2.5,11.4) {\scriptsize $h'_2$};
\node [rotate=0.] at (2.7,12.2) {\scriptsize $g'_1$};
\node [rotate=0.] at (3.,11.1) {\scriptsize $g'_2$};
\end{tikzpicture}
}

\newcommand{\betaByTetrahedraAD}{
\begin{tikzpicture}[baseline={([yshift=-.8ex]current bounding box.center)},scale=1]
\draw [->-] (3.5,10.7) -- (2.5,10.7);
\draw [->-] (2.9,12.8) -- (1.9,12.8);
\draw [->-] (3.5,10.7) -- (2.9,12.8);
\draw [dashed] (2.5,10.7) -- (2.31,11.365);
\draw [dashed] (2.255,11.557) -- (2.242,11.604);
\draw [dashed] (2.187,11.796) -- (1.9,12.8);
\draw [dashed] (3.5,10.7) -- (2.799,11.62);
\draw [dashed] (2.677,11.78) -- (1.9,12.8);
\draw [->-] (2.5,10.7) -- (1.9,11.7);
\draw [->-] (1.9,11.7) -- (1.9,12.8);
\draw [->-] (3.5,10.7) -- (2.9,11.7);
\draw [->-] (2.9,11.7) -- (2.9,12.8);
\draw [->-] (2.9,11.7) -- (1.9,12.8);
\draw [->-] (2.9,11.7) -- (1.9,11.7);
\draw [->-] (3.5,10.7) -- (1.9,11.7);
\node [rotate=0.] at (1.6,12.2) {\scriptsize $g_1$};
\node [rotate=0.] at (1.9,11.1) {\scriptsize $g_2$};
\node [rotate=0.] at (2.7,12.4) {\scriptsize $g_1$};
\node [rotate=0.] at (3.,11.3) {\scriptsize $g_2$};
\node [rotate=0.] at (3.1,10.5) {\scriptsize $\mathrm{hol}$};
\end{tikzpicture}
}

\newcommand{\bilayerModelAA}{
\begin{tikzpicture}[baseline={([yshift=-.8ex]current bounding box.center)},scale=1]
\path [lightgray, fill] (5.5,12.) -- (5.3,11.4) -- (6.3,12.) -- cycle;
\path [lightgray, fill] (5.5,7.3) -- (5.3,6.7) -- (6.3,7.3) -- cycle;
\draw [gray, ->-] (2.,10.8) -- (2.8,10.8);
\draw [gray, ->-] (2.8,10.8) -- (3.6,10.8);
\draw [gray, ->-] (3.6,10.8) -- (4.4,10.8);
\draw [gray, ->-] (2.,10.8) -- (2.9,11.4);
\draw [gray, ->-] (3.6,10.8) -- (4.5,11.4);
\draw [gray, ->-] (4.4,10.8) -- (5.3,11.4);
\draw [gray, ->-] (2.8,10.8) -- (2.9,11.4);
\draw [gray, ->-] (4.4,10.8) -- (4.5,11.4);
\draw [gray, ->-] (4.5,11.4) -- (5.3,11.4);
\draw [gray, ->-] (2.9,11.4) -- (3.9,12.);
\draw [gray, ->-] (4.5,11.4) -- (5.5,12.);
\draw [gray, ->-] (4.5,11.4) -- (4.7,12.);
\draw [gray, ->-] (2.9,11.4) -- (3.1,12.);
\draw [gray, ->-] (3.1,12.) -- (3.9,12.);
\draw [gray, ->-] (3.9,12.) -- (4.7,12.);
\draw [gray, ->-] (4.7,12.) -- (5.5,12.);
\draw [gray, ->-] (3.1,12.) -- (4.,12.6);
\draw [gray, ->-] (3.9,12.) -- (4.8,12.6);
\draw [gray, ->-] (4.7,12.) -- (5.6,12.6);
\draw [gray, ->-] (5.5,12.) -- (6.4,12.6);
\draw [gray, ->-] (3.9,12.) -- (4.,12.6);
\draw [gray, ->-] (4.7,12.) -- (4.8,12.6);
\draw [gray, ->-] (5.5,12.) -- (5.6,12.6);
\draw [gray, ->-] (4.,12.6) -- (4.8,12.6);
\draw [gray, ->-] (4.8,12.6) -- (5.6,12.6);
\draw [gray, ->-] (5.6,12.6) -- (6.4,12.6);
\draw [gray, ->-] (4.4,10.8) -- (5.2,10.8);
\draw [gray, ->-] (5.2,10.8) -- (6.1,11.4);
\draw [gray, ->-] (5.2,10.8) -- (5.3,11.4);
\draw [gray, ->-] (5.3,11.4) -- (6.1,11.4);
\draw [gray, ->-] (6.1,11.4) -- (6.3,12.);
\draw [gray, ->-] (6.3,12.) -- (7.2,12.6);
\draw [gray, ->-] (6.3,12.) -- (6.4,12.6);
\draw [gray, ->-] (6.4,12.6) -- (7.2,12.6);
\draw [->-, thick] (2.9,11.4) -- (3.7,11.4);
\draw [->-, thick] (3.7,11.4) -- (4.5,11.4);
\draw [->-, thick] (2.8,10.8) -- (3.7,11.4);
\draw [->-, thick] (3.7,11.4) -- (4.7,12.);
\draw [->-, thick] (3.6,10.8) -- (3.7,11.4);
\draw [->-, thick] (3.7,11.4) -- (3.9,12.);
\draw [->-, thick] (5.3,11.4) -- (5.5,12.);
\draw [->-, thick] (5.3,11.4) -- (6.3,12.);
\draw [->-, thick] (5.5,12.) -- (6.3,12.);
\draw [] (3.6,12.8) -- (1.6,10.6);
\draw [] (1.6,10.6) -- (5.7,10.6);
\draw [] (5.7,10.6) -- (7.7,12.8);
\draw [] (7.7,12.8) -- (3.6,12.8);
\draw [gray, ->-] (2.,6.1) -- (2.8,6.1);
\draw [gray, ->-] (2.8,6.1) -- (3.6,6.1);
\draw [gray, ->-] (3.6,6.1) -- (4.4,6.1);
\draw [gray, ->-] (2.,6.1) -- (2.9,6.7);
\draw [gray, ->-] (3.6,6.1) -- (4.5,6.7);
\draw [gray, ->-] (4.4,6.1) -- (5.3,6.7);
\draw [gray, ->-] (2.8,6.1) -- (2.9,6.7);
\draw [gray, ->-] (4.4,6.1) -- (4.5,6.7);
\draw [gray, ->-] (4.5,6.7) -- (5.3,6.7);
\draw [gray, ->-] (2.9,6.7) -- (3.9,7.3);
\draw [gray, ->-] (4.5,6.7) -- (5.5,7.3);
\draw [gray, ->-] (4.5,6.7) -- (4.7,7.3);
\draw [gray, ->-] (2.9,6.7) -- (3.1,7.3);
\draw [gray, ->-] (3.1,7.3) -- (3.9,7.3);
\draw [gray, ->-] (3.9,7.3) -- (4.7,7.3);
\draw [gray, ->-] (4.7,7.3) -- (5.5,7.3);
\draw [gray, ->-] (3.1,7.3) -- (4.,7.9);
\draw [gray, ->-] (3.9,7.3) -- (4.8,7.9);
\draw [gray, ->-] (4.7,7.3) -- (5.6,7.9);
\draw [gray, ->-] (5.5,7.3) -- (6.4,7.9);
\draw [gray, ->-] (3.9,7.3) -- (4.,7.9);
\draw [gray, ->-] (4.7,7.3) -- (4.8,7.9);
\draw [gray, ->-] (5.5,7.3) -- (5.6,7.9);
\draw [gray, ->-] (4.,7.9) -- (4.8,7.9);
\draw [gray, ->-] (4.8,7.9) -- (5.6,7.9);
\draw [gray, ->-] (5.6,7.9) -- (6.4,7.9);
\draw [gray, ->-] (4.4,6.1) -- (5.2,6.1);
\draw [gray, ->-] (5.2,6.1) -- (6.1,6.7);
\draw [gray, ->-] (5.2,6.1) -- (5.3,6.7);
\draw [gray, ->-] (5.3,6.7) -- (6.1,6.7);
\draw [gray, ->-] (6.1,6.7) -- (6.3,7.3);
\draw [gray, ->-] (6.3,7.3) -- (7.2,7.9);
\draw [gray, ->-] (6.3,7.3) -- (6.4,7.9);
\draw [gray, ->-] (6.4,7.9) -- (7.2,7.9);
\draw [->-, thick] (2.9,6.7) -- (3.7,6.7);
\draw [->-, thick] (3.7,6.7) -- (4.5,6.7);
\draw [->-, thick] (2.8,6.1) -- (3.7,6.7);
\draw [->-, thick] (3.7,6.7) -- (4.7,7.3);
\draw [->-, thick] (3.6,6.1) -- (3.7,6.7);
\draw [->-, thick] (3.7,6.7) -- (3.9,7.3);
\draw [->-, thick] (5.3,6.7) -- (5.5,7.3);
\draw [->-, thick] (5.3,6.7) -- (6.3,7.3);
\draw [->-, thick] (5.5,7.3) -- (6.3,7.3);
\draw [] (3.6,8.1) -- (1.6,5.9);
\draw [] (1.6,5.9) -- (5.7,5.9);
\draw [] (5.7,5.9) -- (7.7,8.1);
\draw [] (7.7,8.1) -- (3.6,8.1);
\draw [] (5.5,8.8) -- (5.4,8.8);
\draw [] (5.4,8.8) -- (5.6,8.6);
\draw [] (5.6,8.6) -- (5.8,8.8);
\draw [] (5.8,8.8) -- (5.7,8.8);
\draw [] (5.7,8.8) -- (5.7,9.8);
\draw [] (5.7,9.8) -- (5.8,9.8);
\draw [] (5.8,9.8) -- (5.6,10.);
\draw [] (5.6,10.) -- (5.4,9.8);
\draw [] (5.4,9.8) -- (5.5,9.8);
\draw [] (5.5,9.8) -- (5.5,8.8);
\node [rotate=0.] at (3.5,11.6) {\scriptsize $A^a$};
\node [rotate=0.] at (3.5,6.9) {\scriptsize $A^x$};
\node [rotate=0.] at (5.7,11.8) {\scriptsize $B^b$};
\node [rotate=0.] at (5.7,7.1) {\scriptsize $B^y$};
\node [rotate=0.] at (3.7,10.4) {\scriptsize $\text{QD model}(H)$};
\node [rotate=0.] at (4.1,9.7) {\scriptsize $\text{coupling}$};
\node [rotate=0.] at (4.1,9.4) {\scriptsize $\text{characterized}$};
\node [rotate=0.] at (3.7,5.7) {\scriptsize $K-\epsilon\text{ model}$};
\node [rotate=0.] at (4.1,9.1) {\scriptsize $\text{by }F\text{ and }\alpha$};
\end{tikzpicture}
}

\newcommand{\bilayerModelAB}{
\begin{tikzpicture}[baseline={([yshift=-.8ex]current bounding box.center)},scale=1]
\path [lightgray, fill] (3.3,11.1) -- (3.5,10.9) -- (4.1,11.1) -- (4.4,11.5) -- (4.2,11.7) -- (3.6,11.5) -- cycle;
\path [lightgray, fill] (5.4,6.3) -- (6.4,6.9) -- (5.6,6.9) -- cycle;
\draw [] (3.7,12.8) -- (1.6,10.4);
\draw [] (1.6,10.4) -- (5.8,10.4);
\draw [] (5.8,10.4) -- (7.9,12.8);
\draw [] (7.9,12.8) -- (3.7,12.8);
\draw [gray, ->-] (2.1,5.7) -- (2.9,5.7);
\draw [gray, ->-] (2.9,5.7) -- (3.7,5.7);
\draw [gray, ->-] (3.7,5.7) -- (4.5,5.7);
\draw [gray, ->-] (2.1,5.7) -- (3.,6.3);
\draw [gray, ->-] (3.7,5.7) -- (4.6,6.3);
\draw [gray, ->-] (4.5,5.7) -- (5.4,6.3);
\draw [gray, ->-] (2.9,5.7) -- (3.,6.3);
\draw [gray, ->-] (4.5,5.7) -- (4.6,6.3);
\draw [gray, ->-] (4.6,6.3) -- (5.4,6.3);
\draw [gray, ->-] (3.,6.3) -- (4.,6.9);
\draw [gray, ->-] (4.6,6.3) -- (5.6,6.9);
\draw [gray, ->-] (4.6,6.3) -- (4.8,6.9);
\draw [gray, ->-] (3.,6.3) -- (3.2,6.9);
\draw [gray, ->-] (3.2,6.9) -- (4.,6.9);
\draw [gray, ->-] (4.,6.9) -- (4.8,6.9);
\draw [gray, ->-] (4.8,6.9) -- (5.6,6.9);
\draw [gray, ->-] (3.2,6.9) -- (4.1,7.5);
\draw [gray, ->-] (4.,6.9) -- (4.9,7.5);
\draw [gray, ->-] (4.8,6.9) -- (5.7,7.5);
\draw [gray, ->-] (5.6,6.9) -- (6.5,7.5);
\draw [gray, ->-] (4.,6.9) -- (4.1,7.5);
\draw [gray, ->-] (4.8,6.9) -- (4.9,7.5);
\draw [gray, ->-] (5.6,6.9) -- (5.7,7.5);
\draw [gray, ->-] (4.1,7.5) -- (4.9,7.5);
\draw [gray, ->-] (4.9,7.5) -- (5.7,7.5);
\draw [gray, ->-] (5.7,7.5) -- (6.5,7.5);
\draw [gray, ->-] (4.5,5.7) -- (5.3,5.7);
\draw [gray, ->-] (5.3,5.7) -- (6.2,6.3);
\draw [gray, ->-] (5.3,5.7) -- (5.4,6.3);
\draw [gray, ->-] (5.4,6.3) -- (6.2,6.3);
\draw [gray, ->-] (6.2,6.3) -- (6.4,6.9);
\draw [gray, ->-] (6.4,6.9) -- (7.3,7.5);
\draw [gray, ->-] (6.4,6.9) -- (6.5,7.5);
\draw [gray, ->-] (6.5,7.5) -- (7.3,7.5);
\draw [->-, thick] (3.,6.3) -- (3.8,6.3);
\draw [->-, thick] (3.8,6.3) -- (4.6,6.3);
\draw [->-, thick] (2.9,5.7) -- (3.8,6.3);
\draw [->-, thick] (3.8,6.3) -- (4.8,6.9);
\draw [->-, thick] (3.7,5.7) -- (3.8,6.3);
\draw [->-, thick] (3.8,6.3) -- (4.,6.9);
\draw [->-, thick] (5.4,6.3) -- (5.6,6.9);
\draw [->-, thick] (5.4,6.3) -- (6.4,6.9);
\draw [->-, thick] (5.6,6.9) -- (6.4,6.9);
\draw [] (3.5,7.7) -- (1.7,5.5);
\draw [] (1.7,5.5) -- (5.9,5.5);
\draw [] (5.9,5.5) -- (7.7,7.7);
\draw [] (7.7,7.7) -- (3.5,7.7);
\draw [gray, dashed] (2.1,10.7) -- (2.45,10.7);
\draw [gray, dashed] (2.65,10.7) -- (2.9,10.7);
\draw [gray, dashed] (2.9,10.7) -- (3.25,10.7);
\draw [gray, dashed] (3.45,10.7) -- (3.7,10.7);
\draw [gray, dashed] (3.7,10.7) -- (4.05,10.7);
\draw [gray, dashed] (4.25,10.7) -- (4.5,10.7);
\draw [gray, dashed] (2.1,10.7) -- (2.497,10.965);
\draw [gray, dashed] (2.663,11.075) -- (3.,11.3);
\draw [gray, dashed] (3.7,10.7) -- (4.097,10.965);
\draw [gray, dashed] (4.263,11.075) -- (4.6,11.3);
\draw [gray, dashed] (4.5,10.7) -- (4.897,10.965);
\draw [gray, dashed] (5.063,11.075) -- (5.4,11.3);
\draw [gray, dashed] (2.9,10.7) -- (2.931,10.884);
\draw [gray, dashed] (2.963,11.081) -- (3.,11.3);
\draw [gray, dashed] (4.5,10.7) -- (4.531,10.884);
\draw [gray, dashed] (4.563,11.081) -- (4.6,11.3);
\draw [gray, dashed] (4.6,11.3) -- (4.95,11.3);
\draw [gray, dashed] (5.15,11.3) -- (5.4,11.3);
\draw [gray, dashed] (3.,11.3) -- (3.414,11.549);
\draw [gray, dashed] (3.586,11.651) -- (4.,11.9);
\draw [gray, dashed] (4.6,11.3) -- (5.014,11.549);
\draw [gray, dashed] (5.186,11.651) -- (5.6,11.9);
\draw [gray, dashed] (4.6,11.3) -- (4.668,11.505);
\draw [gray, dashed] (4.732,11.695) -- (4.8,11.9);
\draw [gray, dashed] (3.,11.3) -- (3.068,11.505);
\draw [gray, dashed] (3.132,11.695) -- (3.2,11.9);
\draw [gray, dashed] (3.2,11.9) -- (3.45,11.9);
\draw [gray, dashed] (3.65,11.9) -- (4.,11.9);
\draw [gray, dashed] (4.,11.9) -- (4.25,11.9);
\draw [gray, dashed] (4.45,11.9) -- (4.8,11.9);
\draw [gray, dashed] (4.8,11.9) -- (5.05,11.9);
\draw [gray, dashed] (5.25,11.9) -- (5.6,11.9);
\draw [gray, dashed] (3.2,11.9) -- (3.537,12.125);
\draw [gray, dashed] (3.703,12.235) -- (4.1,12.5);
\draw [gray, dashed] (4.,11.9) -- (4.337,12.125);
\draw [gray, dashed] (4.503,12.235) -- (4.9,12.5);
\draw [gray, dashed] (4.8,11.9) -- (5.137,12.125);
\draw [gray, dashed] (5.303,12.235) -- (5.7,12.5);
\draw [gray, dashed] (5.6,11.9) -- (5.937,12.125);
\draw [gray, dashed] (6.103,12.235) -- (6.5,12.5);
\draw [gray, dashed] (4.,11.9) -- (4.037,12.119);
\draw [gray, dashed] (4.069,12.316) -- (4.1,12.5);
\draw [gray, dashed] (4.8,11.9) -- (4.837,12.119);
\draw [gray, dashed] (4.869,12.316) -- (4.9,12.5);
\draw [gray, dashed] (5.6,11.9) -- (5.637,12.119);
\draw [gray, dashed] (5.669,12.316) -- (5.7,12.5);
\draw [gray, dashed] (4.1,12.5) -- (4.35,12.5);
\draw [gray, dashed] (4.55,12.5) -- (4.9,12.5);
\draw [gray, dashed] (4.9,12.5) -- (5.15,12.5);
\draw [gray, dashed] (5.35,12.5) -- (5.7,12.5);
\draw [gray, dashed] (5.7,12.5) -- (5.95,12.5);
\draw [gray, dashed] (6.15,12.5) -- (6.5,12.5);
\draw [gray, dashed] (4.5,10.7) -- (4.85,10.7);
\draw [gray, dashed] (5.05,10.7) -- (5.3,10.7);
\draw [gray, dashed] (5.3,10.7) -- (5.697,10.965);
\draw [gray, dashed] (5.863,11.075) -- (6.2,11.3);
\draw [gray, dashed] (5.3,10.7) -- (5.331,10.884);
\draw [gray, dashed] (5.363,11.081) -- (5.4,11.3);
\draw [gray, dashed] (5.4,11.3) -- (5.75,11.3);
\draw [gray, dashed] (5.95,11.3) -- (6.2,11.3);
\draw [gray, dashed] (6.2,11.3) -- (6.268,11.505);
\draw [gray, dashed] (6.332,11.695) -- (6.4,11.9);
\draw [gray, dashed] (6.4,11.9) -- (6.737,12.125);
\draw [gray, dashed] (6.903,12.235) -- (7.3,12.5);
\draw [gray, dashed] (6.4,11.9) -- (6.437,12.119);
\draw [gray, dashed] (6.469,12.316) -- (6.5,12.5);
\draw [gray, dashed] (6.5,12.5) -- (6.75,12.5);
\draw [gray, dashed] (6.95,12.5) -- (7.3,12.5);
\draw [gray, dashed] (3.,11.3) -- (3.35,11.3);
\draw [gray, dashed] (3.55,11.3) -- (3.8,11.3);
\draw [gray, dashed] (3.8,11.3) -- (4.15,11.3);
\draw [gray, dashed] (4.35,11.3) -- (4.6,11.3);
\draw [gray, dashed] (2.9,10.7) -- (3.297,10.965);
\draw [gray, dashed] (3.463,11.075) -- (3.8,11.3);
\draw [gray, dashed] (3.8,11.3) -- (4.214,11.549);
\draw [gray, dashed] (4.386,11.651) -- (4.8,11.9);
\draw [gray, dashed] (3.7,10.7) -- (3.731,10.884);
\draw [gray, dashed] (3.763,11.081) -- (3.8,11.3);
\draw [gray, dashed] (3.8,11.3) -- (3.868,11.505);
\draw [gray, dashed] (3.932,11.695) -- (4.,11.9);
\draw [gray, dashed] (5.4,11.3) -- (5.468,11.505);
\draw [gray, dashed] (5.532,11.695) -- (5.6,11.9);
\draw [gray, dashed] (5.4,11.3) -- (5.814,11.549);
\draw [gray, dashed] (5.986,11.651) -- (6.4,11.9);
\draw [gray, dashed] (5.6,11.9) -- (5.85,11.9);
\draw [gray, dashed] (6.05,11.9) -- (6.4,11.9);
\draw [] (5.6,8.5) -- (5.5,8.5);
\draw [] (5.5,8.5) -- (5.7,8.3);
\draw [] (5.7,8.3) -- (5.9,8.5);
\draw [] (5.9,8.5) -- (5.8,8.5);
\draw [] (5.8,8.5) -- (5.8,9.5);
\draw [] (5.8,9.5) -- (5.9,9.5);
\draw [] (5.9,9.5) -- (5.7,9.7);
\draw [] (5.7,9.7) -- (5.5,9.5);
\draw [] (5.5,9.5) -- (5.6,9.5);
\draw [] (5.6,9.5) -- (5.6,8.5);
\draw [gray, -<-] (2.4,10.5) -- (2.7,10.9);
\draw [gray, -<-] (3.3,11.1) -- (2.7,10.9);
\draw [gray, -<-] (2.7,10.9) -- (2.5,11.1);
\draw [gray, -<-] (3.2,10.5) -- (3.5,10.9);
\draw [gray, -<-] (4.,10.5) -- (4.3,10.9);
\draw [gray, -<-] (4.9,11.1) -- (4.3,10.9);
\draw [gray, -<-] (4.3,10.9) -- (4.1,11.1);
\draw [gray, -<-] (4.8,10.5) -- (5.1,10.9);
\draw [gray, -<-] (5.7,11.1) -- (5.1,10.9);
\draw [gray, -<-] (5.1,10.9) -- (4.9,11.1);
\draw [gray, -<-] (3.6,11.5) -- (3.4,11.7);
\draw [gray, -<-] (5.,11.7) -- (4.4,11.5);
\draw [gray, -<-] (4.9,11.1) -- (5.2,11.5);
\draw [gray, -<-] (5.2,11.5) -- (5.,11.7);
\draw [gray, -<-] (5.7,11.1) -- (6.,11.5);
\draw [gray, -<-] (6.6,11.7) -- (6.,11.5);
\draw [gray, -<-] (3.4,11.7) -- (3.7,12.1);
\draw [gray, -<-] (4.3,12.3) -- (3.7,12.1);
\draw [gray, -<-] (3.7,12.1) -- (3.5,12.3);
\draw [gray, -<-] (4.2,11.7) -- (4.5,12.1);
\draw [gray, -<-] (5.1,12.3) -- (4.5,12.1);
\draw [gray, -<-] (4.5,12.1) -- (4.3,12.3);
\draw [gray, -<-] (5.,11.7) -- (5.3,12.1);
\draw [gray, -<-] (5.9,12.3) -- (5.3,12.1);
\draw [gray, -<-] (5.3,12.1) -- (5.1,12.3);
\draw [gray, -<-] (6.7,12.3) -- (6.1,12.1);
\draw [gray, -<-] (6.1,12.1) -- (5.9,12.3);
\draw [gray, -<-] (3.4,11.7) -- (2.8,11.5);
\draw [gray, -<-] (6.9,12.1) -- (6.7,12.3);
\draw [gray, -<-] (4.3,12.3) -- (4.6,12.7);
\draw [gray, -<-] (5.1,12.3) -- (5.4,12.7);
\draw [gray, -<-] (5.9,12.3) -- (6.2,12.7);
\draw [gray, -<-] (6.7,12.3) -- (7.,12.7);
\draw [gray, -<-] (5.9,10.9) -- (5.7,11.1);
\draw [-<-, thick] (4.1,11.1) -- (3.5,10.9);
\draw [-<-, thick] (3.5,10.9) -- (3.3,11.1);
\draw [-<-, thick] (3.3,11.1) -- (3.6,11.5);
\draw [-<-, thick] (4.2,11.7) -- (3.6,11.5);
\draw [-<-, thick] (4.1,11.1) -- (4.4,11.5);
\draw [-<-, thick] (4.4,11.5) -- (4.2,11.7);
\draw [-<-, thick] (5.8,11.7) -- (5.2,11.5);
\draw [-<-, thick] (6.,11.5) -- (5.8,11.7);
\draw [-<-, thick] (5.8,11.7) -- (6.1,12.1);
\node [rotate=0.] at (4.,11.4) {\scriptsize $\tilde{B}^\rho$};
\node [rotate=0.] at (5.7,11.4) {\scriptsize $\tilde{A}^\eta$};
\node [rotate=0.] at (3.7,10.2) {\scriptsize $\text{QD model}(\hat{H})$};
\node [rotate=0.] at (4.2,9.4) {\scriptsize $\text{coupling}$};
\node [rotate=0.] at (4.2,9.1) {\scriptsize $\text{characterized}$};
\node [rotate=0.] at (3.6,6.5) {\scriptsize $\tilde{A}^x$};
\node [rotate=0.] at (5.8,6.7) {\scriptsize $\tilde{B}^y$};
\node [rotate=0.] at (3.7,5.3) {\scriptsize $K-\epsilon\text{ model}$};
\node [rotate=0.] at (4.2,8.8) {\scriptsize $\text{by }\hat{F}\text{ and }\alpha'$};
\end{tikzpicture}
}

\newcommand{\bilayerModelAG}{
\begin{tikzpicture}[baseline={([yshift=-.8ex]current bounding box.center)},scale=1]
\path [lightgray, fill] (5.3,8.2) -- (6.3,8.8) -- (5.5,8.8) -- cycle;
\path [lightgray, fill] (5.3,11.4) -- (6.3,12.) -- (5.5,12.) -- cycle;
\draw [] (3.4,9.6) -- (1.6,7.4);
\draw [] (1.6,7.4) -- (5.8,7.4);
\draw [] (5.8,7.4) -- (7.6,9.6);
\draw [] (7.6,9.6) -- (3.4,9.6);
\draw [] (2.,7.6) -- (2.8,7.6);
\draw [] (2.8,7.6) -- (3.6,7.6);
\draw [] (3.6,7.6) -- (4.4,7.6);
\draw [] (2.,7.6) -- (2.9,8.2);
\draw [] (3.6,7.6) -- (4.5,8.2);
\draw [] (4.4,7.6) -- (5.3,8.2);
\draw [] (2.8,7.6) -- (2.9,8.2);
\draw [] (4.4,7.6) -- (4.5,8.2);
\draw [] (4.5,8.2) -- (5.3,8.2);
\draw [] (2.9,8.2) -- (3.9,8.8);
\draw [] (4.5,8.2) -- (5.5,8.8);
\draw [] (4.5,8.2) -- (4.7,8.8);
\draw [] (2.9,8.2) -- (3.1,8.8);
\draw [] (3.1,8.8) -- (3.9,8.8);
\draw [] (3.9,8.8) -- (4.7,8.8);
\draw [] (4.7,8.8) -- (5.5,8.8);
\draw [] (3.1,8.8) -- (4.,9.4);
\draw [] (3.9,8.8) -- (4.8,9.4);
\draw [] (4.7,8.8) -- (5.6,9.4);
\draw [] (5.5,8.8) -- (6.4,9.4);
\draw [] (3.9,8.8) -- (4.,9.4);
\draw [] (4.7,8.8) -- (4.8,9.4);
\draw [] (5.5,8.8) -- (5.6,9.4);
\draw [] (4.,9.4) -- (4.8,9.4);
\draw [] (4.8,9.4) -- (5.6,9.4);
\draw [] (5.6,9.4) -- (6.4,9.4);
\draw [] (4.4,7.6) -- (5.2,7.6);
\draw [] (5.2,7.6) -- (6.1,8.2);
\draw [] (5.2,7.6) -- (5.3,8.2);
\draw [] (5.3,8.2) -- (6.1,8.2);
\draw [] (6.1,8.2) -- (6.3,8.8);
\draw [] (6.3,8.8) -- (7.2,9.4);
\draw [] (6.3,8.8) -- (6.4,9.4);
\draw [] (6.4,9.4) -- (7.2,9.4);
\draw [] (2.9,8.2) -- (3.7,8.2);
\draw [] (3.7,8.2) -- (4.5,8.2);
\draw [] (2.8,7.6) -- (3.7,8.2);
\draw [] (3.7,8.2) -- (4.7,8.8);
\draw [] (3.6,7.6) -- (3.7,8.2);
\draw [] (3.7,8.2) -- (3.9,8.8);
\draw [] (5.3,8.2) -- (5.5,8.8);
\draw [] (5.3,8.2) -- (6.3,8.8);
\draw [] (5.5,8.8) -- (6.3,8.8);
\draw [] (3.4,12.8) -- (1.6,10.6);
\draw [] (1.6,10.6) -- (5.8,10.6);
\draw [] (5.8,10.6) -- (7.6,12.8);
\draw [] (7.6,12.8) -- (3.4,12.8);
\draw [] (2.,10.8) -- (2.8,10.8);
\draw [] (2.8,10.8) -- (3.6,10.8);
\draw [] (3.6,10.8) -- (4.4,10.8);
\draw [] (2.,10.8) -- (2.9,11.4);
\draw [] (3.6,10.8) -- (4.5,11.4);
\draw [] (4.4,10.8) -- (5.3,11.4);
\draw [] (2.8,10.8) -- (2.9,11.4);
\draw [] (4.4,10.8) -- (4.5,11.4);
\draw [] (4.5,11.4) -- (5.3,11.4);
\draw [] (2.9,11.4) -- (3.9,12.);
\draw [] (4.5,11.4) -- (5.5,12.);
\draw [] (4.5,11.4) -- (4.7,12.);
\draw [] (2.9,11.4) -- (3.1,12.);
\draw [] (3.1,12.) -- (3.9,12.);
\draw [] (3.9,12.) -- (4.7,12.);
\draw [] (4.7,12.) -- (5.5,12.);
\draw [] (3.1,12.) -- (4.,12.6);
\draw [] (3.9,12.) -- (4.8,12.6);
\draw [] (4.7,12.) -- (5.6,12.6);
\draw [] (5.5,12.) -- (6.4,12.6);
\draw [] (3.9,12.) -- (4.,12.6);
\draw [] (4.7,12.) -- (4.8,12.6);
\draw [] (5.5,12.) -- (5.6,12.6);
\draw [] (4.,12.6) -- (4.8,12.6);
\draw [] (4.8,12.6) -- (5.6,12.6);
\draw [] (5.6,12.6) -- (6.4,12.6);
\draw [] (4.4,10.8) -- (5.2,10.8);
\draw [] (5.2,10.8) -- (6.1,11.4);
\draw [] (5.2,10.8) -- (5.3,11.4);
\draw [] (5.3,11.4) -- (6.1,11.4);
\draw [] (6.1,11.4) -- (6.3,12.);
\draw [] (6.3,12.) -- (7.2,12.6);
\draw [] (6.3,12.) -- (6.4,12.6);
\draw [] (6.4,12.6) -- (7.2,12.6);
\draw [] (2.9,11.4) -- (3.7,11.4);
\draw [] (3.7,11.4) -- (4.5,11.4);
\draw [] (2.8,10.8) -- (3.7,11.4);
\draw [] (3.7,11.4) -- (4.7,12.);
\draw [] (3.6,10.8) -- (3.7,11.4);
\draw [] (3.7,11.4) -- (3.9,12.);
\draw [] (5.3,11.4) -- (5.5,12.);
\draw [] (5.3,11.4) -- (6.3,12.);
\draw [] (5.5,12.) -- (6.3,12.);
\draw [fill,rotate around={0.:(3.7,8.2)}] (3.7,8.2) ++(0.:0.08 and 0.08) arc(0.:360.:0.08 and 0.08);
\draw [fill,rotate around={0.:(3.7,11.4)}] (3.7,11.4) ++(0.:0.08 and 0.08) arc(0.:360.:0.08 and 0.08);
\node [rotate=0.] at (4.2,8.4) {\scriptsize $K\text{-charge}$};
\node [rotate=0.] at (5.7,8.6) {\scriptsize $K\text{-flux}$};
\node [rotate=0.] at (4.,7.1) {\scriptsize $\text{QD}(K=\mathbb{Z}_2)\text{ model on }\Gamma$};
\node [rotate=0.] at (4.,10.3) {\scriptsize $\text{QD}(N=\mathbb{Z}_2)\text{ model on }{\Gamma}$};
\node [rotate=0.] at (4.2,11.6) {\scriptsize $N\text{-charge}$};
\node [rotate=0.] at (5.7,11.8) {\scriptsize $N\text{-flux}$};
\end{tikzpicture}
}

\newcommand{\bilayerModelAH}{
\begin{tikzpicture}[baseline={([yshift=-.8ex]current bounding box.center)},scale=1]
\path [lightgray, fill] (3.3,11.1) -- (3.5,10.9) -- (4.1,11.1) -- (4.4,11.5) -- (4.2,11.7) -- (3.6,11.5) -- cycle;
\path [lightgray, fill] (5.3,8.2) -- (6.3,8.8) -- (5.5,8.8) -- cycle;
\draw [] (3.7,12.8) -- (1.6,10.4);
\draw [] (1.6,10.4) -- (5.8,10.4);
\draw [] (5.8,10.4) -- (7.9,12.8);
\draw [] (7.9,12.8) -- (3.7,12.8);
\draw [] (3.4,9.6) -- (1.6,7.4);
\draw [] (1.6,7.4) -- (5.8,7.4);
\draw [] (5.8,7.4) -- (7.6,9.6);
\draw [] (7.6,9.6) -- (3.4,9.6);
\draw [] (2.,7.6) -- (2.8,7.6);
\draw [] (2.8,7.6) -- (3.6,7.6);
\draw [] (3.6,7.6) -- (4.4,7.6);
\draw [] (2.,7.6) -- (2.9,8.2);
\draw [] (3.6,7.6) -- (4.5,8.2);
\draw [] (4.4,7.6) -- (5.3,8.2);
\draw [] (2.8,7.6) -- (2.9,8.2);
\draw [] (4.4,7.6) -- (4.5,8.2);
\draw [] (4.5,8.2) -- (5.3,8.2);
\draw [] (2.9,8.2) -- (3.9,8.8);
\draw [] (4.5,8.2) -- (5.5,8.8);
\draw [] (4.5,8.2) -- (4.7,8.8);
\draw [] (2.9,8.2) -- (3.1,8.8);
\draw [] (3.1,8.8) -- (3.9,8.8);
\draw [] (3.9,8.8) -- (4.7,8.8);
\draw [] (4.7,8.8) -- (5.5,8.8);
\draw [] (3.1,8.8) -- (4.,9.4);
\draw [] (3.9,8.8) -- (4.8,9.4);
\draw [] (4.7,8.8) -- (5.6,9.4);
\draw [] (5.5,8.8) -- (6.4,9.4);
\draw [] (3.9,8.8) -- (4.,9.4);
\draw [] (4.7,8.8) -- (4.8,9.4);
\draw [] (5.5,8.8) -- (5.6,9.4);
\draw [] (4.,9.4) -- (4.8,9.4);
\draw [] (4.8,9.4) -- (5.6,9.4);
\draw [] (5.6,9.4) -- (6.4,9.4);
\draw [] (4.4,7.6) -- (5.2,7.6);
\draw [] (5.2,7.6) -- (6.1,8.2);
\draw [] (5.2,7.6) -- (5.3,8.2);
\draw [] (5.3,8.2) -- (6.1,8.2);
\draw [] (6.1,8.2) -- (6.3,8.8);
\draw [] (6.3,8.8) -- (7.2,9.4);
\draw [] (6.3,8.8) -- (6.4,9.4);
\draw [] (6.4,9.4) -- (7.2,9.4);
\draw [] (2.9,8.2) -- (3.7,8.2);
\draw [] (3.7,8.2) -- (4.5,8.2);
\draw [] (2.8,7.6) -- (3.7,8.2);
\draw [] (3.7,8.2) -- (4.7,8.8);
\draw [] (3.6,7.6) -- (3.7,8.2);
\draw [] (3.7,8.2) -- (3.9,8.8);
\draw [] (5.3,8.2) -- (5.5,8.8);
\draw [] (5.3,8.2) -- (6.3,8.8);
\draw [] (5.5,8.8) -- (6.3,8.8);
\draw [lightgray, dashed] (2.1,10.7) -- (2.45,10.7);
\draw [lightgray, dashed] (2.65,10.7) -- (2.9,10.7);
\draw [lightgray, dashed] (2.9,10.7) -- (3.25,10.7);
\draw [lightgray, dashed] (3.45,10.7) -- (3.7,10.7);
\draw [lightgray, dashed] (3.7,10.7) -- (4.05,10.7);
\draw [lightgray, dashed] (4.25,10.7) -- (4.5,10.7);
\draw [lightgray, dashed] (2.1,10.7) -- (2.497,10.965);
\draw [lightgray, dashed] (2.663,11.075) -- (3.,11.3);
\draw [lightgray, dashed] (3.7,10.7) -- (4.097,10.965);
\draw [lightgray, dashed] (4.263,11.075) -- (4.6,11.3);
\draw [lightgray, dashed] (4.5,10.7) -- (4.897,10.965);
\draw [lightgray, dashed] (5.063,11.075) -- (5.4,11.3);
\draw [lightgray, dashed] (2.9,10.7) -- (2.931,10.884);
\draw [lightgray, dashed] (2.963,11.081) -- (3.,11.3);
\draw [lightgray, dashed] (4.5,10.7) -- (4.531,10.884);
\draw [lightgray, dashed] (4.563,11.081) -- (4.6,11.3);
\draw [lightgray, dashed] (4.6,11.3) -- (4.95,11.3);
\draw [lightgray, dashed] (5.15,11.3) -- (5.4,11.3);
\draw [lightgray, dashed] (3.,11.3) -- (3.414,11.549);
\draw [lightgray, dashed] (3.586,11.651) -- (4.,11.9);
\draw [lightgray, dashed] (4.6,11.3) -- (5.014,11.549);
\draw [lightgray, dashed] (5.186,11.651) -- (5.6,11.9);
\draw [lightgray, dashed] (4.6,11.3) -- (4.668,11.505);
\draw [lightgray, dashed] (4.732,11.695) -- (4.8,11.9);
\draw [lightgray, dashed] (3.,11.3) -- (3.068,11.505);
\draw [lightgray, dashed] (3.132,11.695) -- (3.2,11.9);
\draw [lightgray, dashed] (3.2,11.9) -- (3.45,11.9);
\draw [lightgray, dashed] (3.65,11.9) -- (4.,11.9);
\draw [lightgray, dashed] (4.,11.9) -- (4.25,11.9);
\draw [lightgray, dashed] (4.45,11.9) -- (4.8,11.9);
\draw [lightgray, dashed] (4.8,11.9) -- (5.05,11.9);
\draw [lightgray, dashed] (5.25,11.9) -- (5.6,11.9);
\draw [lightgray, dashed] (3.2,11.9) -- (3.537,12.125);
\draw [lightgray, dashed] (3.703,12.235) -- (4.1,12.5);
\draw [lightgray, dashed] (4.,11.9) -- (4.337,12.125);
\draw [lightgray, dashed] (4.503,12.235) -- (4.9,12.5);
\draw [lightgray, dashed] (4.8,11.9) -- (5.137,12.125);
\draw [lightgray, dashed] (5.303,12.235) -- (5.7,12.5);
\draw [lightgray, dashed] (5.6,11.9) -- (5.937,12.125);
\draw [lightgray, dashed] (6.103,12.235) -- (6.5,12.5);
\draw [lightgray, dashed] (4.,11.9) -- (4.037,12.119);
\draw [lightgray, dashed] (4.069,12.316) -- (4.1,12.5);
\draw [lightgray, dashed] (4.8,11.9) -- (4.837,12.119);
\draw [lightgray, dashed] (4.869,12.316) -- (4.9,12.5);
\draw [lightgray, dashed] (5.6,11.9) -- (5.637,12.119);
\draw [lightgray, dashed] (5.669,12.316) -- (5.7,12.5);
\draw [lightgray, dashed] (4.1,12.5) -- (4.35,12.5);
\draw [lightgray, dashed] (4.55,12.5) -- (4.9,12.5);
\draw [lightgray, dashed] (4.9,12.5) -- (5.15,12.5);
\draw [lightgray, dashed] (5.35,12.5) -- (5.7,12.5);
\draw [lightgray, dashed] (5.7,12.5) -- (5.95,12.5);
\draw [lightgray, dashed] (6.15,12.5) -- (6.5,12.5);
\draw [lightgray, dashed] (4.5,10.7) -- (4.85,10.7);
\draw [lightgray, dashed] (5.05,10.7) -- (5.3,10.7);
\draw [lightgray, dashed] (5.3,10.7) -- (5.697,10.965);
\draw [lightgray, dashed] (5.863,11.075) -- (6.2,11.3);
\draw [lightgray, dashed] (5.3,10.7) -- (5.331,10.884);
\draw [lightgray, dashed] (5.363,11.081) -- (5.4,11.3);
\draw [lightgray, dashed] (5.4,11.3) -- (5.75,11.3);
\draw [lightgray, dashed] (5.95,11.3) -- (6.2,11.3);
\draw [lightgray, dashed] (6.2,11.3) -- (6.268,11.505);
\draw [lightgray, dashed] (6.332,11.695) -- (6.4,11.9);
\draw [lightgray, dashed] (6.4,11.9) -- (6.737,12.125);
\draw [lightgray, dashed] (6.903,12.235) -- (7.3,12.5);
\draw [lightgray, dashed] (6.4,11.9) -- (6.437,12.119);
\draw [lightgray, dashed] (6.469,12.316) -- (6.5,12.5);
\draw [lightgray, dashed] (6.5,12.5) -- (6.75,12.5);
\draw [lightgray, dashed] (6.95,12.5) -- (7.3,12.5);
\draw [lightgray, dashed] (3.,11.3) -- (3.35,11.3);
\draw [lightgray, dashed] (3.55,11.3) -- (3.8,11.3);
\draw [lightgray, dashed] (3.8,11.3) -- (4.15,11.3);
\draw [lightgray, dashed] (4.35,11.3) -- (4.6,11.3);
\draw [lightgray, dashed] (2.9,10.7) -- (3.297,10.965);
\draw [lightgray, dashed] (3.463,11.075) -- (3.8,11.3);
\draw [lightgray, dashed] (3.8,11.3) -- (4.214,11.549);
\draw [lightgray, dashed] (4.386,11.651) -- (4.8,11.9);
\draw [lightgray, dashed] (3.7,10.7) -- (3.731,10.884);
\draw [lightgray, dashed] (3.763,11.081) -- (3.8,11.3);
\draw [lightgray, dashed] (3.8,11.3) -- (3.868,11.505);
\draw [lightgray, dashed] (3.932,11.695) -- (4.,11.9);
\draw [lightgray, dashed] (5.4,11.3) -- (5.468,11.505);
\draw [lightgray, dashed] (5.532,11.695) -- (5.6,11.9);
\draw [lightgray, dashed] (5.4,11.3) -- (5.814,11.549);
\draw [lightgray, dashed] (5.986,11.651) -- (6.4,11.9);
\draw [lightgray, dashed] (5.6,11.9) -- (5.85,11.9);
\draw [lightgray, dashed] (6.05,11.9) -- (6.4,11.9);
\draw [] (2.4,10.5) -- (2.7,10.9);
\draw [] (3.3,11.1) -- (2.7,10.9);
\draw [] (2.7,10.9) -- (2.5,11.1);
\draw [] (3.2,10.5) -- (3.5,10.9);
\draw [] (4.,10.5) -- (4.3,10.9);
\draw [] (4.9,11.1) -- (4.3,10.9);
\draw [] (4.3,10.9) -- (4.1,11.1);
\draw [] (4.8,10.5) -- (5.1,10.9);
\draw [] (5.7,11.1) -- (5.1,10.9);
\draw [] (5.1,10.9) -- (4.9,11.1);
\draw [] (3.6,11.5) -- (3.4,11.7);
\draw [] (5.,11.7) -- (4.4,11.5);
\draw [] (4.9,11.1) -- (5.2,11.5);
\draw [] (5.2,11.5) -- (5.,11.7);
\draw [] (5.7,11.1) -- (6.,11.5);
\draw [] (6.6,11.7) -- (6.,11.5);
\draw [] (3.4,11.7) -- (3.7,12.1);
\draw [] (4.3,12.3) -- (3.7,12.1);
\draw [] (3.7,12.1) -- (3.5,12.3);
\draw [] (4.2,11.7) -- (4.5,12.1);
\draw [] (5.1,12.3) -- (4.5,12.1);
\draw [] (4.5,12.1) -- (4.3,12.3);
\draw [] (5.,11.7) -- (5.3,12.1);
\draw [] (5.9,12.3) -- (5.3,12.1);
\draw [] (5.3,12.1) -- (5.1,12.3);
\draw [] (6.7,12.3) -- (6.1,12.1);
\draw [] (6.1,12.1) -- (5.9,12.3);
\draw [] (3.4,11.7) -- (2.8,11.5);
\draw [] (6.9,12.1) -- (6.7,12.3);
\draw [] (4.3,12.3) -- (4.6,12.7);
\draw [] (5.1,12.3) -- (5.4,12.7);
\draw [] (5.9,12.3) -- (6.2,12.7);
\draw [] (6.7,12.3) -- (7.,12.7);
\draw [] (5.9,10.9) -- (5.7,11.1);
\draw [] (4.1,11.1) -- (3.5,10.9);
\draw [] (3.5,10.9) -- (3.3,11.1);
\draw [] (3.3,11.1) -- (3.6,11.5);
\draw [] (4.2,11.7) -- (3.6,11.5);
\draw [] (4.1,11.1) -- (4.4,11.5);
\draw [] (4.4,11.5) -- (4.2,11.7);
\draw [] (5.8,11.7) -- (5.2,11.5);
\draw [] (6.,11.5) -- (5.8,11.7);
\draw [] (5.8,11.7) -- (6.1,12.1);
\draw [fill,rotate around={0.:(5.8,11.7)}] (5.8,11.7) ++(0.:0.08 and 0.08) arc(0.:360.:0.08 and 0.08);
\draw [fill,rotate around={0.:(3.7,8.2)}] (3.7,8.2) ++(0.:0.08 and 0.08) arc(0.:360.:0.08 and 0.08);
\node [rotate=0.] at (4.2,8.4) {\scriptsize $K\text{-charge}$};
\node [rotate=0.] at (5.7,8.6) {\scriptsize $K\text{-flux}$};
\node [rotate=0.] at (4.,7.1) {\scriptsize $\text{QD}(K=\mathbb{Z}_2)\text{ model on }\Gamma$};
\node [rotate=0.] at (4.,10.1) {\scriptsize $\text{QD}(\hat{N}=\mathbb{Z}_2)\text{ model on }\tilde{\Gamma}$};
\node [rotate=0.] at (6.2,11.) {\scriptsize $\hat{N}\text{-charge}$};
\node [rotate=0.] at (3.9,11.4) {\scriptsize $\hat{N}\text{-flux}$};
\end{tikzpicture}
}

\newcommand{\configurationKAA}{
\begin{tikzpicture}[baseline={([yshift=-.8ex]current bounding box.center)},scale=1]
\draw [->-] (2.2,10.4) -- (3.4,10.4);
\draw [->-] (3.4,10.4) -- (4.,11.6);
\draw [->-] (4.,11.6) -- (3.4,12.8);
\draw [->-] (2.2,10.4) -- (1.6,11.6);
\draw [->-] (1.6,11.6) -- (2.2,12.8);
\draw [->-] (2.2,12.8) -- (3.4,12.8);
\draw [->-] (2.2,10.4) -- (2.8,11.6);
\draw [->-] (2.8,11.6) -- (3.4,12.8);
\draw [->-] (1.6,11.6) -- (2.8,11.6);
\draw [->-] (2.8,11.6) -- (4.,11.6);
\draw [->-] (3.4,10.4) -- (2.8,11.6);
\draw [->-] (2.8,11.6) -- (2.2,12.8);
\node [rotate=0.] at (2.2,11.8) {\scriptsize $k_{1}$};
\node [rotate=0.] at (2.3,11.) {\scriptsize $k_{2}$};
\node [rotate=0.] at (3.,10.8) {\scriptsize $k_{3}$};
\node [rotate=0.] at (3.5,11.4) {\scriptsize $k_{4}$};
\node [rotate=0.] at (3.3,12.2) {\scriptsize $k_{5}$};
\node [rotate=0.] at (2.6,12.4) {\scriptsize $k_{6}$};
\end{tikzpicture}
}

\newcommand{\configurationKAB}{
\begin{tikzpicture}[baseline={([yshift=-.8ex]current bounding box.center)},scale=1]
\draw [->-] (2.2,10.4) -- (3.4,10.4);
\draw [->-] (3.4,10.4) -- (4.,11.6);
\draw [->-] (4.,11.6) -- (3.4,12.8);
\draw [->-] (2.2,10.4) -- (1.6,11.6);
\draw [->-] (1.6,11.6) -- (2.2,12.8);
\draw [->-] (2.2,12.8) -- (3.4,12.8);
\draw [->-] (2.2,10.4) -- (2.8,11.6);
\draw [->-] (2.8,11.6) -- (3.4,12.8);
\draw [->-] (1.6,11.6) -- (2.8,11.6);
\draw [->-] (2.8,11.6) -- (4.,11.6);
\draw [->-] (3.4,10.4) -- (2.8,11.6);
\draw [->-] (2.8,11.6) -- (2.2,12.8);
\node [rotate=0.] at (2.2,11.8) {\scriptsize $yk_{1}$};
\node [rotate=0.] at (2.3,11.) {\scriptsize $yk_{2}$};
\node [rotate=0.] at (3.,10.8) {\scriptsize $yk_{3}$};
\node [rotate=0.] at (3.5,11.4) {\scriptsize $k_{4}y^{-1}$};
\node [rotate=0.] at (3.3,12.2) {\scriptsize $k_{5}y^{-1}$};
\node [rotate=0.] at (2.6,12.4) {\scriptsize $k_{6}y^{-1}$};
\end{tikzpicture}
}

\newcommand{\DoubleLayerGraph}{
\begin{tikzpicture}[baseline={([yshift=-.8ex]current bounding box.center)},scale=1]
\draw [dashed] (2.,10.6) -- (2.,10.7);
\draw [dashed] (2.,10.9) -- (2.,11.);
\draw [dashed] (2.8,11.8) -- (2.689,11.745);
\draw [dashed] (2.511,11.655) -- (2.4,11.6);
\draw [dashed] (2.8,11.8) -- (2.8,11.9);
\draw [dashed] (2.8,12.1) -- (2.8,12.2);
\draw [dashed] (3.2,11.6) -- (3.089,11.655);
\draw [dashed] (2.911,11.745) -- (2.8,11.8);
\draw [dashed] (3.2,11.2) -- (3.2,11.3);
\draw [dashed] (3.2,11.5) -- (3.2,11.6);
\draw [dashed] (3.2,11.2) -- (3.089,11.145);
\draw [dashed] (2.911,11.055) -- (2.8,11.);
\draw [dashed] (2.4,11.2) -- (2.4,11.3);
\draw [dashed] (2.4,11.5) -- (2.4,11.6);
\draw [dashed] (2.8,11.) -- (2.689,11.055);
\draw [dashed] (2.511,11.145) -- (2.4,11.2);
\draw [dashed] (2.4,11.2) -- (2.289,11.145);
\draw [dashed] (2.111,11.055) -- (2.,11.);
\draw [dashed] (2.8,10.6) -- (2.8,10.7);
\draw [dashed] (2.8,10.9) -- (2.8,11.);
\draw [dashed] (3.6,11.8) -- (3.489,11.745);
\draw [dashed] (3.311,11.655) -- (3.2,11.6);
\draw [dashed] (3.6,11.) -- (3.489,11.055);
\draw [dashed] (3.311,11.145) -- (3.2,11.2);
\draw [dashed] (2.4,11.6) -- (2.289,11.655);
\draw [dashed] (2.111,11.745) -- (2.,11.8);
\draw [dashed] (4.4,11.8) -- (4.4,11.9);
\draw [dashed] (4.4,12.1) -- (4.4,12.2);
\draw [dashed] (4.4,11.8) -- (4.289,11.745);
\draw [dashed] (4.111,11.655) -- (4.,11.6);
\draw [dashed] (4.8,11.6) -- (4.689,11.655);
\draw [dashed] (4.511,11.745) -- (4.4,11.8);
\path [lightgray, fill] (4.,12.) -- (4.8,12.) -- (4.4,11.4) -- cycle;
\draw [dashed] (2.,11.) -- (1.889,11.055);
\draw [dashed] (1.711,11.145) -- (1.6,11.2);
\draw [dashed] (3.6,10.6) -- (3.6,10.7);
\draw [dashed] (3.6,10.9) -- (3.6,11.);
\draw [dashed] (4.,11.2) -- (3.889,11.145);
\draw [dashed] (3.711,11.055) -- (3.6,11.);
\draw [dashed] (4.4,10.6) -- (4.4,10.7);
\draw [dashed] (4.4,10.9) -- (4.4,11.);
\draw [dashed] (4.8,11.2) -- (4.689,11.145);
\draw [dashed] (4.511,11.055) -- (4.4,11.);
\draw [dashed] (4.4,11.) -- (4.289,11.055);
\draw [dashed] (4.111,11.145) -- (4.,11.2);
\draw [dashed] (4.,11.2) -- (4.,11.3);
\draw [dashed] (4.,11.5) -- (4.,11.6);
\draw [dashed] (4.,11.6) -- (3.889,11.655);
\draw [dashed] (3.711,11.745) -- (3.6,11.8);
\draw [dashed] (4.8,11.2) -- (4.8,11.3);
\draw [dashed] (4.8,11.5) -- (4.8,11.6);
\draw [dashed] (5.2,11.8) -- (5.089,11.745);
\draw [dashed] (4.911,11.655) -- (4.8,11.6);
\draw [dashed] (2.,11.8) -- (2.,11.9);
\draw [dashed] (2.,12.1) -- (2.,12.2);
\draw [dashed] (2.4,12.4) -- (2.289,12.345);
\draw [dashed] (2.111,12.255) -- (2.,12.2);
\draw [dashed] (2.,12.2) -- (1.889,12.255);
\draw [dashed] (1.711,12.345) -- (1.6,12.4);
\draw [dashed] (3.2,12.4) -- (3.089,12.345);
\draw [dashed] (2.911,12.255) -- (2.8,12.2);
\draw [dashed] (2.8,12.2) -- (2.689,12.255);
\draw [dashed] (2.511,12.345) -- (2.4,12.4);
\draw [dashed] (3.6,11.8) -- (3.6,11.9);
\draw [dashed] (3.6,12.1) -- (3.6,12.2);
\draw [dashed] (4.,12.4) -- (3.889,12.345);
\draw [dashed] (3.711,12.255) -- (3.6,12.2);
\draw [dashed] (3.6,12.2) -- (3.489,12.255);
\draw [dashed] (3.311,12.345) -- (3.2,12.4);
\draw [dashed] (4.8,12.4) -- (4.689,12.345);
\draw [dashed] (4.511,12.255) -- (4.4,12.2);
\draw [dashed] (4.4,12.2) -- (4.289,12.255);
\draw [dashed] (4.111,12.345) -- (4.,12.4);
\draw [dashed] (2.,11.8) -- (1.889,11.745);
\draw [dashed] (1.711,11.655) -- (1.6,11.6);
\draw [dashed] (5.2,12.2) -- (5.089,12.255);
\draw [dashed] (4.911,12.345) -- (4.8,12.4);
\draw [dashed] (2.4,12.4) -- (2.4,12.5);
\draw [dashed] (2.4,12.7) -- (2.4,12.8);
\draw [dashed] (3.2,12.4) -- (3.2,12.5);
\draw [dashed] (3.2,12.7) -- (3.2,12.8);
\draw [dashed] (4.,12.4) -- (4.,12.5);
\draw [dashed] (4.,12.7) -- (4.,12.8);
\draw [dashed] (4.8,12.4) -- (4.8,12.5);
\draw [dashed] (4.8,12.7) -- (4.8,12.8);
\draw [dashed] (5.2,11.) -- (5.089,11.055);
\draw [dashed] (4.911,11.145) -- (4.8,11.2);
\draw [] (4.4,11.4) -- (4.8,12.);
\draw [] (4.,12.) -- (4.8,12.);
\draw [] (1.6,10.8) -- (2.4,10.8);
\draw [] (2.4,10.8) -- (3.2,10.8);
\draw [] (3.2,10.8) -- (4.,10.8);
\draw [] (1.6,10.8) -- (2.,11.4);
\draw [] (3.2,10.8) -- (3.6,11.4);
\draw [] (4.,10.8) -- (4.4,11.4);
\draw [] (2.4,10.8) -- (2.,11.4);
\draw [] (4.,10.8) -- (3.6,11.4);
\draw [] (3.6,11.4) -- (4.4,11.4);
\draw [] (2.,11.4) -- (2.4,12.);
\draw [] (3.6,11.4) -- (4.,12.);
\draw [] (3.6,11.4) -- (3.2,12.);
\draw [] (2.,11.4) -- (1.6,12.);
\draw [] (1.6,12.) -- (2.4,12.);
\draw [] (2.4,12.) -- (3.2,12.);
\draw [] (3.2,12.) -- (4.,12.);
\draw [] (1.6,12.) -- (2.,12.6);
\draw [] (2.4,12.) -- (2.8,12.6);
\draw [] (3.2,12.) -- (3.6,12.6);
\draw [] (4.,12.) -- (4.4,12.6);
\draw [] (2.4,12.) -- (2.,12.6);
\draw [] (3.2,12.) -- (2.8,12.6);
\draw [] (4.,12.) -- (3.6,12.6);
\draw [] (2.,12.6) -- (2.8,12.6);
\draw [] (2.8,12.6) -- (3.6,12.6);
\draw [] (3.6,12.6) -- (4.4,12.6);
\draw [] (4.,10.8) -- (4.8,10.8);
\draw [] (4.8,10.8) -- (5.2,11.4);
\draw [] (4.8,10.8) -- (4.4,11.4);
\draw [] (4.4,11.4) -- (5.2,11.4);
\draw [] (5.2,11.4) -- (4.8,12.);
\draw [] (4.8,12.) -- (5.2,12.6);
\draw [] (4.8,12.) -- (4.4,12.6);
\draw [] (4.4,12.6) -- (5.2,12.6);
\draw [] (2.,11.4) -- (2.8,11.4);
\draw [] (2.8,11.4) -- (3.6,11.4);
\draw [] (2.4,10.8) -- (2.8,11.4);
\draw [] (2.8,11.4) -- (3.2,12.);
\draw [] (3.2,10.8) -- (2.8,11.4);
\draw [] (2.8,11.4) -- (2.4,12.);
\draw [] (4.4,11.4) -- (4.,12.);
\draw [fill,rotate around={0.:(2.8,11.4)}] (2.8,11.4) ++(0.:0.08 and 0.08) arc(0.:360.:0.08 and 0.08);
\node [rotate=0.] at (3.1,11.5) {\scriptsize $\tilde{A}_v^x$};
\node [rotate=0.] at (4.5,11.8) {\scriptsize $\tilde{B}_p^{y}$};
\end{tikzpicture}
}

\newcommand{\DoubleLayerGraphAA}{
\begin{tikzpicture}[baseline={([yshift=-.8ex]current bounding box.center)},scale=1]
\draw [fill,rotate around={0.:(4.4,11.8)}] (4.4,11.8) ++(0.:0.08 and 0.08) arc(0.:360.:0.08 and 0.08);
\path [lightgray, fill] (2.4,11.2) -- (2.8,11.) -- (3.2,11.2) -- (3.2,11.6) -- (2.8,11.8) -- (2.4,11.6) -- cycle;
\draw [dashed] (2.,10.6) -- (2.,11.);
\draw [dashed] (2.4,11.2) -- (2.,11.);
\draw [dashed] (2.,11.) -- (1.6,11.2);
\draw [dashed] (2.8,10.6) -- (2.8,11.);
\draw [dashed] (3.6,10.6) -- (3.6,11.);
\draw [dashed] (4.,11.2) -- (3.6,11.);
\draw [dashed] (3.6,11.) -- (3.2,11.2);
\draw [dashed] (4.4,10.6) -- (4.4,11.);
\draw [dashed] (4.8,11.2) -- (4.4,11.);
\draw [dashed] (4.4,11.) -- (4.,11.2);
\draw [dashed] (2.4,11.6) -- (2.,11.8);
\draw [dashed] (3.6,11.8) -- (3.2,11.6);
\draw [dashed] (4.,11.2) -- (4.,11.6);
\draw [dashed] (4.,11.6) -- (3.6,11.8);
\draw [dashed] (4.8,11.2) -- (4.8,11.6);
\draw [dashed] (5.2,11.8) -- (4.8,11.6);
\draw [dashed] (2.,11.8) -- (2.,12.2);
\draw [dashed] (2.4,12.4) -- (2.,12.2);
\draw [dashed] (2.,12.2) -- (1.6,12.4);
\draw [dashed] (2.8,11.8) -- (2.8,12.2);
\draw [dashed] (3.2,12.4) -- (2.8,12.2);
\draw [dashed] (2.8,12.2) -- (2.4,12.4);
\draw [dashed] (3.6,11.8) -- (3.6,12.2);
\draw [dashed] (4.,12.4) -- (3.6,12.2);
\draw [dashed] (3.6,12.2) -- (3.2,12.4);
\draw [dashed] (4.8,12.4) -- (4.4,12.2);
\draw [dashed] (4.4,12.2) -- (4.,12.4);
\draw [dashed] (2.,11.8) -- (1.6,11.6);
\draw [dashed] (5.2,12.2) -- (4.8,12.4);
\draw [dashed] (2.4,12.4) -- (2.4,12.8);
\draw [dashed] (3.2,12.4) -- (3.2,12.8);
\draw [dashed] (4.,12.4) -- (4.,12.8);
\draw [dashed] (4.8,12.4) -- (4.8,12.8);
\draw [dashed] (5.2,11.) -- (4.8,11.2);
\draw [dashed] (3.2,11.2) -- (2.8,11.);
\draw [dashed] (2.8,11.) -- (2.4,11.2);
\draw [dashed] (2.4,11.2) -- (2.4,11.6);
\draw [dashed] (2.8,11.8) -- (2.4,11.6);
\draw [dashed] (3.2,11.2) -- (3.2,11.6);
\draw [dashed] (3.2,11.6) -- (2.8,11.8);
\draw [dashed] (4.4,11.8) -- (4.,11.6);
\draw [dashed] (4.8,11.6) -- (4.4,11.8);
\draw [dashed] (4.4,11.8) -- (4.4,12.2);
\draw [fill,rotate around={0.:(4.4,11.8)}] (4.4,11.8) ++(0.:0.08 and 0.08) arc(0.:360.:0.08 and 0.08);
\node [rotate=0.] at (2.8,11.4) {\scriptsize $\tilde{B}_p^{\eta}$};
\node [rotate=0.] at (4.6,11.9) {\scriptsize $\tilde{A}_{\tilde{v}}^{\rho}$};
\end{tikzpicture}
}

\newcommand{\DoubleLayerGraphAB}{
\begin{tikzpicture}[baseline={([yshift=-.8ex]current bounding box.center)},scale=1]
\draw [dashed] (2.,10.6) -- (2.,10.7);
\draw [dashed] (2.,10.9) -- (2.,11.);
\draw [dashed] (2.8,11.8) -- (2.689,11.745);
\draw [dashed] (2.511,11.655) -- (2.4,11.6);
\draw [dashed] (2.8,11.8) -- (2.8,11.9);
\draw [dashed] (2.8,12.1) -- (2.8,12.2);
\draw [dashed] (3.2,11.6) -- (3.089,11.655);
\draw [dashed] (2.911,11.745) -- (2.8,11.8);
\draw [dashed] (3.2,11.2) -- (3.2,11.3);
\draw [dashed] (3.2,11.5) -- (3.2,11.6);
\draw [dashed] (3.2,11.2) -- (3.089,11.145);
\draw [dashed] (2.911,11.055) -- (2.8,11.);
\draw [dashed] (2.4,11.2) -- (2.4,11.3);
\draw [dashed] (2.4,11.5) -- (2.4,11.6);
\draw [dashed] (2.8,11.) -- (2.689,11.055);
\draw [dashed] (2.511,11.145) -- (2.4,11.2);
\draw [dashed] (2.4,11.2) -- (2.289,11.145);
\draw [dashed] (2.111,11.055) -- (2.,11.);
\draw [dashed] (2.8,10.6) -- (2.8,10.7);
\draw [dashed] (2.8,10.9) -- (2.8,11.);
\draw [dashed] (3.6,11.8) -- (3.489,11.745);
\draw [dashed] (3.311,11.655) -- (3.2,11.6);
\draw [dashed] (3.6,11.) -- (3.489,11.055);
\draw [dashed] (3.311,11.145) -- (3.2,11.2);
\draw [dashed] (2.4,11.6) -- (2.289,11.655);
\draw [dashed] (2.111,11.745) -- (2.,11.8);
\draw [dashed] (4.4,11.8) -- (4.4,11.9);
\draw [dashed] (4.4,12.1) -- (4.4,12.2);
\draw [dashed] (4.4,11.8) -- (4.289,11.745);
\draw [dashed] (4.111,11.655) -- (4.,11.6);
\draw [dashed] (4.8,11.6) -- (4.689,11.655);
\draw [dashed] (4.511,11.745) -- (4.4,11.8);
\path [lightgray, fill] (4.,12.) -- (4.8,12.) -- (4.4,11.4) -- cycle;
\draw [dashed] (2.,11.) -- (1.889,11.055);
\draw [dashed] (1.711,11.145) -- (1.6,11.2);
\draw [dashed] (3.6,10.6) -- (3.6,10.7);
\draw [dashed] (3.6,10.9) -- (3.6,11.);
\draw [dashed] (4.,11.2) -- (3.889,11.145);
\draw [dashed] (3.711,11.055) -- (3.6,11.);
\draw [dashed] (4.4,10.6) -- (4.4,10.7);
\draw [dashed] (4.4,10.9) -- (4.4,11.);
\draw [dashed] (4.8,11.2) -- (4.689,11.145);
\draw [dashed] (4.511,11.055) -- (4.4,11.);
\draw [dashed] (4.4,11.) -- (4.289,11.055);
\draw [dashed] (4.111,11.145) -- (4.,11.2);
\draw [dashed] (4.,11.2) -- (4.,11.3);
\draw [dashed] (4.,11.5) -- (4.,11.6);
\draw [dashed] (4.,11.6) -- (3.889,11.655);
\draw [dashed] (3.711,11.745) -- (3.6,11.8);
\draw [dashed] (4.8,11.2) -- (4.8,11.3);
\draw [dashed] (4.8,11.5) -- (4.8,11.6);
\draw [dashed] (5.2,11.8) -- (5.089,11.745);
\draw [dashed] (4.911,11.655) -- (4.8,11.6);
\draw [dashed] (2.,11.8) -- (2.,11.9);
\draw [dashed] (2.,12.1) -- (2.,12.2);
\draw [dashed] (2.4,12.4) -- (2.289,12.345);
\draw [dashed] (2.111,12.255) -- (2.,12.2);
\draw [dashed] (2.,12.2) -- (1.889,12.255);
\draw [dashed] (1.711,12.345) -- (1.6,12.4);
\draw [dashed] (3.2,12.4) -- (3.089,12.345);
\draw [dashed] (2.911,12.255) -- (2.8,12.2);
\draw [dashed] (2.8,12.2) -- (2.689,12.255);
\draw [dashed] (2.511,12.345) -- (2.4,12.4);
\draw [dashed] (3.6,11.8) -- (3.6,11.9);
\draw [dashed] (3.6,12.1) -- (3.6,12.2);
\draw [dashed] (4.,12.4) -- (3.889,12.345);
\draw [dashed] (3.711,12.255) -- (3.6,12.2);
\draw [dashed] (3.6,12.2) -- (3.489,12.255);
\draw [dashed] (3.311,12.345) -- (3.2,12.4);
\draw [dashed] (4.8,12.4) -- (4.689,12.345);
\draw [dashed] (4.511,12.255) -- (4.4,12.2);
\draw [dashed] (4.4,12.2) -- (4.289,12.255);
\draw [dashed] (4.111,12.345) -- (4.,12.4);
\draw [dashed] (2.,11.8) -- (1.889,11.745);
\draw [dashed] (1.711,11.655) -- (1.6,11.6);
\draw [dashed] (5.2,12.2) -- (5.089,12.255);
\draw [dashed] (4.911,12.345) -- (4.8,12.4);
\draw [dashed] (2.4,12.4) -- (2.4,12.5);
\draw [dashed] (2.4,12.7) -- (2.4,12.8);
\draw [dashed] (3.2,12.4) -- (3.2,12.5);
\draw [dashed] (3.2,12.7) -- (3.2,12.8);
\draw [dashed] (4.,12.4) -- (4.,12.5);
\draw [dashed] (4.,12.7) -- (4.,12.8);
\draw [dashed] (4.8,12.4) -- (4.8,12.5);
\draw [dashed] (4.8,12.7) -- (4.8,12.8);
\draw [dashed] (5.2,11.) -- (5.089,11.055);
\draw [dashed] (4.911,11.145) -- (4.8,11.2);
\draw [] (4.4,11.4) -- (4.8,12.);
\draw [] (4.,12.) -- (4.8,12.);
\draw [] (1.6,10.8) -- (2.4,10.8);
\draw [] (2.4,10.8) -- (3.2,10.8);
\draw [] (3.2,10.8) -- (4.,10.8);
\draw [] (1.6,10.8) -- (2.,11.4);
\draw [] (3.2,10.8) -- (3.6,11.4);
\draw [] (4.,10.8) -- (4.4,11.4);
\draw [] (2.4,10.8) -- (2.,11.4);
\draw [] (4.,10.8) -- (3.6,11.4);
\draw [] (3.6,11.4) -- (4.4,11.4);
\draw [] (2.,11.4) -- (2.4,12.);
\draw [] (3.6,11.4) -- (4.,12.);
\draw [] (3.6,11.4) -- (3.2,12.);
\draw [] (2.,11.4) -- (1.6,12.);
\draw [] (1.6,12.) -- (2.4,12.);
\draw [] (2.4,12.) -- (3.2,12.);
\draw [] (3.2,12.) -- (4.,12.);
\draw [] (1.6,12.) -- (2.,12.6);
\draw [] (2.4,12.) -- (2.8,12.6);
\draw [] (3.2,12.) -- (3.6,12.6);
\draw [] (4.,12.) -- (4.4,12.6);
\draw [] (2.4,12.) -- (2.,12.6);
\draw [] (3.2,12.) -- (2.8,12.6);
\draw [] (4.,12.) -- (3.6,12.6);
\draw [] (2.,12.6) -- (2.8,12.6);
\draw [] (2.8,12.6) -- (3.6,12.6);
\draw [] (3.6,12.6) -- (4.4,12.6);
\draw [] (4.,10.8) -- (4.8,10.8);
\draw [] (4.8,10.8) -- (5.2,11.4);
\draw [] (4.8,10.8) -- (4.4,11.4);
\draw [] (4.4,11.4) -- (5.2,11.4);
\draw [] (5.2,11.4) -- (4.8,12.);
\draw [] (4.8,12.) -- (5.2,12.6);
\draw [] (4.8,12.) -- (4.4,12.6);
\draw [] (4.4,12.6) -- (5.2,12.6);
\draw [] (2.,11.4) -- (2.8,11.4);
\draw [] (2.8,11.4) -- (3.6,11.4);
\draw [] (2.4,10.8) -- (2.8,11.4);
\draw [] (2.8,11.4) -- (3.2,12.);
\draw [] (3.2,10.8) -- (2.8,11.4);
\draw [] (2.8,11.4) -- (2.4,12.);
\draw [] (4.4,11.4) -- (4.,12.);
\draw [fill,rotate around={0.:(2.8,11.4)}] (2.8,11.4) ++(0.:0.08 and 0.08) arc(0.:360.:0.08 and 0.08);
\node [rotate=0.] at (3.1,11.6) {\scriptsize $\text{charge}$};
\node [rotate=0.] at (4.5,11.8) {\scriptsize $\text{flux}$};
\end{tikzpicture}
}

\newcommand{\DoubleLayerGraphAC}{
\begin{tikzpicture}[baseline={([yshift=-.8ex]current bounding box.center)},scale=1]
\draw [fill,rotate around={0.:(4.4,11.8)}] (4.4,11.8) ++(0.:0.08 and 0.08) arc(0.:360.:0.08 and 0.08);
\path [lightgray, fill] (2.4,11.2) -- (2.8,11.) -- (3.2,11.2) -- (3.2,11.6) -- (2.8,11.8) -- (2.4,11.6) -- cycle;
\draw [dashed] (2.,10.6) -- (2.,11.);
\draw [dashed] (2.4,11.2) -- (2.,11.);
\draw [dashed] (2.,11.) -- (1.6,11.2);
\draw [dashed] (2.8,10.6) -- (2.8,11.);
\draw [dashed] (3.6,10.6) -- (3.6,11.);
\draw [dashed] (4.,11.2) -- (3.6,11.);
\draw [dashed] (3.6,11.) -- (3.2,11.2);
\draw [dashed] (4.4,10.6) -- (4.4,11.);
\draw [dashed] (4.8,11.2) -- (4.4,11.);
\draw [dashed] (4.4,11.) -- (4.,11.2);
\draw [dashed] (2.4,11.6) -- (2.,11.8);
\draw [dashed] (3.6,11.8) -- (3.2,11.6);
\draw [dashed] (4.,11.2) -- (4.,11.6);
\draw [dashed] (4.,11.6) -- (3.6,11.8);
\draw [dashed] (4.8,11.2) -- (4.8,11.6);
\draw [dashed] (5.2,11.8) -- (4.8,11.6);
\draw [dashed] (2.,11.8) -- (2.,12.2);
\draw [dashed] (2.4,12.4) -- (2.,12.2);
\draw [dashed] (2.,12.2) -- (1.6,12.4);
\draw [dashed] (2.8,11.8) -- (2.8,12.2);
\draw [dashed] (3.2,12.4) -- (2.8,12.2);
\draw [dashed] (2.8,12.2) -- (2.4,12.4);
\draw [dashed] (3.6,11.8) -- (3.6,12.2);
\draw [dashed] (4.,12.4) -- (3.6,12.2);
\draw [dashed] (3.6,12.2) -- (3.2,12.4);
\draw [dashed] (4.8,12.4) -- (4.4,12.2);
\draw [dashed] (4.4,12.2) -- (4.,12.4);
\draw [dashed] (2.,11.8) -- (1.6,11.6);
\draw [dashed] (5.2,12.2) -- (4.8,12.4);
\draw [dashed] (2.4,12.4) -- (2.4,12.8);
\draw [dashed] (3.2,12.4) -- (3.2,12.8);
\draw [dashed] (4.,12.4) -- (4.,12.8);
\draw [dashed] (4.8,12.4) -- (4.8,12.8);
\draw [dashed] (5.2,11.) -- (4.8,11.2);
\draw [dashed] (3.2,11.2) -- (2.8,11.);
\draw [dashed] (2.8,11.) -- (2.4,11.2);
\draw [dashed] (2.4,11.2) -- (2.4,11.6);
\draw [dashed] (2.8,11.8) -- (2.4,11.6);
\draw [dashed] (3.2,11.2) -- (3.2,11.6);
\draw [dashed] (3.2,11.6) -- (2.8,11.8);
\draw [dashed] (4.4,11.8) -- (4.,11.6);
\draw [dashed] (4.8,11.6) -- (4.4,11.8);
\draw [dashed] (4.4,11.8) -- (4.4,12.2);
\draw [fill,rotate around={0.:(4.4,11.8)}] (4.4,11.8) ++(0.:0.08 and 0.08) arc(0.:360.:0.08 and 0.08);
\node [rotate=0.] at (4.7,12.) {\scriptsize $\tilde{\text{charge}}$};
\node [rotate=0.] at (2.8,11.4) {\scriptsize $\tilde{\text{flux}}$};
\end{tikzpicture}
}

\newcommand{\dualGraphAA}{
\begin{tikzpicture}[baseline={([yshift=-.8ex]current bounding box.center)},scale=1]
\draw [->-] (4.8,12.4) -- (3.2,10.3);
\draw [->-] (3.2,10.3) -- (3.2,11.5);
\draw [->-] (1.6,12.4) -- (3.2,11.5);
\draw [->-] (4.8,12.4) -- (3.2,11.5);
\draw [->-] (1.6,12.4) -- (3.2,10.3);
\draw [->-] (4.8,12.4) -- (1.6,12.4);
\node [rotate=0.] at (3.1,12.8) {\scriptsize $ $};
\end{tikzpicture}
}

\newcommand{\dualGraphAB}{
\begin{tikzpicture}[baseline={([yshift=-.8ex]current bounding box.center)},scale=1]
\draw [gray, ->-, dotted] (4.8,12.4) -- (3.888,11.203);
\draw [gray, ->-, dotted] (3.767,11.044) -- (3.2,10.3);
\draw [gray, ->-, dotted] (3.2,10.3) -- (3.2,11.1);
\draw [gray, ->-, dotted] (3.2,11.3) -- (3.2,11.5);
\draw [gray, ->-, dotted] (1.6,12.4) -- (2.84,11.703);
\draw [gray, ->-, dotted] (3.014,11.605) -- (3.2,11.5);
\draw [gray, ->-, dotted] (4.8,12.4) -- (3.56,11.703);
\draw [gray, ->-, dotted] (3.386,11.605) -- (3.2,11.5);
\draw [gray, ->-, dotted] (1.6,12.4) -- (2.512,11.203);
\draw [gray, ->-, dotted] (2.633,11.044) -- (3.2,10.3);
\draw [gray, ->-, dotted] (4.8,12.4) -- (3.3,12.4);
\draw [gray, ->-, dotted] (3.1,12.4) -- (1.6,12.4);
\draw [->-] (3.2,12.2) -- (3.2,12.8);
\draw [->-] (3.2,12.2) -- (2.7,11.2);
\draw [->-] (2.7,11.2) -- (3.7,11.2);
\draw [->-] (3.7,11.2) -- (3.2,12.2);
\draw [->-] (2.7,11.2) -- (1.7,10.6);
\draw [->-] (3.7,11.2) -- (4.7,10.6);
\end{tikzpicture}
}

\newcommand{\dualGraphAC}{
\begin{tikzpicture}[baseline={([yshift=-.8ex]current bounding box.center)},scale=1]
\draw [-<-] (3.2,10.3) -- (3.2,11.5);
\draw [-<-] (1.6,12.4) -- (3.2,11.5);
\draw [-<-] (4.8,12.4) -- (3.2,11.5);
\draw [-<-] (1.6,12.4) -- (3.2,10.3);
\draw [-<-] (4.8,12.4) -- (1.6,12.4);
\draw [-<-] (4.8,12.4) -- (3.2,10.3);
\draw [gray, dotted, ->-] (3.2,12.2) -- (2.972,11.743);
\draw [gray, dotted, ->-] (2.882,11.564) -- (2.7,11.2);
\draw [gray, dotted, ->-] (2.7,11.2) -- (3.1,11.2);
\draw [gray, dotted, ->-] (3.3,11.2) -- (3.7,11.2);
\draw [gray, dotted, ->-] (3.7,11.2) -- (3.518,11.564);
\draw [gray, dotted, ->-] (3.428,11.743) -- (3.2,12.2);
\draw [gray, dotted, ->-] (2.7,11.2) -- (2.658,11.175);
\draw [gray, dotted, ->-] (2.487,11.072) -- (1.7,10.6);
\draw [gray, dotted, ->-] (3.7,11.2) -- (3.742,11.175);
\draw [gray, dotted, ->-] (3.913,11.072) -- (4.7,10.6);
\draw [gray, dotted, ->-] (3.2,12.2) -- (3.2,12.3);
\draw [gray, dotted, ->-] (3.2,12.5) -- (3.2,12.8);
\end{tikzpicture}
}

\newcommand{\dualLatticeIsing}{
\begin{tikzpicture}[baseline={([yshift=-.8ex]current bounding box.center)},scale=1]
\draw [dashed] (3.7,10.7) -- (4.2,10.7);
\draw [dashed] (2.5,10.3) -- (2.5,10.7);
\draw [dashed] (2.5,11.9) -- (2.5,12.3);
\draw [dashed] (3.7,10.3) -- (3.7,10.7);
\draw [dashed] (3.7,10.7) -- (3.7,11.2);
\draw [dashed] (3.7,11.4) -- (3.7,11.9);
\draw [dashed] (3.7,11.9) -- (3.7,12.3);
\draw [dashed] (2.5,10.7) -- (3.,10.7);
\draw [dashed] (3.2,10.7) -- (3.7,10.7);
\draw [dashed] (2.,10.7) -- (2.5,10.7);
\draw [dashed] (3.7,11.9) -- (4.2,11.9);
\draw [dashed] (2.5,11.9) -- (3.,11.9);
\draw [dashed] (3.2,11.9) -- (3.7,11.9);
\draw [dashed] (2.,11.9) -- (2.5,11.9);
\draw [] (1.6,10.1) -- (1.9,10.1);
\draw [] (1.9,10.1) -- (3.1,10.1);
\draw [] (3.1,10.1) -- (4.3,10.1);
\draw [] (1.9,9.8) -- (1.9,10.1);
\draw [] (1.9,10.1) -- (1.9,11.3);
\draw [] (1.9,11.3) -- (1.9,12.5);
\draw [] (1.9,12.5) -- (1.9,12.8);
\draw [] (1.6,11.3) -- (1.9,11.3);
\draw [] (3.1,11.3) -- (4.3,11.3);
\draw [] (1.6,12.5) -- (1.9,12.5);
\draw [] (1.9,12.5) -- (3.1,12.5);
\draw [] (3.1,12.5) -- (4.3,12.5);
\draw [] (3.1,9.8) -- (3.1,10.1);
\draw [] (3.1,10.1) -- (3.1,11.3);
\draw [] (3.1,11.3) -- (3.1,12.5);
\draw [] (3.1,12.5) -- (3.1,12.8);
\draw [] (4.3,9.8) -- (4.3,10.1);
\draw [] (4.3,10.1) -- (4.3,11.3);
\draw [] (4.3,11.3) -- (4.3,12.5);
\draw [] (4.3,12.5) -- (4.3,12.8);
\draw [thick, dashed] (2.5,10.7) -- (2.5,11.2);
\draw [thick, dashed] (2.5,11.4) -- (2.5,11.9);
\draw [] (5.5,10.1) -- (5.5,11.3);
\draw [] (5.5,11.3) -- (5.5,12.5);
\draw [] (5.5,12.5) -- (5.5,12.8);
\draw [] (4.3,12.5) -- (5.5,12.5);
\draw [] (5.5,12.5) -- (5.8,12.5);
\draw [] (4.3,11.3) -- (5.5,11.3);
\draw [] (5.5,11.3) -- (5.8,11.3);
\draw [] (4.3,10.1) -- (5.5,10.1);
\draw [] (5.5,10.1) -- (5.8,10.1);
\draw [] (5.5,9.8) -- (5.5,10.1);
\draw [thick] (1.9,11.3) -- (3.1,11.3);
\draw [fill=white,draw,rotate around={0.:(2.5,11.9)}] (2.5,11.9) ++(0.:0.08 and 0.08) arc(0.:360.:0.08 and 0.08);
\draw [fill=white,draw,rotate around={0.:(3.7,11.9)}] (3.7,11.9) ++(0.:0.08 and 0.08) arc(0.:360.:0.08 and 0.08);
\draw [fill=white,draw,rotate around={0.:(3.7,10.7)}] (3.7,10.7) ++(0.:0.08 and 0.08) arc(0.:360.:0.08 and 0.08);
\draw [fill=white,draw,rotate around={0.:(2.5,10.7)}] (2.5,10.7) ++(0.:0.08 and 0.08) arc(0.:360.:0.08 and 0.08);
\draw [fill,rotate around={0.:(1.9,12.5)}] (1.9,12.5) ++(0.:0.08 and 0.08) arc(0.:360.:0.08 and 0.08);
\draw [fill,rotate around={0.:(3.1,12.5)}] (3.1,12.5) ++(0.:0.08 and 0.08) arc(0.:360.:0.08 and 0.08);
\draw [fill,rotate around={0.:(4.3,12.5)}] (4.3,12.5) ++(0.:0.08 and 0.08) arc(0.:360.:0.08 and 0.08);
\draw [fill,rotate around={0.:(4.3,11.3)}] (4.3,11.3) ++(0.:0.08 and 0.08) arc(0.:360.:0.08 and 0.08);
\draw [fill,rotate around={0.:(5.5,11.3)}] (5.5,11.3) ++(0.:0.08 and 0.08) arc(0.:360.:0.08 and 0.08);
\draw [fill,rotate around={0.:(5.5,12.5)}] (5.5,12.5) ++(0.:0.08 and 0.08) arc(0.:360.:0.08 and 0.08);
\draw [fill,rotate around={0.:(5.5,10.1)}] (5.5,10.1) ++(0.:0.08 and 0.08) arc(0.:360.:0.08 and 0.08);
\draw [fill,rotate around={0.:(4.3,10.1)}] (4.3,10.1) ++(0.:0.08 and 0.08) arc(0.:360.:0.08 and 0.08);
\draw [fill,rotate around={0.:(3.1,10.1)}] (3.1,10.1) ++(0.:0.08 and 0.08) arc(0.:360.:0.08 and 0.08);
\draw [fill,rotate around={0.:(1.9,11.3)}] (1.9,11.3) ++(0.:0.08 and 0.08) arc(0.:360.:0.08 and 0.08);
\draw [fill,rotate around={0.:(3.1,11.3)}] (3.1,11.3) ++(0.:0.08 and 0.08) arc(0.:360.:0.08 and 0.08);
\draw [fill,rotate around={0.:(1.9,10.1)}] (1.9,10.1) ++(0.:0.08 and 0.08) arc(0.:360.:0.08 and 0.08);
\node [rotate=0.] at (1.7,11.5) {\scriptsize $\sigma^x_{v_1}$};
\node [rotate=0.] at (3.3,11.5) {\scriptsize $\sigma^x_{v_2}$};
\node [rotate=0.] at (2.7,12.1) {\scriptsize $\sigma^z_{p_1}$};
\node [rotate=0.] at (2.7,10.5) {\scriptsize $\sigma^z_{p_2}$};
\end{tikzpicture}
}

\newcommand{\PEMdualGraphAA}{
\begin{tikzpicture}[baseline={([yshift=-.8ex]current bounding box.center)},scale=1]
\draw [->-] (3.2,9.8) -- (4.,10.8);
\draw [->-] (2.4,10.8) -- (3.2,9.8);
\draw [->-] (3.2,9.8) -- (4.8,9.8);
\draw [->-] (1.6,9.8) -- (3.2,9.8);
\draw [->-] (1.6,9.8) -- (2.4,10.8);
\draw [->-] (1.6,9.8) -- (2.4,8.8);
\draw [->-] (2.4,8.8) -- (4.,8.8);
\draw [->-] (2.4,8.8) -- (3.2,9.8);
\draw [->-] (3.2,9.8) -- (4.,8.8);
\draw [->-] (2.4,12.8) -- (4.,12.8);
\draw [->-] (3.2,11.8) -- (4.,12.8);
\draw [->-] (2.4,12.8) -- (3.2,11.8);
\draw [->-] (3.2,11.8) -- (4.8,11.8);
\draw [->-] (1.6,11.8) -- (3.2,11.8);
\draw [->-] (1.6,11.8) -- (2.4,12.8);
\draw [->-] (1.6,11.8) -- (2.4,10.8);
\draw [->-] (2.4,10.8) -- (3.2,11.8);
\draw [->-] (3.2,11.8) -- (4.,10.8);
\draw [->-] (2.4,10.8) -- (4.,10.8);
\draw [->-] (5.6,10.8) -- (6.4,11.8);
\draw [->-] (4.8,11.8) -- (5.6,10.8);
\draw [->-] (5.6,10.8) -- (7.2,10.8);
\draw [->-] (6.4,11.8) -- (7.2,10.8);
\draw [->-] (4.,10.8) -- (5.6,10.8);
\draw [->-] (4.,10.8) -- (4.8,11.8);
\draw [->-] (4.,10.8) -- (4.8,9.8);
\draw [->-] (6.4,9.8) -- (7.2,10.8);
\draw [->-] (4.8,9.8) -- (5.6,10.8);
\draw [->-] (5.6,10.8) -- (6.4,9.8);
\draw [->-] (5.6,12.8) -- (7.2,12.8);
\draw [->-] (4.,12.8) -- (5.6,12.8);
\draw [->-] (4.,12.8) -- (4.8,11.8);
\draw [->-] (6.4,11.8) -- (7.2,12.8);
\draw [->-] (4.8,11.8) -- (5.6,12.8);
\draw [->-] (5.6,12.8) -- (6.4,11.8);
\draw [->-] (4.8,11.8) -- (6.4,11.8);
\draw [->-] (5.6,8.8) -- (6.4,9.8);
\draw [->-] (4.8,9.8) -- (5.6,8.8);
\draw [->-] (5.6,8.8) -- (7.2,8.8);
\draw [->-] (6.4,9.8) -- (7.2,8.8);
\draw [->-] (4.,8.8) -- (5.6,8.8);
\draw [->-] (4.,8.8) -- (4.8,9.8);
\draw [->-] (4.8,9.8) -- (6.4,9.8);
\node [gray, rotate=0.] at (3.2,8.3) {\scriptsize $ $};
\node [gray, rotate=0.] at (3.2,13.3) {\scriptsize $ $};
\end{tikzpicture}
}

\newcommand{\PEMdualGraphAB}{
\begin{tikzpicture}[baseline={([yshift=-.8ex]current bounding box.center)},scale=1]
\draw [gray, ->-, dotted] (3.2,9.3) -- (3.568,9.76);
\draw [gray, ->-, dotted] (3.693,9.917) -- (4.,10.3);
\draw [gray, ->-, dotted] (2.4,10.3) -- (2.707,9.917);
\draw [gray, ->-, dotted] (2.832,9.76) -- (3.2,9.3);
\draw [gray, ->-, dotted] (3.2,9.3) -- (3.9,9.3);
\draw [gray, ->-, dotted] (4.1,9.3) -- (4.8,9.3);
\draw [gray, ->-, dotted] (1.6,9.3) -- (2.3,9.3);
\draw [gray, ->-, dotted] (2.5,9.3) -- (3.2,9.3);
\draw [gray, ->-, dotted] (1.6,9.3) -- (1.968,9.76);
\draw [gray, ->-, dotted] (2.093,9.917) -- (2.4,10.3);
\draw [gray, ->-, dotted] (1.6,9.3) -- (1.938,8.878);
\draw [gray, ->-, dotted] (2.062,8.722) -- (2.4,8.3);
\draw [gray, ->-, dotted] (2.4,8.3) -- (3.1,8.3);
\draw [gray, ->-, dotted] (3.3,8.3) -- (4.,8.3);
\draw [gray, ->-, dotted] (2.4,8.3) -- (2.738,8.722);
\draw [gray, ->-, dotted] (2.862,8.878) -- (3.2,9.3);
\draw [gray, ->-, dotted] (3.2,9.3) -- (3.538,8.878);
\draw [gray, ->-, dotted] (3.662,8.722) -- (4.,8.3);
\draw [gray, ->-, dotted] (2.4,12.3) -- (3.1,12.3);
\draw [gray, ->-, dotted] (3.3,12.3) -- (4.,12.3);
\draw [gray, ->-, dotted] (3.2,11.3) -- (3.63,11.837);
\draw [gray, ->-, dotted] (3.755,11.993) -- (4.,12.3);
\draw [gray, ->-, dotted] (2.4,12.3) -- (2.645,11.993);
\draw [gray, ->-, dotted] (2.77,11.837) -- (3.2,11.3);
\draw [gray, ->-, dotted] (3.2,11.3) -- (3.9,11.3);
\draw [gray, ->-, dotted] (4.1,11.3) -- (4.8,11.3);
\draw [gray, ->-, dotted] (1.6,11.3) -- (2.3,11.3);
\draw [gray, ->-, dotted] (2.5,11.3) -- (3.2,11.3);
\draw [gray, ->-, dotted] (1.6,11.3) -- (2.03,11.837);
\draw [gray, ->-, dotted] (2.155,11.993) -- (2.4,12.3);
\draw [gray, ->-, dotted] (1.6,11.3) -- (1.871,10.961);
\draw [gray, ->-, dotted] (1.996,10.805) -- (2.4,10.3);
\draw [gray, ->-, dotted] (2.4,10.3) -- (2.804,10.805);
\draw [gray, ->-, dotted] (2.929,10.961) -- (3.2,11.3);
\draw [gray, ->-, dotted] (3.2,11.3) -- (3.471,10.961);
\draw [gray, ->-, dotted] (3.596,10.805) -- (4.,10.3);
\draw [gray, ->-, dotted] (2.4,10.3) -- (3.1,10.3);
\draw [gray, ->-, dotted] (3.3,10.3) -- (4.,10.3);
\draw [gray, ->-, dotted] (5.6,10.3) -- (5.968,10.76);
\draw [gray, ->-, dotted] (6.093,10.917) -- (6.4,11.3);
\draw [gray, ->-, dotted] (4.8,11.3) -- (5.107,10.917);
\draw [gray, ->-, dotted] (5.232,10.76) -- (5.6,10.3);
\draw [gray, ->-, dotted] (5.6,10.3) -- (6.3,10.3);
\draw [gray, ->-, dotted] (6.5,10.3) -- (7.2,10.3);
\draw [gray, ->-, dotted] (6.4,11.3) -- (6.738,10.878);
\draw [gray, ->-, dotted] (6.862,10.722) -- (7.2,10.3);
\draw [gray, ->-, dotted] (4.,10.3) -- (4.7,10.3);
\draw [gray, ->-, dotted] (4.9,10.3) -- (5.6,10.3);
\draw [gray, ->-, dotted] (4.,10.3) -- (4.368,10.76);
\draw [gray, ->-, dotted] (4.493,10.917) -- (4.8,11.3);
\draw [gray, ->-, dotted] (4.,10.3) -- (4.338,9.878);
\draw [gray, ->-, dotted] (4.462,9.722) -- (4.8,9.3);
\draw [gray, ->-, dotted] (6.4,9.3) -- (6.707,9.683);
\draw [gray, ->-, dotted] (6.832,9.84) -- (7.2,10.3);
\draw [gray, ->-, dotted] (4.8,9.3) -- (5.138,9.722);
\draw [gray, ->-, dotted] (5.262,9.878) -- (5.6,10.3);
\draw [gray, ->-, dotted] (5.6,10.3) -- (5.938,9.878);
\draw [gray, ->-, dotted] (6.062,9.722) -- (6.4,9.3);
\draw [gray, ->-, dotted] (5.6,12.3) -- (6.3,12.3);
\draw [gray, ->-, dotted] (6.5,12.3) -- (7.2,12.3);
\draw [gray, ->-, dotted] (4.,12.3) -- (4.7,12.3);
\draw [gray, ->-, dotted] (4.9,12.3) -- (5.6,12.3);
\draw [gray, ->-, dotted] (4.,12.3) -- (4.271,11.961);
\draw [gray, ->-, dotted] (4.396,11.805) -- (4.8,11.3);
\draw [gray, ->-, dotted] (6.4,11.3) -- (6.768,11.76);
\draw [gray, ->-, dotted] (6.893,11.917) -- (7.2,12.3);
\draw [gray, ->-, dotted] (4.8,11.3) -- (5.204,11.805);
\draw [gray, ->-, dotted] (5.329,11.961) -- (5.6,12.3);
\draw [gray, ->-, dotted] (5.6,12.3) -- (5.871,11.961);
\draw [gray, ->-, dotted] (5.996,11.805) -- (6.4,11.3);
\draw [gray, ->-, dotted] (4.8,11.3) -- (5.5,11.3);
\draw [gray, ->-, dotted] (5.7,11.3) -- (6.4,11.3);
\draw [gray, ->-, dotted] (5.6,8.3) -- (5.907,8.683);
\draw [gray, ->-, dotted] (6.032,8.84) -- (6.4,9.3);
\draw [gray, ->-, dotted] (4.8,9.3) -- (5.168,8.84);
\draw [gray, ->-, dotted] (5.293,8.683) -- (5.6,8.3);
\draw [gray, ->-, dotted] (5.6,8.3) -- (6.3,8.3);
\draw [gray, ->-, dotted] (6.5,8.3) -- (7.2,8.3);
\draw [gray, ->-, dotted] (6.4,9.3) -- (6.804,8.795);
\draw [gray, ->-, dotted] (6.929,8.639) -- (7.2,8.3);
\draw [gray, ->-, dotted] (4.,8.3) -- (4.7,8.3);
\draw [gray, ->-, dotted] (4.9,8.3) -- (5.6,8.3);
\draw [gray, ->-, dotted] (4.,8.3) -- (4.307,8.683);
\draw [gray, ->-, dotted] (4.432,8.84) -- (4.8,9.3);
\draw [gray, ->-, dotted] (4.8,9.3) -- (5.5,9.3);
\draw [gray, ->-, dotted] (5.7,9.3) -- (6.4,9.3);
\draw [->-] (2.4,8.9) -- (1.6,8.7);
\draw [->-] (2.4,9.7) -- (2.4,8.9);
\draw [->-] (1.6,10.) -- (2.4,9.7);
\draw [->-] (3.2,10.) -- (2.4,9.7);
\draw [->-] (3.2,10.) -- (4.,9.7);
\draw [->-] (4.,9.7) -- (4.,8.9);
\draw [->-] (4.,8.9) -- (3.2,8.7);
\draw [->-] (2.4,8.9) -- (3.2,8.7);
\draw [->-] (3.2,8.7) -- (3.2,7.8);
\draw [->-] (4.8,9.9) -- (4.,9.7);
\draw [->-] (4.8,10.7) -- (4.8,9.9);
\draw [->-] (4.,11.) -- (4.8,10.7);
\draw [->-] (5.6,11.) -- (4.8,10.7);
\draw [->-] (5.6,11.) -- (6.4,10.7);
\draw [->-] (6.4,10.7) -- (6.4,9.9);
\draw [->-] (6.4,9.9) -- (5.6,9.7);
\draw [->-] (4.8,9.9) -- (5.6,9.7);
\draw [->-] (5.6,9.7) -- (5.6,8.9);
\draw [->-] (4.8,8.6) -- (4.8,7.8);
\draw [->-] (4.,8.9) -- (4.8,8.6);
\draw [->-] (5.6,8.9) -- (4.8,8.6);
\draw [->-] (5.6,8.9) -- (6.4,8.6);
\draw [->-] (6.4,8.6) -- (6.4,7.8);
\draw [->-] (2.4,11.) -- (1.6,10.8);
\draw [->-] (2.4,11.8) -- (2.4,11.);
\draw [->-] (1.6,12.1) -- (2.4,11.8);
\draw [->-] (3.2,12.1) -- (2.4,11.8);
\draw [->-] (3.2,12.1) -- (4.,11.8);
\draw [->-] (4.,11.8) -- (4.,11.);
\draw [->-] (4.,11.) -- (3.2,10.8);
\draw [->-] (2.4,11.) -- (3.2,10.8);
\draw [->-] (3.2,10.8) -- (3.2,10.);
\draw [->-] (4.8,12.) -- (4.,11.8);
\draw [->-] (4.8,12.8) -- (4.8,12.);
\draw [->-] (6.4,12.8) -- (6.4,12.);
\draw [->-] (6.4,12.) -- (5.6,11.8);
\draw [->-] (4.8,12.) -- (5.6,11.8);
\draw [->-] (5.6,11.8) -- (5.6,11.);
\draw [->-] (7.2,10.9) -- (6.4,10.7);
\draw [->-] (7.2,11.7) -- (7.2,10.9);
\draw [->-] (6.4,12.) -- (7.2,11.7);
\draw [->-] (7.2,8.8) -- (6.4,8.6);
\draw [->-] (7.2,9.6) -- (7.2,8.8);
\draw [->-] (6.4,9.9) -- (7.2,9.6);
\draw [->-] (3.2,12.8) -- (3.2,12.1);
\end{tikzpicture}
}

\newcommand{\plaquetteActionAA}{
\begin{tikzpicture}[baseline={([yshift=-.8ex]current bounding box.center)},scale=1]
\draw [->-] (2.,11.2) -- (3.2,11.2);
\draw [->-] (3.2,11.2) -- (3.2,12.4);
\draw [->-] (3.2,12.4) -- (2.,12.4);
\draw [->-] (2.,12.4) -- (2.,11.2);
\draw [] (1.6,11.2) -- (2.,11.2);
\draw [] (2.,11.2) -- (2.,10.8);
\draw [] (3.2,10.8) -- (3.2,11.2);
\draw [] (3.2,11.2) -- (3.6,11.2);
\draw [] (1.6,12.4) -- (2.,12.4);
\draw [] (2.,12.4) -- (2.,12.8);
\draw [] (3.2,12.8) -- (3.2,12.4);
\draw [] (3.2,12.4) -- (3.6,12.4);
\node [rotate=0.] at (2.3,11.8) {\scriptsize $a$};
\node [rotate=0.] at (2.6,12.7) {\scriptsize $b$};
\node [rotate=0.] at (3.5,11.8) {\scriptsize $c$};
\node [rotate=0.] at (2.6,10.9) {\scriptsize $d$};
\end{tikzpicture}
}

\newcommand{\starActionAA}{
\begin{tikzpicture}[baseline={([yshift=-.8ex]current bounding box.center)},scale=1]
\draw [->-] (1.6,12.) -- (2.4,12.);
\draw [->-] (2.4,12.8) -- (2.4,12.);
\draw [->-] (2.4,11.2) -- (2.4,12.);
\draw [->-] (3.2,12.) -- (2.4,12.);
\node [rotate=0.] at (1.8,12.3) {\scriptsize $a$};
\node [rotate=0.] at (2.6,11.5) {\scriptsize $b$};
\node [rotate=0.] at (3.,12.3) {\scriptsize $c$};
\node [rotate=0.] at (2.6,12.6) {\scriptsize $d$};
\end{tikzpicture}
}

\newcommand{\starActionAB}{
\begin{tikzpicture}[baseline={([yshift=-.8ex]current bounding box.center)},scale=1]
\draw [->-] (4.,11.2) -- (4.8,11.2);
\draw [->-] (4.8,12.) -- (4.8,11.2);
\draw [->-] (4.8,10.4) -- (4.8,11.2);
\draw [->-] (5.6,11.2) -- (4.8,11.2);
\node [rotate=0.] at (4.2,11.5) {\scriptsize $ga$};
\node [rotate=0.] at (5.,10.7) {\scriptsize $gb$};
\node [rotate=0.] at (5.4,11.5) {\scriptsize $gc$};
\node [rotate=0.] at (5.,11.8) {\scriptsize $gd$};
\end{tikzpicture}
}

\newcommand{\stateIdenticalAA}{
\begin{tikzpicture}[baseline={([yshift=-.8ex]current bounding box.center)},scale=1]
\draw [->-] (4.,12.) -- (5.6,12.);
\draw [] (3.6,12.) -- (4.,12.);
\draw [] (4.,12.) -- (4.,12.4);
\draw [] (4.,11.6) -- (4.,12.);
\draw [] (5.6,11.6) -- (5.6,12.);
\draw [] (5.6,12.) -- (6.,12.);
\draw [] (5.6,12.4) -- (5.6,12.);
\node [rotate=0.] at (4.8,12.3) {\scriptsize $g$};
\end{tikzpicture}
}

\newcommand{\stateIdenticalAB}{
\begin{tikzpicture}[baseline={([yshift=-.8ex]current bounding box.center)},scale=1]
\draw [] (1.6,12.4) -- (2.,12.4);
\draw [] (2.,12.4) -- (2.,12.8);
\draw [] (2.,12.) -- (2.,12.4);
\draw [] (3.6,12.) -- (3.6,12.4);
\draw [] (3.6,12.4) -- (4.,12.4);
\draw [] (3.6,12.8) -- (3.6,12.4);
\draw [-<-] (2.,12.4) -- (3.6,12.4);
\node [rotate=0.] at (2.8,12.7) {\scriptsize $\bar g$};
\end{tikzpicture}
}

\newcommand{\triangleDiagram}{
\begin{tikzpicture}[baseline={([yshift=-.8ex]current bounding box.center)},scale=1]
\draw [->-] (3.3,7.8) -- (1.6,7.8);
\draw [->-] (1.6,7.8) -- (2.4,6.6);
\draw [->-] (3.3,7.8) -- (2.4,6.6);
\node [rotate=-55.] at (1.8,7.1) {\scriptsize $(x_1,\rho_1)$};
\node [rotate=55.] at (3.,7.1) {\scriptsize $(x_2,\rho_2)$};
\node [rotate=0.] at (2.5,8.) {\scriptsize $(x_{12},\rho_{12})$};
\end{tikzpicture}
}

\newcommand{\triangleDiagramAA}{
\begin{tikzpicture}[baseline={([yshift=-.8ex]current bounding box.center)},scale=1]
\draw [->-] (3.3,12.6) -- (1.6,12.6);
\draw [->-] (1.6,12.6) -- (2.4,11.4);
\draw [->-] (3.3,12.6) -- (2.4,11.4);
\node [rotate=-55.] at (1.8,11.9) {\scriptsize $(x_1,\bar\eta\rho_1)$};
\node [rotate=55.] at (3.,11.9) {\scriptsize $(x_2, {\eta}^{y}\rho_2)$};
\node [rotate=0.] at (2.5,12.8) {\scriptsize $(x_{12},\bar\eta^{x_1}\rho_{12})$};
\end{tikzpicture}
}

\newcommand{\triangleDiagramAB}{
\begin{tikzpicture}[baseline={([yshift=-.8ex]current bounding box.center)},scale=1]
\draw [->-] (3.3,12.6) -- (1.6,12.6);
\draw [->-] (1.6,12.6) -- (2.4,11.4);
\draw [->-] (3.3,12.6) -- (2.4,11.4);
\node [rotate=-55.] at (1.8,11.9) {\scriptsize $(x_1,\rho'_1)$};
\node [rotate=55.] at (3.,11.9) {\scriptsize $(x_2,\rho'_2)$};
\node [rotate=0.] at (2.5,12.8) {\scriptsize $(x_{12},\rho'_{12})$};
\end{tikzpicture}
}

\newcommand{\triangulationFigureAA}{
\begin{tikzpicture}[baseline={([yshift=-.8ex]current bounding box.center)},scale=1]
\draw [->-] (1.6,10.2) -- (3.,10.2);
\draw [->-] (3.,10.2) -- (4.5,10.2);
\draw [->-] (1.6,10.2) -- (2.3,11.3);
\draw [->-] (3.,10.2) -- (3.8,11.3);
\draw [->-] (4.5,10.2) -- (5.2,11.3);
\draw [->-] (3.,10.2) -- (2.3,11.3);
\draw [->-] (4.5,10.2) -- (3.8,11.3);
\draw [->-] (2.3,11.3) -- (3.8,11.3);
\draw [->-] (3.8,11.3) -- (5.2,11.3);
\draw [->-] (2.3,11.3) -- (3.,12.4);
\draw [->-] (3.8,11.3) -- (4.5,12.4);
\draw [->-] (3.8,11.3) -- (3.,12.4);
\draw [->-] (5.2,11.3) -- (4.5,12.4);
\draw [->-] (2.3,11.3) -- (1.6,12.4);
\draw [->-] (1.6,12.4) -- (3.,12.4);
\draw [->-] (3.,12.4) -- (4.5,12.4);
\node [rotate=0.] at (3.8,12.8) {\scriptsize $g_{12}$};
\node [rotate=0.] at (2.3,12.8) {\scriptsize $g_{23}$};
\node [rotate=0.] at (5.,12.) {\scriptsize $g_{14}$};
\node [gray, rotate=0.] at (4.5,12.7) {\scriptsize ${1}$};
\node [gray, rotate=0.] at (3.,12.7) {\scriptsize ${2}$};
\node [gray, rotate=0.] at (1.6,12.7) {\scriptsize ${3}$};
\node [gray, rotate=0.] at (5.5,11.3) {\scriptsize ${4}$};
\node [gray, rotate=0.] at (4.1,11.5) {\scriptsize ${5}$};
\node [gray, rotate=0.] at (2.7,11.5) {\scriptsize ${6}$};
\node [gray, rotate=0.] at (4.8,10.2) {\scriptsize ${7}$};
\node [gray, rotate=0.] at (3.,10.) {\scriptsize ${8}$};
\node [gray, rotate=0.] at (1.6,10.) {\scriptsize ${9}$};
\end{tikzpicture}
}

\newcommand{\triangulationFigureAB}{
\begin{tikzpicture}[baseline={([yshift=-.8ex]current bounding box.center)},scale=1]
\draw [->-] (1.6,9.8) -- (2.4,9.8);
\draw [->-] (2.4,9.8) -- (3.2,9.8);
\draw [->-] (3.2,9.8) -- (4.,9.8);
\draw [->-] (1.6,9.8) -- (2.,10.4);
\draw [->-] (2.4,9.8) -- (2.8,10.4);
\draw [->-] (3.2,9.8) -- (3.6,10.4);
\draw [->-] (4.,9.8) -- (4.4,10.4);
\draw [->-] (2.4,9.8) -- (2.,10.4);
\draw [->-] (3.2,9.8) -- (2.8,10.4);
\draw [->-] (4.,9.8) -- (3.6,10.4);
\draw [->-] (2.,10.4) -- (2.8,10.4);
\draw [->-] (2.8,10.4) -- (3.6,10.4);
\draw [->-] (3.6,10.4) -- (4.4,10.4);
\draw [->-] (2.,10.4) -- (2.4,11.);
\draw [->-] (2.8,10.4) -- (3.2,11.);
\draw [->-] (3.6,10.4) -- (4.,11.);
\draw [->-] (2.8,10.4) -- (2.4,11.);
\draw [->-] (3.6,10.4) -- (3.2,11.);
\draw [->-] (4.4,10.4) -- (4.,11.);
\draw [->-] (2.,10.4) -- (1.6,11.);
\draw [->-] (1.6,11.) -- (2.4,11.);
\draw [->-] (2.4,11.) -- (3.2,11.);
\draw [->-] (3.2,11.) -- (4.,11.);
\draw [->-] (1.6,11.) -- (2.,11.6);
\draw [->-] (2.4,11.) -- (2.8,11.6);
\draw [->-] (3.2,11.) -- (3.6,11.6);
\draw [->-] (4.,11.) -- (4.4,11.6);
\draw [->-] (2.4,11.) -- (2.,11.6);
\draw [->-] (3.2,11.) -- (2.8,11.6);
\draw [->-] (4.,11.) -- (3.6,11.6);
\draw [->-] (2.,11.6) -- (2.8,11.6);
\draw [->-] (2.8,11.6) -- (3.6,11.6);
\draw [->-] (3.6,11.6) -- (4.4,11.6);
\draw [->-] (2.,11.6) -- (2.4,12.2);
\draw [->-] (2.8,11.6) -- (3.2,12.2);
\draw [->-] (3.6,11.6) -- (4.,12.2);
\draw [->-] (2.8,11.6) -- (2.4,12.2);
\draw [->-] (3.6,11.6) -- (3.2,12.2);
\draw [->-] (4.4,11.6) -- (4.,12.2);
\draw [->-] (2.,11.6) -- (1.6,12.2);
\draw [->-] (1.6,12.2) -- (2.4,12.2);
\draw [->-] (2.4,12.2) -- (3.2,12.2);
\draw [->-] (3.2,12.2) -- (4.,12.2);
\draw [->-] (1.6,12.2) -- (2.,12.8);
\draw [->-] (2.4,12.2) -- (2.8,12.8);
\draw [->-] (3.2,12.2) -- (3.6,12.8);
\draw [->-] (4.,12.2) -- (4.4,12.8);
\draw [->-] (2.4,12.2) -- (2.,12.8);
\draw [->-] (3.2,12.2) -- (2.8,12.8);
\draw [->-] (4.,12.2) -- (3.6,12.8);
\draw [->-] (2.,12.8) -- (2.8,12.8);
\draw [->-] (2.8,12.8) -- (3.6,12.8);
\draw [->-] (3.6,12.8) -- (4.4,12.8);
\draw [white,rotate around={0.:(2.8,9.6)}] (2.8,9.6) ++(0.:0.01 and 0.01) arc(0.:360.:0.01 and 0.01);
\end{tikzpicture}
}

\newcommand{\unitCellAA}{
\begin{tikzpicture}[baseline={([yshift=-.8ex]current bounding box.center)},scale=1]
\draw [->-] (4.,12.5) -- (2.8,10.7);
\draw [->-] (2.8,10.7) -- (2.8,11.7);
\draw [->-] (1.6,12.5) -- (2.8,11.7);
\draw [->-] (4.,12.5) -- (2.8,11.7);
\draw [->-] (1.6,12.5) -- (2.8,10.7);
\draw [->-] (4.,12.5) -- (1.6,12.5);
\node [rotate=0.] at (2.5,12.2) {\scriptsize $g_1$};
\node [rotate=0.] at (3.1,12.2) {\scriptsize $g_2$};
\node [rotate=0.] at (2.6,11.5) {\scriptsize $g_3$};
\node [rotate=0.] at (2.8,12.8) {\scriptsize $g_{12}$};
\node [rotate=0.] at (3.7,11.5) {\scriptsize $g_{23}$};
\node [lightgray, rotate=0.] at (3.,11.6) {\scriptsize $v$};
\node [lightgray, rotate=0.] at (2.8,12.2) {\scriptsize $p$};
\node [rotate=0.] at (1.9,11.5) {\scriptsize $g_{13}$};
\end{tikzpicture}
}

\newcommand{\unitCellAB}{
\begin{tikzpicture}[baseline={([yshift=-.8ex]current bounding box.center)},scale=1]
\draw [->-] (2.8,10.7) -- (1.6,12.5);
\draw [->-] (4.,12.5) -- (2.8,10.7);
\draw [->-] (2.8,10.7) -- (2.8,11.7);
\draw [->-] (1.6,12.5) -- (2.8,11.7);
\draw [->-] (4.,12.5) -- (2.8,11.7);
\draw [->-] (4.,12.5) -- (1.6,12.5);
\node [rotate=0.] at (2.5,12.2) {\scriptsize $g_1$};
\node [rotate=0.] at (3.1,12.2) {\scriptsize $g_2$};
\node [rotate=0.] at (2.6,11.5) {\scriptsize $g_3$};
\node [rotate=0.] at (2.8,12.8) {\scriptsize $g_{12}$};
\node [rotate=0.] at (3.7,11.5) {\scriptsize $g_{23}$};
\node [lightgray, rotate=0.] at (3.,11.6) {\scriptsize $v$};
\node [rotate=0.] at (1.9,11.5) {\scriptsize $g_{13}$};
\end{tikzpicture}
}

\newcommand{\unitCellAD}{
\begin{tikzpicture}[baseline={([yshift=-.8ex]current bounding box.center)},scale=1]
\draw [->-] (4.,12.5) -- (2.8,10.7);
\draw [->-] (1.6,12.5) -- (2.8,11.7);
\draw [->-] (4.,12.5) -- (2.8,11.7);
\draw [->-] (4.,12.5) -- (1.6,12.5);
\draw [->-] (2.8,11.7) -- (2.8,10.7);
\draw [->-] (1.6,12.5) -- (2.8,10.7);
\node [rotate=0.] at (2.5,12.2) {\scriptsize $g_1$};
\node [rotate=0.] at (3.1,12.2) {\scriptsize $g_2$};
\node [rotate=0.] at (2.6,11.5) {\scriptsize $g_3$};
\node [rotate=0.] at (2.8,12.8) {\scriptsize $g_{12}$};
\node [rotate=0.] at (3.7,11.5) {\scriptsize $g_{23}$};
\node [rotate=0.] at (1.9,11.5) {\scriptsize $g_{13}$};
\end{tikzpicture}
}

\newcommand{\unitCellAE}{
\begin{tikzpicture}[baseline={([yshift=-.8ex]current bounding box.center)},scale=1]
\draw [->-] (4.,12.5) -- (2.8,10.7);
\draw [->-] (1.6,12.5) -- (2.8,11.7);
\draw [->-] (4.,12.5) -- (2.8,11.7);
\draw [->-] (4.,12.5) -- (1.6,12.5);
\draw [->-] (2.8,11.7) -- (2.8,10.7);
\draw [->-] (1.6,12.5) -- (2.8,10.7);
\node [rotate=0.] at (2.8,12.8) {\scriptsize $g_{12}$};
\node [rotate=0.] at (3.7,11.5) {\scriptsize $g_{23}$};
\node [rotate=0.] at (1.9,11.5) {\scriptsize $g_{13}$};
\node [rotate=0.] at (2.5,12.2) {\scriptsize $gg_1$};
\node [rotate=0.] at (3.1,12.2) {\scriptsize $gg_2$};
\node [rotate=0.] at (2.6,11.5) {\scriptsize $g_3\bar{g}$};
\end{tikzpicture}
}

\newcommand{\unitCellAF}{
\begin{tikzpicture}[baseline={([yshift=-.8ex]current bounding box.center)},scale=1]
\draw [->-] (2.8,11.2) ..controls (2.977,11.408) and (3.154,11.615) .. (3.2,11.8)
			 ..controls (3.246,11.985) and (3.161,12.146) .. (3.,12.2)
			 ..controls (2.839,12.254) and (2.603,12.199) .. (2.5,12.1)
			 ..controls (2.397,12.001) and (2.428,11.857) .. (2.5,11.7)
			 ..controls (2.572,11.543) and (2.686,11.371) .. (2.8,11.2);
\draw [->-] (1.6,12.5) -- (2.8,11.2);
\draw [->-] (4.,12.5) -- (2.8,11.2);
\draw [->-] (4.,12.5) -- (1.6,12.5);
\draw [fill,rotate around={0.:(2.8,11.2)}] (2.8,11.2) ++(0.:0.08 and 0.08) arc(0.:360.:0.08 and 0.08);
\node [rotate=0.] at (2.2,11.6) {\scriptsize $g_1$};
\node [rotate=0.] at (3.4,11.6) {\scriptsize $g_2$};
\node [rotate=0.] at (2.8,12.8) {\scriptsize $g_{12}$};
\node [rotate=0.] at (2.8,12.) {\scriptsize $\mathrm{hol}$};
\node [gray, rotate=0.] at (2.9,11.) {\scriptsize $v$};
\end{tikzpicture}
}

\newcommand{\unitCellAG}{
\begin{tikzpicture}[baseline={([yshift=-.8ex]current bounding box.center)},scale=1]
\draw [->-] (2.8,11.7) ..controls (3.067,11.9) and (3.333,12.1) .. (3.2,12.2)
			 ..controls (3.067,12.3) and (2.533,12.3) .. (2.4,12.2)
			 ..controls (2.267,12.1) and (2.533,11.9) .. (2.8,11.7);
\draw [] (2.8,10.7) ..controls (2.893,10.86) and (2.987,11.02) .. (3.,11.1)
			 ..controls (3.013,11.18) and (2.947,11.18) .. (2.9,11.1)
			 ..controls (2.853,11.02) and (2.827,10.86) .. (2.8,10.7);
\draw [] (2.8,10.7) ..controls (2.773,10.9) and (2.747,11.1) .. (2.7,11.2)
			 ..controls (2.653,11.3) and (2.587,11.3) .. (2.6,11.2)
			 ..controls (2.613,11.1) and (2.707,10.9) .. (2.8,10.7);
\draw [->-] (4.,12.5) -- (2.8,10.7);
\draw [->-] (1.6,12.5) -- (2.8,11.7);
\draw [->-] (4.,12.5) -- (2.8,11.7);
\draw [->-] (4.,12.5) -- (1.6,12.5);
\draw [->-] (2.8,11.7) -- (2.8,10.7);
\draw [->-] (1.6,12.5) -- (2.8,10.7);
\node [rotate=0.] at (2.3,11.9) {\scriptsize $g_1$};
\node [rotate=0.] at (3.3,11.9) {\scriptsize $g_2$};
\node [rotate=0.] at (2.6,11.5) {\scriptsize $g_3$};
\node [rotate=0.] at (2.8,12.8) {\scriptsize $g_{12}$};
\node [rotate=0.] at (3.7,11.5) {\scriptsize $g_{23}$};
\node [rotate=0.] at (1.9,11.5) {\scriptsize $g_{13}$};
\node [rotate=0.] at (2.8,12.1) {\scriptsize $\mathrm{hol}$};
\node [gray, rotate=0.] at (3.,11.6) {\scriptsize $v$};
\end{tikzpicture}
}

\newcommand{\unitCellAH}{
\begin{tikzpicture}[baseline={([yshift=-.8ex]current bounding box.center)},scale=1]
\draw [->-] (2.8,11.7) ..controls (3.067,11.9) and (3.333,12.1) .. (3.2,12.2)
			 ..controls (3.067,12.3) and (2.533,12.3) .. (2.4,12.2)
			 ..controls (2.267,12.1) and (2.533,11.9) .. (2.8,11.7);
\draw [] (2.8,10.7) ..controls (2.893,10.86) and (2.987,11.02) .. (3.,11.1)
			 ..controls (3.013,11.18) and (2.947,11.18) .. (2.9,11.1)
			 ..controls (2.853,11.02) and (2.827,10.86) .. (2.8,10.7);
\draw [] (2.8,10.7) ..controls (2.773,10.9) and (2.747,11.1) .. (2.7,11.2)
			 ..controls (2.653,11.3) and (2.587,11.3) .. (2.6,11.2)
			 ..controls (2.613,11.1) and (2.707,10.9) .. (2.8,10.7);
\draw [->-] (4.,12.5) -- (2.8,10.7);
\draw [->-] (1.6,12.5) -- (2.8,11.7);
\draw [->-] (4.,12.5) -- (2.8,11.7);
\draw [->-] (4.,12.5) -- (1.6,12.5);
\draw [->-] (2.8,11.7) -- (2.8,10.7);
\draw [->-] (1.6,12.5) -- (2.8,10.7);
\node [rotate=0.] at (2.8,12.8) {\scriptsize $g_{12}$};
\node [rotate=0.] at (3.7,11.5) {\scriptsize $g_{23}$};
\node [rotate=0.] at (1.9,11.5) {\scriptsize $g_{13}$};
\node [rotate=0.] at (2.3,11.9) {\scriptsize $gg_1$};
\node [rotate=0.] at (3.3,11.9) {\scriptsize $gg_2$};
\node [rotate=0.] at (2.6,11.5) {\scriptsize $g_3\bar{g}$};
\node [rotate=0.] at (2.8,12.1) {\scriptsize $g\mathrm{hol}\bar{g}$};
\node [gray, rotate=0.] at (3.,11.6) {\scriptsize $v$};
\end{tikzpicture}
}

\newcommand{\unitCellAI}{
\begin{tikzpicture}[baseline={([yshift=-.8ex]current bounding box.center)},scale=1]
\path [lightgray, fill] (2.8,11.) -- (4.,12.8) -- (1.6,12.8) -- cycle;
\draw [white, fill] (2.8,12.) ..controls (3.067,12.2) and (3.333,12.4) .. (3.2,12.5)
			 ..controls (3.067,12.6) and (2.533,12.6) .. (2.4,12.5)
			 ..controls (2.267,12.4) and (2.533,12.2) .. (2.8,12.);
\draw [->-] (2.8,12.) ..controls (3.067,12.2) and (3.333,12.4) .. (3.2,12.5)
			 ..controls (3.067,12.6) and (2.533,12.6) .. (2.4,12.5)
			 ..controls (2.267,12.4) and (2.533,12.2) .. (2.8,12.);
\draw [] (2.8,11.) ..controls (2.893,11.16) and (2.987,11.32) .. (3.,11.4)
			 ..controls (3.013,11.48) and (2.947,11.48) .. (2.9,11.4)
			 ..controls (2.853,11.32) and (2.827,11.16) .. (2.8,11.);
\draw [] (2.8,11.) ..controls (2.773,11.2) and (2.747,11.4) .. (2.7,11.5)
			 ..controls (2.653,11.6) and (2.587,11.6) .. (2.6,11.5)
			 ..controls (2.613,11.4) and (2.707,11.2) .. (2.8,11.);
\draw [->-] (4.,12.8) -- (2.8,11.);
\draw [->-] (1.6,12.8) -- (2.8,12.);
\draw [->-] (4.,12.8) -- (2.8,12.);
\draw [->-] (4.,12.8) -- (1.6,12.8);
\draw [->-] (2.8,12.) -- (2.8,11.);
\draw [->-] (1.6,12.8) -- (2.8,11.);
\node [rotate=0.] at (2.9,11.9) {\scriptsize $v$};
\end{tikzpicture}
}

\newcommand{\unitCellAJ}{
\begin{tikzpicture}[baseline={([yshift=-.8ex]current bounding box.center)},scale=1]
\draw [->-] (2.8,10.6) ..controls (3.067,10.8) and (3.333,11.) .. (3.2,11.1)
			 ..controls (3.067,11.2) and (2.533,11.2) .. (2.4,11.1)
			 ..controls (2.267,11.) and (2.533,10.8) .. (2.8,10.6);
\draw [->-] (4.,11.4) -- (2.8,9.6);
\draw [->-] (1.6,11.4) -- (2.8,10.6);
\draw [->-] (4.,11.4) -- (2.8,10.6);
\draw [->-] (4.,11.4) -- (1.6,11.4);
\draw [->-] (2.8,10.6) -- (2.8,9.6);
\draw [->-] (1.6,11.4) -- (2.8,9.6);
\draw [->-] (5.9,11.9) -- (7.3,10.5);
\draw [->-] (8.7,12.1) -- (7.3,10.5);
\draw [->-] (8.7,12.1) -- (5.9,11.9);
\draw [->-] (9.7,11.8) -- (8.5,10.4);
\draw [->-] (8.5,10.4) -- (7.8,10.7);
\draw [->-] (9.7,11.8) -- (7.8,10.7);
\draw [->-] (5.6,11.1) -- (7.6,10.);
\draw [->-] (7.6,10.) -- (7.4,9.);
\draw [->-] (5.6,11.1) -- (7.4,9.);
\draw [->-] (9.7,10.6) -- (8.2,10.);
\draw [->-] (8.2,10.) -- (8.,9.);
\draw [->-] (9.7,10.6) -- (8.,9.);
\draw [fill,rotate around={0.:(7.3,10.5)}] (7.3,10.5) ++(0.:0.08 and 0.08) arc(0.:360.:0.08 and 0.08);
\draw [fill,rotate around={0.:(7.6,10.)}] (7.6,10.) ++(0.:0.08 and 0.08) arc(0.:360.:0.08 and 0.08);
\draw [fill,rotate around={0.:(8.2,10.)}] (8.2,10.) ++(0.:0.08 and 0.08) arc(0.:360.:0.08 and 0.08);
\draw [fill,rotate around={0.:(8.5,10.4)}] (8.5,10.4) ++(0.:0.08 and 0.08) arc(0.:360.:0.08 and 0.08);
\draw [fill,rotate around={0.:(7.8,10.7)}] (7.8,10.7) ++(0.:0.08 and 0.08) arc(0.:360.:0.08 and 0.08);
\draw [fill,rotate around={0.:(2.8,10.6)}] (2.8,10.6) ++(0.:0.08 and 0.08) arc(0.:360.:0.08 and 0.08);
\node [rotate=0.] at (6.8,10.3) {\scriptsize $g_1$};
\node [rotate=0.] at (8.6,10.) {\scriptsize $g_2$};
\node [rotate=0.] at (7.3,9.6) {\scriptsize $g_3$};
\node [rotate=0.] at (2.8,11.7) {\scriptsize $g_{12}$};
\node [rotate=0.] at (8.8,9.2) {\scriptsize $g_{23}$};
\node [rotate=0.] at (6.5,9.6) {\scriptsize $g_{13}$};
\node [rotate=0.] at (2.8,11.) {\scriptsize $\mathrm{hol}$};
\node [gray, rotate=0.] at (3.,10.5) {\scriptsize $v$};
\node [rotate=0.] at (6.7,11.4) {\scriptsize $g_1$};
\node [rotate=0.] at (7.2,12.2) {\scriptsize $g_{12}$};
\node [rotate=0.] at (9.3,11.) {\scriptsize $g_2$};
\node [rotate=0.] at (7.9,9.4) {\scriptsize $g_3$};
\node [rotate=-15.] at (8.,10.4) {\scriptsize $\mathrm{hol}$};
\node [rotate=0.] at (4.8,10.7) {\scriptsize $\Rightarrow$};
\node [rotate=0.] at (2.3,10.8) {\scriptsize $g_1$};
\node [rotate=0.] at (3.4,10.9) {\scriptsize $g_3$};
\node [rotate=0.] at (3.4,10.2) {\scriptsize $g_{23}$};
\node [rotate=0.] at (2.1,10.2) {\scriptsize $g_{13}$};
\node [rotate=0.] at (2.6,10.2) {\scriptsize $g_2$};
\node [rotate=45.] at (7.8,11.4) {\scriptsize $g_1g_{12}$};
\node [rotate=30.] at (8.8,11.5) {\scriptsize $g_1g_{12}$};
\node [rotate=0.] at (5.,12.8) {\scriptsize $ $};
\node [rotate=0.] at (4.8,8.3) {\scriptsize $ $};
\end{tikzpicture}
}

\newcommand{\unitCellAK}{
\begin{tikzpicture}[baseline={([yshift=-.8ex]current bounding box.center)},scale=1]
\draw [->-] (1.7,12.4) -- (3.1,11.);
\draw [->-] (4.5,12.6) -- (3.1,11.);
\draw [->-] (4.5,12.6) -- (1.7,12.4);
\draw [->-] (6.8,12.5) -- (5.6,11.1);
\draw [->-] (6.8,12.5) -- (5.688,11.888);
\draw [->-] (5.512,11.792) -- (5.327,11.69);
\draw [->-] (5.151,11.593) -- (4.8,11.4);
\draw [->-] (1.5,10.8) -- (3.215,9.857);
\draw [->-] (3.39,9.76) -- (3.5,9.7);
\draw [->-] (1.5,10.8) -- (2.9,8.4);
\draw [->-] (6.2,10.4) -- (4.7,9.8);
\draw [->-] (4.7,9.8) -- (4.4,8.7);
\draw [->-] (6.2,10.4) -- (4.4,8.7);
\draw [->-] (3.5,9.7) -- (3.5,10.5);
\draw [->-] (2.9,8.4) -- (3.5,10.5);
\draw [->-] (4.4,8.7) -- (4.7,10.5);
\draw [->-] (6.2,10.4) -- (4.7,10.5);
\draw [->-] (3.1,11.) -- (3.1,11.8);
\draw [->-] (1.7,12.4) -- (3.1,11.8);
\draw [->-] (4.5,12.6) -- (3.1,11.8);
\draw [->-] (4.8,11.4) -- (4.8,12.3);
\draw [->-] (3.5,9.7) -- (2.9,8.4);
\draw [->-] (4.7,9.8) -- (4.7,10.5);
\draw [->-] (5.6,11.1) -- (5.6,12.);
\draw [->-] (5.6,12.) -- (4.8,12.3);
\draw [->-] (6.8,12.5) -- (5.6,12.);
\draw [->-] (5.6,11.1) -- (4.8,11.4);
\draw [->-] (5.6,11.1) -- (4.8,12.3);
\draw [->-] (1.5,10.8) -- (3.5,10.5);
\draw [->-] (6.8,12.5) -- (4.8,12.3);
\draw [fill,rotate around={0.:(3.1,11.)}] (3.1,11.) ++(0.:0.08 and 0.08) arc(0.:360.:0.08 and 0.08);
\draw [fill,rotate around={0.:(3.5,9.7)}] (3.5,9.7) ++(0.:0.08 and 0.08) arc(0.:360.:0.08 and 0.08);
\draw [fill,rotate around={0.:(4.7,9.8)}] (4.7,9.8) ++(0.:0.08 and 0.08) arc(0.:360.:0.08 and 0.08);
\draw [fill,rotate around={0.:(5.6,11.1)}] (5.6,11.1) ++(0.:0.08 and 0.08) arc(0.:360.:0.08 and 0.08);
\draw [fill,rotate around={0.:(4.8,11.4)}] (4.8,11.4) ++(0.:0.08 and 0.08) arc(0.:360.:0.08 and 0.08);
\draw [fill=white,draw,rotate around={0.:(3.1,11.8)}] (3.1,11.8) ++(0.:0.08 and 0.08) arc(0.:360.:0.08 and 0.08);
\draw [fill=white,draw,rotate around={0.:(4.8,12.3)}] (4.8,12.3) ++(0.:0.08 and 0.08) arc(0.:360.:0.08 and 0.08);
\draw [fill=white,draw,rotate around={0.:(5.6,12.)}] (5.6,12.) ++(0.:0.08 and 0.08) arc(0.:360.:0.08 and 0.08);
\draw [fill=white,draw,rotate around={0.:(3.5,10.5)}] (3.5,10.5) ++(0.:0.08 and 0.08) arc(0.:360.:0.08 and 0.08);
\draw [fill=white,draw,rotate around={0.:(4.7,10.5)}] (4.7,10.5) ++(0.:0.08 and 0.08) arc(0.:360.:0.08 and 0.08);
\node [rotate=0.] at (2.7,10.) {\scriptsize $g_1$};
\node [rotate=0.] at (5.1,9.8) {\scriptsize $g_2$};
\node [rotate=0.] at (3.4,8.9) {\scriptsize $g_3$};
\node [rotate=0.] at (5.3,9.) {\scriptsize $g_{23}$};
\node [rotate=0.] at (2.3,9.2) {\scriptsize $g_{13}$};
\node [rotate=0.] at (2.6,11.7) {\scriptsize $g_1$};
\node [rotate=0.] at (3.,12.7) {\scriptsize $g_{12}$};
\node [rotate=0.] at (6.3,11.7) {\scriptsize $g_2$};
\node [rotate=0.] at (4.7,9.3) {\scriptsize $g_3$};
\node [rotate=-15.] at (5.1,11.1) {\scriptsize $\mathrm{hol}$};
\node [rotate=45.] at (3.9,11.6) {\scriptsize $g_1g_{12}$};
\node [rotate=30.] at (6.,12.) {\scriptsize $g_1g_{12}$};
\node [rotate=0.] at (3.2,11.5) {\scriptsize $g$};
\node [rotate=0.] at (5.7,11.5) {\scriptsize $g$};
\node [rotate=0.] at (4.7,11.7) {\scriptsize $g$};
\node [rotate=0.] at (3.7,10.1) {\scriptsize $g$};
\node [rotate=0.] at (4.8,10.1) {\scriptsize $g$};
\end{tikzpicture}
}

\newcommand{\unitCellAL}{
\begin{tikzpicture}[baseline={([yshift=-.8ex]current bounding box.center)},scale=1]
\draw [->-] (2.8,10.7) -- (1.6,12.5);
\draw [->-] (4.,12.5) -- (2.8,10.7);
\draw [->-] (2.8,10.7) -- (2.8,11.7);
\draw [->-] (1.6,12.5) -- (2.8,11.7);
\draw [->-] (4.,12.5) -- (2.8,11.7);
\draw [->-] (4.,12.5) -- (1.6,12.5);
\node [rotate=0.] at (2.8,12.8) {\scriptsize $g_{12}$};
\node [rotate=0.] at (3.7,11.5) {\scriptsize $g_{23}$};
\node [lightgray, rotate=0.] at (3.,11.6) {\scriptsize $v$};
\node [rotate=0.] at (1.9,11.5) {\scriptsize $g_{13}$};
\node [rotate=0.] at (2.5,12.2) {\scriptsize $gg_1$};
\node [rotate=0.] at (3.1,12.2) {\scriptsize $gg_2$};
\node [rotate=0.] at (2.6,11.5) {\scriptsize $gg_3$};
\end{tikzpicture}
}

\newcommand{\unitCellAM}{
\begin{tikzpicture}[baseline={([yshift=-.8ex]current bounding box.center)},scale=1]
\draw [->-] (1.6,12.5) -- (2.8,11.2);
\draw [->-] (4.,12.5) -- (2.8,11.2);
\draw [->-] (4.,12.5) -- (1.6,12.5);
\node [rotate=0.] at (2.2,11.6) {\scriptsize $g_1$};
\node [rotate=0.] at (3.4,11.6) {\scriptsize $g_2$};
\node [rotate=0.] at (2.8,12.8) {\scriptsize $g_{12}$};
\node [gray, rotate=0.] at (2.8,12.) {\scriptsize $p$};
\end{tikzpicture}
}

\newcommand{\vertexDiagramAA}{
\begin{tikzpicture}[baseline={([yshift=-.8ex]current bounding box.center)},scale=1]
\draw [->-] (5.,12.8) -- (3.4,10.8);
\draw [->-] (3.4,10.8) -- (3.4,12.);
\draw [->-] (1.6,12.8) -- (3.4,12.);
\draw [->-] (5.,12.8) -- (3.4,12.);
\draw [->-] (1.6,12.8) -- (3.4,10.8);
\draw [->-] (5.,12.8) -- (1.6,12.8);
\node [lightgray, rotate=0.] at (3.2,11.9) {\scriptsize $v$};
\node [rotate=-25.] at (2.9,12.5) {\scriptsize $(x_1,\rho_1)$};
\node [rotate=25.] at (4.,12.5) {\scriptsize $(x_2,\rho_2)$};
\node [rotate=-90.] at (3.6,11.5) {\scriptsize $(x_3,\rho_3)$};
\end{tikzpicture}
}

\newcommand{\vertexDiagramAB}{
\begin{tikzpicture}[baseline={([yshift=-.8ex]current bounding box.center)},scale=1]
\draw [->-] (5.2,13.) -- (3.4,10.6);
\draw [->-] (3.4,10.6) -- (3.4,12.);
\draw [->-] (1.4,13.) -- (3.4,12.);
\draw [->-] (5.2,13.) -- (3.4,12.);
\draw [->-] (1.4,13.) -- (3.4,10.6);
\draw [->-] (5.2,13.) -- (1.4,13.);
\node [lightgray, rotate=0.] at (3.2,11.9) {\scriptsize $v$};
\node [rotate=29.] at (4.,12.6) {\scriptsize $(xx_2,\rho^{\bar x}_2)$};
\node [rotate=-90.] at (3.6,11.5) {\scriptsize $(xx_3,\rho^{\bar x}_3)$};
\node [rotate=-25.] at (2.8,12.5) {\scriptsize $(xx_1,\rho^{\bar x}_1)$};
\end{tikzpicture}
}

\title{Electric-Magnetic duality in twisted quantum double model of topological orders}
\date{\today}
\author[a,b]{Yuting Hu}
\author[b,c,a]{Yidun Wan}
\affiliation[a]{Department of Physics and Institute for Quantum Science and Engineering, South University of Science and Technology, Shenzhen 518055, China}
\affiliation[b]{State Key Laboratory of Surface Physics, Fudan University, Shanghai 200433, China}
\affiliation[c]{Department of Physics, Center for Field Theory and Particle Physics, and Institute for Nanoelectronic devices and Quantum computing, Fudan University, Shanghai 200433, China}
\emailAdd{yuting.phys@gmail.com, ydwan@fudan.edu.cn}

\abstract{
We derive a partial electric-magnetic (PEM) duality transformation of the twisted quantum double (TQD) model TQD$(G,\alpha)$---discrete Dijkgraaf-Witten model---with a finite gauge group $G$, which has an Abelian normal subgroup $N$, and a three-cocycle $\alpha \in H^3(G,U(1))$. Any equivalence between two TQD models, say, TQD$(G,\alpha)$ and TQD$(G',\alpha')$, can be realized as a PEM duality transformation, which exchanges the $N$-charges and $N$-fluxes only. Via the PEM duality, we construct an explicit isomorphism between the corresponding TQD algebras $D^\alpha(G)$ and $D^{\alpha'}(G')$ and derive the map between the anyons of one model and those of the other. 
}

\begin{document}
	
	\maketitle
	\flushbottom

	\section{Introduction}
	As an important theme in modern physics, dualities weave together apparently different physical theories, such that not only the theories can be understood from each other's perspective but also they combined can deepen our understanding of the fundamental physics underneath. Needless to mention the popular gauge-gravity duality and the dualities between gauge theories in different dimensions, in this work, we shall construct a duality between certain types of lattice gauge theories in $2+1$ dimensions, which can serve as effective models of topological orders in two spatial dimensions. We shall name this duality a partial electric-magnetic (PEM) duality for reasons to be explained shortly. 
	
\begin{figure}[!ht]
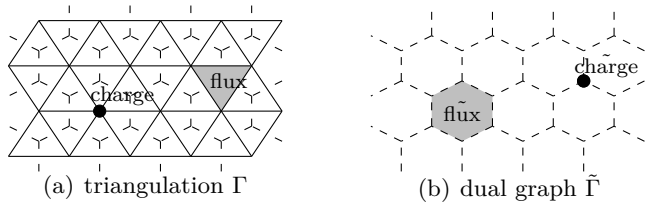

		\centering\subfigure[triangulation $\Gamma$]{\DoubleLayerGraphAB}
		\qquad
		\subfigure[dual graph $\tilde{\Gamma}$]{\DoubleLayerGraphAC}
		\caption{EM duality between an Abelian gauge theory defined on the triangulation lattice $\Gamma$ and its dual model on the dual graph $\tilde\Gamma$. Gauge charges on vertices and gauge fluxes on plaquettes for gauge group $G$ are mapped to dual fluxes and dual charges for the dual gauge group $\hat{G}$ on the dual graph.}
		\label{fig:DualityAbelian}
	\end{figure}

	A lattice gauge theory has EM duality if the gauge group $G$ is Abelian. A well known example is the Ising model\cite{Kramers1941StatisticsI}. Such duality is important to understand the matter phases and phase transitions. Under the EM duality, the gauge charges and gauge fluxes are exchanged in the dual theory. See Fig. \ref{fig:DualityAbelian} for an example. The dual gauge group is $\hat{G}=Irrep(G)$ whose elements are unitary irreducible representations of $G$. 
	
	When the gauge group $G$ is non-Abelian, there is no such EM duality for the entire gauge group $G$ because the irreducible representations of $G$ do not form a group. Nevertheless, if a normal Abelian subgroup $N\subseteq G$ exists, there could be partial EM duality which exchanges $N$-charges and $N$-fluxes only. Such a duality is what we mean by a PEM. 
	
	A large class of topological phases can be described by discrete topological gauge field theories, such as the quantum double (QD) models\cite{Kitaev2003a}, the twisted quantum double (TQD) models\cite{Dijkgraaf1990,Hu2012a,Mesaros2011}, and the Levin-Wen model\cite{Levin2004}.
	When QD models are extended (with the defining finite groups generalized to Hopf algebras), EM dualities (that exchange charges and fluxes for the gauge Hopf algebras) can be realized (\cite{Buerschaper2009,Buerschaper2013,Hu2018,Wang2020Electric-magneticBoundaries}).
	
	On the other hand, two TQD models with different input data (groups and 3-cocycles) may be equivalent. An example is the equivalence between the QD model with $G=D_4$ and a TQD model with $G'=\mathbb{Z}_{2} \times \mathbb{Z}_{2} \times \mathbb{Z}_{2}$ and some nontrivial 3-cocycle over $G'$ (\cite{Propitius1995}). The topological quantum numbers (modular $S,T$ matrices) are identical in both models. 
	
	In this paper, we show that such equivalences can be constructed via the PEM duality in TQD models. In general, it is known whether two TQD models are equivalent [\cite{Davydov2012,Naidu2007,UribeJongbloed2017}, in the sense that their corresponding quantum double categories, which characterize the topological phases of the the corresponding TQD models, are equivalent. We find that for every such an equivalence of two TQD models, there exists a PEM duality transformation between the two models. 
	
	Given an existing normal Abelian subgroup $N\subset G$, the PEM duality (if exists) maps $N$-charges/fluxes to $\hat{N}$-fluxes/charges in the dual model, while $K$-charges/fluxes remain unchanged, where $K\equiv N\backslash G$ is the right quotient group. Such a PEM duality should be formulated by a Fourier transform over $N$ and $\hat{N}$, which we call a partial Fourier transform.

	To derive the PEM dulaity, we require $\alpha|_N=1$, such that the TQD$(G,\alpha)$ model contains a QD$(N)$ model, which is mapped to the QD$(\hat{N})$ model on the dual graph $\tilde \Gamma$. Hence the PEM dualtiy maps $N$-charges to $\hat{N}$-fluxes (and vice versa). With some extra condition on $\alpha$, we show that the dual operators generate the TQD algebra $D^{\alpha'}({G'})$ \cite{Dijkgraaf1991}. 
	
In general, a TQD model is not self-dual under the PEM duality transformation: Not only the lattice structure is changed (the triangulation is mapped to a reciprocal bilayer graph) but also the gauge group $G$ is mapped to a dual group $G'$, while the algebra of the local operators are mapped from $D^{\alpha } G$ to $D^{\alpha '} G'$. Every equivalence between two TQD models, say,  TQD$(G,\alpha)$ and TQD$(G',\alpha')$, can be realized by a PEM duality transformation.

We derive an isomorphism between the TQD algebras $D^\alpha(G)$ and $D^{\alpha'}(G)$, which is the mathematics behind the PEM duality. With such an isomorphism, we can explicitly construct the duality transformation of the anyons in a TQD model to those of the dual model, in terms of the representations of the corresponding TQD algebras.

The PEM duality begs to reformulate the  TQD$(G,\alpha)$ model as a bilayer model, namely a coupling of a QD$(N)$ model on the upper layer and a model on the lower layer. The model on the lower layer may not be a TQD model in general because its input data consists of the group $K$ and the 3-cochain $\epsilon$. Under the PEM duality, the upper layer model is mapped to the QD$(\hat{N})$ model on the dual graph $\tilde\Gamma$. The lower layer remains unchanged, as defined on $\Gamma$. We call the dual model a \textbf{reciprocal bilayer model}.

	
	\section{Review of EM-duality in discrete Abelian gauge theories}

	In this section, we briefly review some known examples of EM-duality in discrete gauge theories.
	
\begin{figure}[htb]
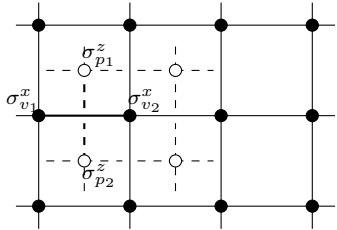

		\begin{center}
			\dualLatticeIsing
		\end{center}
		\caption{The EM duality between the Ising model on the lattice (with solid lines), and the dual model (with dashed lines) on dual lattice. The plaquettes $p_1,p_2,\dots$ in the original lattices are identified as vertices on the dual lattice. The EM duality leads to the correspondence $\{\sigma^x_v\}\leftrightarrow\{\sigma^z_p\}$. }
		\label{fig:dualLatticeIsing}
	\end{figure}

	We focus on the EM-duality in the two-dimensional Ising model\cite{Kramers1941StatisticsI}, which is defined on a square lattice, while the dual model is defined on the dual lattice (see Fig. \ref{fig:dualLatticeIsing}). The duality transformation relates the observables in the Ising model at high temperature to their counterparts in the dual model at low temperature.
	
	Such an EM-duality transformation is essentially a Fourier transform, which is better understood by treating the Ising model as a discrete gauge theory. In the original Ising model, the spins $\sigma^x_v$ at the vertices $v$ of the lattice are the Pauli $x$-matrix. Each spin yields a $\mathbb{Z}_2$ group because $(\sigma^x)^2 =1$, where $1$ is the $2\times 2$ identity matrix, such that $\{\sigma^x, (\sigma^x)^2\} \cong \mathbb{Z}_2$. To understand the Ising model as a gauge theory, we can gauge the Ising model by trading the spins at the vertices by gauge degrees of freedom on the links. For example, two neighbouring spins $\sigma_{v_1}^x$ and $\sigma_{v_2}^x$ are traded with the degree of freedom $\sigma^x_{v_1}\sigma^x_{v_2} := \sigma^x_{v_1}\otimes\sigma^x_{v_2}$, which is an abbreviation of the full tensor product that involves the identity operators at all the other vertices of the lattice, on the link between $v_1$ and $v_2$. Clearly, the gauge group is also $\mathbf{Z}_2$: If we define $g=\sigma^x_{v_1}\sigma^x_{v_2}$, then $g^2 = 1$, the $4\times 4$ identity matrix as a submatrix of the full identity matrix. 
	
	In the dual Ising model, however, the spins at the vertices $p$ are the Pauli $z$-matrix $\sigma_p^z$. The gauge degrees of freedom on the dual links (dashed lines in Fig. \ref{fig:dualLatticeIsing}) are thus $\sigma^z_{p_1}\sigma^z_{p_2}:=\sigma^z_{p_1}\otimes\sigma^z_{p_2}$. The elements $\sigma^x_{v_1}\sigma^x_{v_2}$ on the links and $\sigma^z_{p_1}\sigma^z_{p_2}$ on the dual links are related by a Fourier transform in the Hilbert space as follows. If we define $s=(1+\sigma^z_{p_1}\sigma^z_{p_2})/2$, then we can specify the local basis $\ket{g}$ on a link in the Ising model (such that $g\ket{g'}=\ket{gg'}$), and the local basis $\ket{s}$ in the dual Ising model respectively. The EM duality transformation is then a Fourier transform between these two bases. Such a Fourier transform can be conveniently formulated in the language of group representation theory. In the above local basis, $g$ effectively takes values in $\mathbb{Z}_2=\{+1,-1\}$, and $s$ takes values of $0,1$ that label the two irreducible representations of $\mathbb{Z}_2$. The Fourier transform on the local basis of the Hilbert space reads
	\begin{equation}
	\label{eq:AbelianFT}
	|s\rangle =\frac{1}{\sqrt{|G |}}\sum _{g} \overline{\rho _{s}( g) }|g\rangle.
	\end{equation}
	The $\rho _{s}(g)$ is the representation matrix of $g$ in the irreducible representation $s$. 
	
	To study the duality transformation of the observables, we examine a Fourier transform on the statistical weight of the model. For simplicity, we assume the absence of external magnetic field. Each link contributes to the statistical weight  a factor $\lambda ( g)$, with  $\lambda ( \pm ) =\exp( \pm \beta )$ at the inverse temperature $\beta $. Similarly, in the dual model, each link contributes $\tilde{\lambda }( s) =\exp\left((-1)^s\tilde{\beta }\right)$. The Fourier transform \eqref{eq:AbelianFT} induces the transformation of the statistical weight:
	\begin{equation}
	\tilde{\lambda }( s) =\frac{1}{|\mathbb{Z}_{2} |}\sum _{g} \rho _{s}( g) \lambda ( g),
	\end{equation}
	which reads $\tilde{\lambda }( 0) =\sinh \beta$, $\tilde{\lambda }( 1) =\cosh \beta $ (For derivation, e.g., see \cite{Itzykson1989StatisticalTheory}). The relation between $\beta$ and $\tilde{\beta}$ is derived from the identification $\tilde{\lambda }( s) =\exp\left((-1)^s\tilde{\beta }\right)$:
	\begin{equation}
	\sinh 2\beta \sinh 2\tilde{\beta } =1,
	\end{equation}
	which is known as the Kramers-Wannier duality.
	
	In the above example, We formulate the EM duality in the Ising model by a Fourier transform on the local Hilbert space and on the observables of the model. Such an EM duality can be generalized to Abelian gauge theories defined on a lattice (or on a simplicial complex in general) and described by a similar Fourier transform (for example, see \cite{Itzykson1989StatisticalTheory}). Under the Fourier transform in Eq. \eqref{eq:AbelianFT}, the dual group is formed by all unitary one-dimensional irreducible representations, denoted by $G'$, albeit $G'\cong G$ in Abelian cases.
	
	The EM duality discussed above can not be generalized directly such that the dual model is a gauge theory for certain group $G'$ because the irreducible representations in such cases do not form a group\footnote{In non-Abelian cases, the EM duality discussed above can be generalized in the framework of generalized gauge theory with gauge quantum groups. Then the EM duality following the above approach maps a non-Abelian gauge group to a gauge quantum group.}.  In this paper, we propose a partial EM duality to solve this problem. If there exists a normal Abelian subgroup $N$ of the gauge group $G$,
	then the EM duality could still exist via the partial Fourier transform over $N$, which is the reason we call such duality a partial EM duality.
	

	\section{Conditions on PEM duality}
	
	In this section, we formulate the conditions on the existence of PEM duality. 
	
	Two TQD models with different input data (groups and 3-cocycles) may be equivalent. For example, the QD model with $G=D_4$ is equivalent to the TQD model with $G'=\mathbb{Z}_{2} \times \mathbb{Z}_{2} \times \mathbb{Z}_{2}$ and certain nontrivial 3-cocycle $\alpha'$ over $G'$. Two TQD models, TQD$(G,\alpha)$ and TQD$(G',\alpha')$, are equivalent if they yield the same set of topological quantum numbers, characterizing the same topological order. Mathematically, this happens if the representation categories $Rep_{(D^\alpha G)}$ and $Rep_{(D^{\alpha'} G')}$ of the TQD algebras $D^\alpha(G)$ and $D^{\alpha'}(G')$. The existence conditions of such equivalences as found by Naidu and others \cite{Davydov2012,Naidu2007} are reviewed briefly as follows.
	
	We follow the language in Ref\cite{Naidu2007}. Let $Vec_G^\alpha$ be the fusion category whose objects are vector spaces graded by $G$ and associativity dictated by $\alpha$. Two fusion categories $Vec_G^\alpha$ and $Vec^{\alpha'}_{G'}$ are weakly Morita-equivalent (\cite{Muger2003}) if  their categories centers are equivalent as braided tensor categories, i.e.,
	\begin{equation}
	\mathcal{Z} (Vec^{\alpha }_{G} )\cong \mathcal{Z} (Vec^{\alpha '}_{G'} ).
	\end{equation}
	Note that $\mathcal{Z} (Vec^{\alpha }_{G} )\cong Rep_{(D^\alpha G)} $ as braided tensor categories. Given a right-module category $\mathcal{M}$ over the fusion category $\mathcal{C}=Vec^{\alpha }_{G}$, we denote the dual category (see \cite{Ostrik2003a} ) of $\mathcal{C}$ by $\mathcal{C}^*_{\mathcal{M}}:=Fun_\mathcal{C}(\mathcal M,\mathcal M)$, whose objects are  the $\mathcal{C}$-module functors from $\mathcal{M}$ to itself and morphisms are natural module transformations.
	
	The weakly Morita-equivalence holds if there exists a right-module category $\mathcal{M}$ such that the dual category $\mathcal{C}^*_{\mathcal{M}}$ is equivalent to $Vec^{\alpha '}_{G'}$ for certain $G'$ and $\alpha'$. According to Ref\cite{Naidu2007} such an $\mathcal{M}$ exists if
	\begin{enumerate}
		\item $G$ contains a normal Abelian subgroup $N$, such that $\alpha|_N$ is trivial in the third cohomology group $H^3(N,U(1))$.
		\item There is a 2-cochain $\mu\in H^2(G,Map(K,U(1)))$, such that $\delta^2 \mu=\alpha$ with $\alpha\in H^3(G, Map(K, \mathbb{C}))$ (where $\alpha$ is viewed as a constant valued 3-cocycle), and the cohomology class $[\mu^y/\mu]$ is trivial in $H^2(G,Map(K,U(1)))$ $\forall y\in K$.
	\end{enumerate}
	Here $Map(K,U(1))$ is a function space with the left $G$-actions defined by $(g\triangleright f)(k)=f(k\triangleleft g)$, where $f:K\rightarrow U(1)$,  $g\in G$,  $k\in K$, and $k\triangleleft g$ is the right $G$-action on the quotient group $K$. When these conditions are met, we can construct a module category $\mathcal{M}(K,\mu)$ whose simple objects are given by elements in $K$ and associativity  given by $\mu$. The category $\mathcal{C}^*_{\mathcal{M}(K,\mu)}$ is equivalent to $Vec^{\alpha '}_{G'}$  for certain $G'$ and $\alpha'$.
	
	Later, Uribe \cite{UribeJongbloed2017} formulated the conditions on the equivalence in terms of explicit representatives of $\alpha$ and $\mu$. In this formulation, the categories $Vec^{\alpha }_{G}$ and $Vec^{\alpha '}_{G'}$ are weakly Morita-equivalent if and only if
	\begin{enumerate}
		\item There exists a normal Abelian subgroup $N\subset G$. Then, there exists certain $2$-cocycle $F\in H^{2}( K,N)$  $G$ can be written as a semidirect product $G=N\rtimes _{F} K$.
		\item There exists a 2-cocycle $\hat{F} \in H^{2}( K,\hat{N})$, such that the 4-cochain $\hat{F} \land F$ defined by $(\hat{F} \land F)( k_{1} ,k_{2} ,k_{3} ,k_{4}) :=\hat{F}( k_{1} ,k_{2})( F( k_{3} ,k_{4}))$ is cohomologically trivial in $H^{4}( K,U( 1))$, where $\hat{N}$ is the Abelian group whose elements are the unitary irreducible representations of $N$, and $k_i\in K$. In other words, there exists a 3-cochain $\epsilon \in C^{3}( K,U( 1))$, such that $\delta _{K} \epsilon =\hat{F} \land F$, i.e., 
		\begin{equation}
		\label{eq:FFcondition}
		\delta_K\epsilon ( k_{1} ,k_{2} ,k_{3} ,k_{4}) 
		=\frac{\epsilon ( k_{2} ,k_{3} ,k_{4}) \epsilon ( k_{1} ,k_{2} k_{3} ,k_{4}) \epsilon ( k_{1} ,k_{2} ,k_{3} ) }{\epsilon ( k_{1} k_{2} ,k_{3} ,k_{4}) \epsilon ( k_{1} ,k_{2} ,k_{3} k_{4}) }
		=\hat{F}( k_{1} ,k_{2})( F( k_{3} ,k_{4})).
		\end{equation}
	\end{enumerate}
	When there exists $F$ and $\hat{F}$ satisfying these two conditions, there is a weakly Morita equivalence $Vec_G^\alpha \cong Vec_{G'}^{\alpha'}$, with $G=N\rtimes _{F} K$, $G'=K\ltimes _{\hat{F}}\hat{N}$, and the 3-cocycles are (up to a coboundary)
	\begin{equation}
	\label{eq:omegaFepsilon}
	\alpha ((a_{1} ,k_{1} ),(a_{2} ,k_{2} ),(a_{3} ,k_{3} ))=\hat{F} (k_{1} ,k_{2} )(a_{3} )\epsilon (k_{1} ,k_{2} ,k_{3} ),
	\end{equation}
	\begin{equation}
	\label{eq:omegahatFepsilon}
	\alpha '((x_{1} ,\rho _{1} ),(x_{2} ,\rho _{2} ),(x_{3} ,\rho _{3} ))=\rho _{1} (F(x_{2} ,x_{3} ))\epsilon (x_{1} ,x_{2} ,x_{3} ).
	\end{equation}
	Note that $\alpha |_{N} =1$ and $\alpha '|_{\hat{N}} =1$.

	In this paper, we will adapt the explicit representatives of $\alpha$ and $\alpha'$ in Eqs. \eqref{eq:omegaFepsilon} and \eqref{eq:omegahatFepsilon}. We show that for every weakly Mortia equivalence $Vec_G^\alpha \cong Vec_{G'}^{\alpha'}$, we can derive a PEM dulaity by a partial Fourier transform to be defined. Note that since our PEM duality is a local duality transformation, it does not induce any topological phase transition.

	\section{Main results}
	
	In this section, we summarize the main result of this paper. We consider the quantum double (QD) models and the more general twisted quantum double (TQD) models defined on a graph $\Gamma$ as a triangulation of a closed surface. These models are discrete topological gauge theories describing time-reversal invariant topological orders. 
	
\begin{figure}[htb]
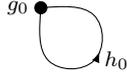

		\begin{center}
			\algebraQD
		\end{center}
		\caption{The $g_0$ at the vertex and $h_0$ at the loop $p$.}
		\label{fig:algebraQD}
	\end{figure}

	The elementary excitations in these models are gauge charges, gauge fluxes, and dyons (charge-flux composites) living on $\Gamma$. Charges (fluxes) are local excitations at the vertices (plaquettes) violating the Gauss law (flatness condition), which is implemented by local gauge-transformation operators at the vertices (holonomy measurement operators). In this section, we will explain these operators in a minimal setting, and leave the detailed discussion in the later sections. In this minimal setting, we only consider one vertex $v$ and one plaquette adjacent to $v$ in $\Gamma$. The plaquette can be homeomorphically minimized to a disk bounded by a single loop $p$ attached to $v$ (See Fig. \ref{fig:algebraQD}). Then we can define the Hilbert space and the local operators at $v$ and on $p$ as follows. The local Hilbert space is spanned by two group elements, $g_0$ at $v$, and $h_0$ on $p$. The $h_0$ is the \textit{holonomy} along the loop. We define the local gauge transformation operator by
	\begin{equation}
	A^{g} |g_0,h_0\rangle =|gg_0,gh_0\bar g \rangle,
	\end{equation}
	and the holonomy measurement operator $B^h$ by
	\begin{equation}
	B^{h} |g_0,h_0\rangle =\delta _{h,h_0} |g_0,h_0\rangle ,
	\end{equation}
	where $g,h,g_0,h_0\in G$. 	Throughout this paper, we will denote the inversed group element by a bar: $\bar a=a^{-1}$

	The operators $A^g$ and $B^h$ form the Drinfeld's quantum double algebra $D(G)$, whose multiplication rule is
	\begin{equation}
	(A^{g_{1}} B^{h_{1}})( A^{g_{2}} B^{h_{2}}) =\delta _{h_{1} ,g_{2} h_{2} \bar{g_{2}}} (A^{g_{1} g_{2}} B^{h_{2}}).
	\end{equation}
	It is the algebra of local  observables in the QD models, i.e.,  for all $g,h\in G$, the $A^g$ and $B^h$ commute with the Hamiltonian (will be defined in later sections). In the TQD models, we can generalize the operators $A^g$ and $B^h$, which give rise to a TQD algebra $D^{\alpha} G$, where $\alpha \in H^3[G, U(1)]$. The algebra multiplication rule becomes
	\begin{equation}
	(A^{g_{1}} B^{h_{1}} )(A^{g_{2}} B^{h_{2}} ):=\delta _{h_{1} ,g_{2} h_{2} \bar{g_2}} \beta _{h_{2}} (g_{1} ,g_{2} )A^{g_{1} g_{2}} B^{h_{2}},
	\end{equation}
	where $\beta _{h_{2}} (g_{1} ,g_{2} )$ is given by
	\begin{equation}
	\beta _{h_{2}} (g_{1} ,g_{2} ):=\frac{\alpha (g_{1} ,g_{2} ,h_{2} )\alpha \left( g_{1} g_{2} h_{2} \bar{g_{2}}\bar{g_1} ,g_{1} ,g_{2}\right)}{\alpha \left( g_{1} ,g_{2} h_{2} \bar{g_2} ,g_{2}\right)}.
	\end{equation}
	When $\alpha =1$, $D^\alpha G$ reduces to a QD algebra. (See the detailed definition of TQD algebras, see Appendix \ref{sec:semidirect}.)
	
	We already know the EM duality transformation in Abelian gauge theories\cite{Kramers1941StatisticsI,Itzykson1989StatisticalTheory}. In the QD models with Abelian group $G$, the EM duality is essentially a Fourier transform on $G$ and maps charges to dual fluxes and vice versa. When the group $G$ is non-Abelian, no gauge group structure survive under the Fourier transform on $G$.  
	
	Nevertheless, if $G$ has a normal Abelian subgroup $N$, we may apply a partial Fourier transform on $N$. This partial Fourier transform is called a PEM duality transformation if the dual model can be identified as a topological gauge theory with observables forming the TQD algebra $D^{\alpha'}(G')$ for certain finite gauge group $G'$. Nevertheless, in general, a partial Fourier transform cannot be a PEM duality because a PEM duality is supposed to realize a Morita equivalence $Vec_G^{\alpha}\cong Vec_{G'}^{\alpha'}$, which is possible if and only if the conditions \eqref{eq:FFcondition}\eqref{eq:omegaFepsilon} and \eqref{eq:omegahatFepsilon} are met. We assume hereafter these conditions are fulfilled.
    
Let $N\subset G$ , such that $G$ is a semidirect product $G=N\rtimes_F K$ where $K=N\backslash G$ is the quotient group, and $F\in H^2(K,N)$ is a 2-cocycle . The 3-cocycle $\alpha \in H^3(G,U(1))$ has the form in Eq. \eqref{eq:omegaFepsilon} for some certain 2-cocycle $\hat{F}\in H^2(K,\hat{N})$.  For the semidirect product structure, see Appendix \ref{sec:semidirectStructure}. For $(a,x) \in N\rtimes_F K$, we define the partial Fourier transform:
	\begin{equation}\label{eq:FTtransformation}
	|x,\rho \rangle =\frac{1}{\sqrt{|N|}}\sum _{a\in N} \overline{\rho ( a) }|a,x\rangle .
	\end{equation}
	In the pair $( x,\rho )$, $\rho\in \hat{N}$, where $\hat{N}$ is the Abelian group whose elements are the unitary irreducible representations of $N$.

   In what follows, we will define the dual local operators. The pairs $(x,\rho)$ form the dual group $G'=K\ltimes_{\hat{F}} \hat{N}$ with the semidirect product structure specified by $\hat{F}$. (See Appendix \ref{sec:semidirectStructure} for the details about the semidirect product structures of $G$ and $G'$.) We will define the partial Fourier transform on the local operators by
	\begin{equation}
    \label{eq:BAtildeAB}
\tilde{B}^{\left(xy\bar{x} ,\rho \right)}\tilde{A}^{( x,\eta )} =	\frac{1}{|N|}\sum _{a,b\in N} \overline{\rho (a)} \eta (b) A^{( a,x)} B^{( b,y)}.
	\end{equation}
where $\rho,\eta\in \hat{N}$. The the dual local operators in the dual model are 
\begin{equation}
\tilde{A}^{(x,\eta)}=\sum_{y,\rho}\tilde{B}^{\left(xy\bar{x} ,\rho \right)}\tilde{A}^{( x,\eta )},
\end{equation}
\begin{equation}
\tilde{B}^{(y,\rho)}=\tilde{B}^{\left(xy\bar{x} ,\rho \right)}\tilde{A}^{(1_K,1_{\hat{N}} )}.
\end{equation}
In later sections, we show that the dual operators generate the TQD algebra $D^{\alpha'}(G')$, where $\alpha'\in H^3(G',U(1))$ takes the form in Eq. \eqref{eq:omegahatFepsilon}, i.e., they satisfy
    	\begin{equation}
    	\label{eq:BABATQDalgebra}
	\tilde{B}_{\tilde p}^{h'_{1}} \tilde{A}_{\tilde v}^{g'_{1}}\tilde{B}_{\tilde p}^{h'_{2}} \tilde{A}_{\tilde v}^{g'_{2}} =\delta _{h'_{1} ,g'_{1} h'_{2}\bar{g'_{1}}} \beta '_{h'_{1}} (g'_{1} ,g'_{2} )\tilde{B}_{\tilde p}^{h'_{1}} \tilde{A}_{\tilde v}^{g'_1g'_2},
	\end{equation}
	where\begin{equation}
	\beta '_{h'_{1}} (g'_{1} ,g'_{2} ):=\frac{\alpha '(h'_{1} ,g'_{1} ,g'_{2} )\alpha '\left( g'_{1} ,g'_{2} ,\bar{g'_2}\bar{g'_1} h'_{1} g'_{1} g'_{2}\right)}{\alpha '\left( g'_{1} ,\bar{g'_1} h'_{1} g'_{1} ,g'_{2}\right)},
	\end{equation}
	with $\alpha' \in H^3[G',U(1)]$ given by Eq. \eqref{eq:omegahatFepsilon}. Here $h_i', g_i' \in G'$.
	
	By definition, the partial Fourier transform acts on the Hilbert subspace spanned by the elements of $N$. It is convenient to factorize the local operators into $N$ part and $K$ part by setting
	\begin{equation}
	\label{eq:AaAx} A^{a} =A^{(a,1)} ,A^{x} =A^{(1,x)} ,B^{b} =\sum _{y\in K} B^{(b,y)} ,B^{y} =\sum _{b\in N} B^{(b,y)},
	\end{equation}
	and 
	\begin{equation}
	\tilde{A}^{\eta } =\tilde{A}^{(1,\eta )} ,\tilde{A}^{x} =\tilde{A}^{(1,x)} ,\tilde{B}^{\rho } =\sum _{y\in K}\tilde{B}^{(y,\rho )} ,\tilde{B}^{y} =\sum _{\rho \in \tilde{N}}\tilde{B}^{(y,\rho )}.
	\end{equation}
As consequence of the partial Fourier transform \eqref{eq:BAtildeAB}, we have 
	\begin{equation}
	\label{eq:mapABa}
	A^{a}  =\sum _{\rho \in \hat{N}}{\rho (a)}\tilde{B}^{\rho }, B^{b} =\frac{1}{|N|}\sum _{\eta \in \hat{N}}\overline{\eta (b)} \ \tilde{A}^{\eta }
	\end{equation}
	\begin{equation}
	    \tilde{A}^x=A^x,\tilde{B}^y=B^y.
	\end{equation}	
	In later sections, we show that the matrices of $\tilde{A}^\eta$ and $\tilde{B}^\rho$ in the dual basis of the Hilbert space define a QD$(\hat{N})$ model on the dual graph $\tilde \Gamma$ of the original graph $\Gamma$, where the Hilbert subspace spanned by $\hat{N}$ elements. See Fig. \ref{fig:dualGraph} for the details how we define $\tilde \Gamma$. The identification of the dual model on $\tilde \Gamma$ with the QD$(\hat{N})$ model, together with that  $N$-charges are mapped to dual $\hat{N}$-fluxes (and vice versa),  justifies the terminology of the PEM duality, whose definition is summarized as follows and to be derived in later sections. 
    
\begin{figure}[!ht]
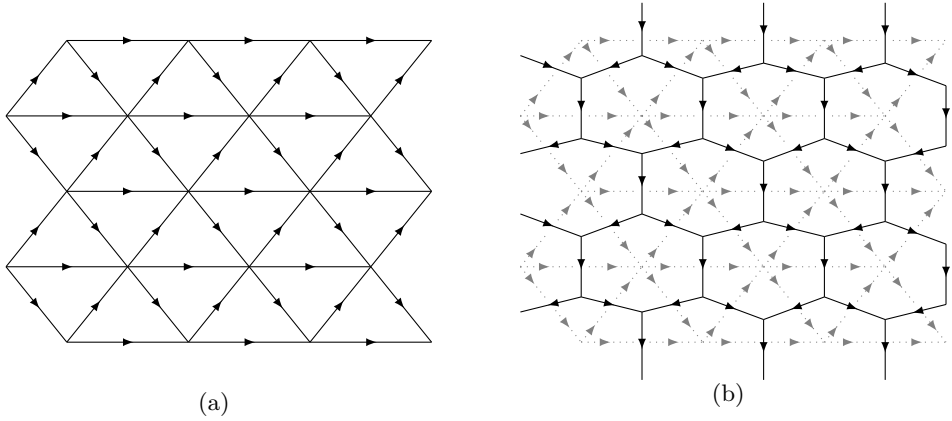

		\begin{center}
			\subfigure[]{\PEMdualGraphAA}
			\qquad
			\subfigure[]{\PEMdualGraphAB}
		\end{center}
		\caption{For the triangulation $\Gamma$ in (a), we define the dual graph $\tilde \Gamma$ in (b) in which the direction of each dual edge is a $\pi/2$ clockwise rotation of the corresponding edge in $\Gamma$. }
		\label{fig:PEMdualGraph}
	\end{figure}

The \textbf{PEM duality transformation} consists of three maps: 
\begin{enumerate}
\item A partial Fourier transform \eqref{eq:FTtransformation} on local basis of Hilbert space, 
\item a partial Fourier transform \eqref{eq:BAtildeAB} on the local operators,
\item and a map from $\Gamma$ to $\tilde{\Gamma}$. For every triangulation $\Gamma$, we define the dual graph $\tilde{\Gamma}$ in which the direction of each dual edge is a $\pi/2$ clockwise rotation of the corresponding edge in $\Gamma$. See Fig. \ref{fig:PEMdualGraph} for an illustration,
\end{enumerate}
such that
\begin{enumerate}
    \item[a.] the matrices of the dual local operators $\tilde{A}^\eta,\tilde{B}^\rho$ in the dual basis define the QD$(\hat{N})$ Hamiltonian on the dual graph $\Tilde{\Gamma}$,
    \item[b.] and the dual operators $\tilde{A}^{(x,\eta)}$ and $\tilde{B}^{(y,\rho)}$ generate the TQD algebra $D^{\alpha'}({G'})$.
\end{enumerate}

We will show that the partial EM duality has the following features:
	\begin{itemize}
		\item In general, a TQD model is not self-dual under the PEM duality transformation: Not only the lattice structure is changed (the triangulation is mapped to a reciprocal bilayer graph) but also the gauge group $G$ is mapped to a dual group $G'$, while the algebra of the local operators are mapped from $D^{\alpha } G$ to $D^{\alpha '} G'$.
		\item The charges and fluxes are exchanged with respect to the subgroup $N$ of the gauge group $G$, via the duality transformation $\{A^a\}\leftrightarrow \{\tilde{B}^\rho\}$.
		
		\item Every weakly Morita equivalence $Vec_G^\alpha\equiv Vec_{G'}^{\alpha'}$  can be realized as a PEM duality on the TQD$(G,\alpha)$ model.
	\end{itemize}
	
As alluded to earlier, we are able to derive the PEM dulaity based on the following three conditions: 
\begin{enumerate}
    \item[c1.] There exists a normal Abelian subgroup $N$ of $G$, such that the elements of $G$ can be written as pairs $(a,x)$, and thus we can define the partial Fourier transform over $a\in N$; 
    \item[c2.] $\alpha|_N=1$, such that the TQD$(G,\alpha)$ model contains a QD$(N)$ model, which is mapped to the QD$(\hat{N})$ model on the dual graph $\tilde \Gamma$; 
    \item[c3.] an extra condition \eqref{eq:FFcondition} such that as in Eq. \eqref{eq:omegaFepsilon}, $\alpha((a_1,x_1),(a_2,x_2),(a_3,x_3))$ can be factorized into an $N$-$K$-mixed-factor $\hat{F}(x_1,x_2)(a)$ and a $K$-factor $\epsilon(x_1,x_2,x_3)$. Under the partial Fourier transform, the $K$-factor is intact, while the $N$-$K$-mixed-part is mapped to the $\hat{N}$-$K$-mixed-factor $\rho_1(F(x_2,x_3))$ as in Eq. \eqref{eq:omegahatFepsilon}. Then the $\hat{N}$-$K$-mixed-factor renders the algebra of the dual local operators a TQD algebra $D^{\alpha'}(G')$.
    
\end{enumerate}

The above conditions urge the TQD$(G,\alpha)$ model to be factorized into an $N$-part and a $K$-part as well. Such factorization is manifest in a reformulation of the  TQD$(G,\alpha)$ model as a bilayer model.

Both layers have the same graph structure as $\Gamma$. The upper layer accommodates a QD$( N)$ model because $\alpha|_N=1$; however, the model that inhabits on the lower layer may not be a TQD model in general because its input data consists of the group $K$ and the 3-cochain $\epsilon$. The original TQD$(G,\alpha)$ model is viewed as a coupling of the two models on the two layers. Under the PEM duality, the upper layer model is mapped to the QD$(\hat{N})$ model on the dual graph $\tilde\Gamma$. The lower layer remains unchanged, as defined on $\Gamma$. We call the dual model a \textbf{reciprocal bilayer model}. 

\begin{figure}[htb]
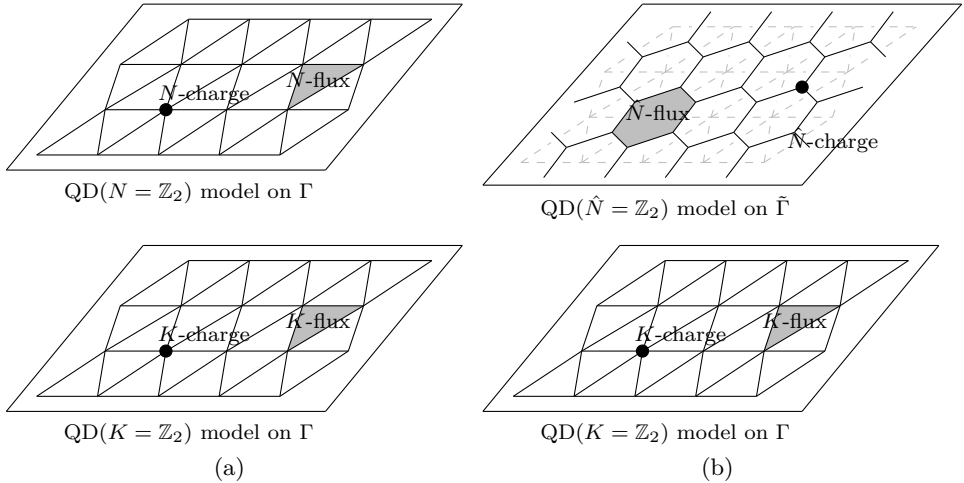

		\begin{center}
			\subfigure[]
			\bilayerModelAG
			\subfigure[]
			\bilayerModelAH
		\end{center}
		\caption{(a) The QD$( \mathbb{Z}_4 )$ model can be understood as a bilayer system. The both layers are QD$( \mathbb{Z}_2)$ models on $\Gamma$. (b) The TQD$( \mathbb{Z}_2\times \mathbb{Z}_2 )$ model can be understood as a bilayer system. on the upper layer is a QD$( \mathbb{Z}_2)$ model On $\Gamma$, while on the lower layer a QD$( \mathbb{Z}_2)$ models on $\tilde\Gamma$.}
		\label{fig:bilayerModelZtwo}
	\end{figure}

We illustrate the bilayer structure and the PEM duality by the QD$(Z_4)$ model as a quick example. Seen in Fig. \ref{fig:bilayerModelZtwo}, the QD$(Z_4)$ model is a bilayer model with the QD$(\mathbb{Z}_2)$ model on both layers. The QD$(Z_4)$ model has a trivial $\alpha=1$, but the nontrivial semidirect product structure in $\mathbb{Z}_4=\mathbb{Z}_2\rtimes \mathbb{Z}_2$ leads to a nontrivial coupling between the two layers. Under the partial Fourier transform, the upper layer is mapped to the QD$(\mathbb{Z}_2)$ on the dual graph. The direct product structure in the dual group $G'=\mathbb{Z}_2\times\mathbb{Z}_2$ has no contribution to the coupling; however, the Fourier transform generates a nontrivial $\hat{N}$-$K$-mixed factor in $\alpha'$ that leads to the nontrivial coupling.
    
	\section{PEM duality in twisted quantum double models}
	
	In this section, we shall derive the PEM duality via a partial Fourier transform in TQD models, which are exactly solvable Hamiltonian models of discrete topological gauge field theories. 
	
	\subsection{Quantum double model as discrete gauge field theories}
	
	We begin with a quick review of of the Kitaev (QD) model \cite{Kitaev2003a}.
	
	The original QD model with a finite gauge group $G$ is defined on a 2D directed graph that is embedded in an oriented closed surface. The model can be extended to open surface with boundaries, but we do not consider such cases in this paper\cite{Beigi2011,Bullivant2017}.
	
	The Hilbert space of the model is spanned by the group elements of $G$ on the edges of the graph. Every local basis vector with $g\in G$ on an edge $e$ is invariant under simultaneous reversion of the direction of $e$ and the inversion of $g$ as $\bar g$. That is,
	
	\begin{equation}
	\bket{\stateIdenticalAA}\equiv
	\bket{\stateIdenticalAB},
	\end{equation}
	and hence the Hilbert space does not depend on the edge directions. For simplicity, here we consider a square lattice, but the physics does not depend on the valence of the lattice. 
	
	The Hamiltonian consists of local gauge transformation and local holonomy measurement operators. A local gauge transformation operator $A_v^g$ at a vertex $v$ is defined by
	\begin{equation}\label{eq:AvTransform}
	A_v^g\bket\starActionAA
	=\bket\starActionAB.
	\end{equation}
The local holonomy measurement operator $B_{p}$ is an projection operator defined as
	\begin{equation}\label{eq:BpOperator}
	B_p\bket\plaquetteActionAA=\delta_{abcd}\bket\plaquetteActionAA,
	\end{equation}
	where $\delta _{abcd} =1$ if $abcd=1$, the unit element of $G,$ and $\delta _{abcd} =0$ otherwise. Here  the product $abcd$ of group elements around the plaquette $p$ is called the holonomy around $p$. Hence, $B_{p}$ projects onto the states with trivial holonomy around $p$. 
The Hamiltonian of the model reads
	\begin{equation}
	H=-\sum _{v} A_{v} -\sum _{p} B_{p},
	\end{equation}
	where $A_{v}$ is a projection operator
	\begin{equation}
	A_{v} =\frac{1}{|G|}\sum _{g\in G} A^{g}_{v},
	\end{equation}
	which projects onto the states that are gauge invariant at vertex $v$. 
	
	\subsection{Twisted quantum double algebra in twisted quantum double model}
	
\begin{figure}[htb]
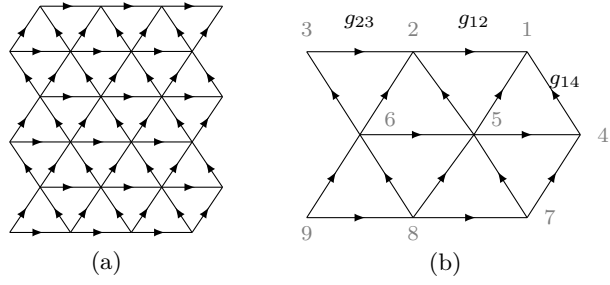

		\begin{center}
			\subfigure[]{
				\triangulationFigureAB}
			\quad
			\subfigure[]{
				\triangulationFigureAA}
		\end{center}
		\caption{(a) An example of (one part of) a triangulation of the surface. (b) Ordering labels are assigned to vertices such that the arrows are directed from the greater number to the smaller one.  }
		\label{fig:triangulationFigure}
	\end{figure}

	The QD model with a finite group $G$ can be extended to the TQD model\cite{Hu2012a} with the input data $( G,\alpha )$, where $\alpha\in H^3(G,U(1)) $. The TQD model is defined on a particular graph as a triangulation of an oriented closed surface (extension to open surfaces can be found in \cite{Bullivant2017}). The edges on the graph are directed, and the directions of the three edges bounding every triangle can not be the same. Practically, we assign ordered numbers to the vertices such that each edge is directed from the larger end to the smaller end of the edge. See Fig. \ref{fig:triangulationFigure} for example.

	Similar to that in the QD model, the Hilbert space of the TQD model is spanned by the configurations of group elements on all edges. The Hamiltonian consists of two terms defined at the vertices $v$ and on the triangles $p$, as defined by
	\begin{equation}
    \label{eq:QDHam}
	H =-\sum _{v} Q_{v} -\sum _{p} B_{p},
	\end{equation}
	\begin{equation}
	\label{eq:TQDHam}
	Q_{v} =\frac{1}{|G|}\sum _{g\in G} Q^{g}_{v}, \quad B_p=B_p^{1},
	\end{equation}
where $B_p^1$ is the special case of $B_p^{h=1}$, where $B^{h}_{p}$ is a local holonomy measurement operator defined by
	\begin{equation}
	B_p^h\bket\unitCellAF=\delta_{\mathrm{hol},h}\bket\unitCellAF
	\end{equation}
	where $\mathrm{hol}=g_1g_{12}\bar{g_2}$ is the holonomy of the triangle $p$, defined as follows.  On each triangle, we choose the vertex $v$ labeled by the smallest number as the base point of the triangle. We define the holonomy (e.g., $\mathrm{hol}=g_1g_{12}\bar{g_2}$ in the above equation) by the product of group elements along the boundary edges of $p$ in the counterclockwise direction starting from and ending at $v$. We draw a loop inside the triangle to label the holonomy.
    
  The operator $Q_v^g$ acts on the local states at $v$. For a simple example, 
    \begin{equation}
    \begin{aligned}
  Q_v^g\bket{\unitCellAD}
    =\frac{\alpha(g_3\bar{g},g_1)\alpha(g,g_1,g_{12})}{\alpha(g_3\bar{g},g,g_2)}\delta_{g_1g_{12}\bar{g_2}}\bket{\unitCellAE}.
    \end{aligned}
    \end{equation}
    For a detailed definition of $Q_v^g$, see Ref\cite{Hu2012a}. 	
	
	The algebra of local observables is the TQD algebra $D^\alpha(G)$. To construct such an algebra, we will first extend the operators $Q_v^g$ to $A_v^g$ as follows. The definition of $A_v^g$ depends on the holonomy of each neighboring triangle. Every neighboring triangle that has the base point at $v$ contributes to a coefficient to the action of $A_v^g$. For simplicity, we assume only one neighboring triangle that has the base point at $v$. Such extension in the situation with many such triangles will be straightforward. We define
	
\begin{equation}
\label{eq:extendAvgTQD}
    \begin{aligned}
&A_v^g\bket{\unitCellAG}\\
    =&\frac{\alpha(g_3\bar{g},g_1)\alpha(g,g_1,g_{12})}{\alpha(g_3\bar{g},g,g_2)}
    \frac{\alpha(g\mathrm{hol}\bar{g},g,g_2)}{\alpha(g,\mathrm{hol},g_2)}
    \bket{\unitCellAH}.
    \end{aligned}
    \end{equation}
See Appendix \ref{sec:TQDinTQD} for a generic definition of $A_v^g$.

	\begin{figure}[!ht]
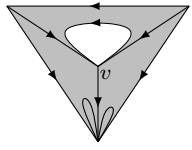

		\begin{center}
			{\unitCellAI}
\end{center}
		\caption{The $A_v^g$ acts on a vertex connecting 3 triangles. There is one triangle has the base point at $v$. We draw a loop inside the triangle starting from and ending at $v$. We treat this loop as the boundary with the trivial boundary condition. The $A_v^g$ is defined as the boundary operator on this boundary.}
		\label{fig:asBoundary}
	\end{figure}

The operator $Q_v^g$ is related to $A_v^g$ by
\begin{equation}
Q_v^g=A_v^gB_p^1
\end{equation}
where $p$ refers to the triangle that has the base point at $v$ in the above example. Throughout the paper we assume that all 3-cocycles $\alpha\in H^3(G,U(1))$ are normalized  such that $\alpha(g,h,1)=\alpha(g,1,h)=\alpha(g,1,h)=1$ for $g,h\in G$.

The operators $A_v^g$ and $B_p^h$ generate the TQD algebra $D^\alpha G$, with the multiplication rule
	\begin{equation}
	A^{g_{1}}_{v} B^{h_{1}}_{p} A^{g_{2}}_{v} B^{h_{2}}_{p} =\delta _{h_{1} ,g_{1} h_{2} \bar{g}_{1}} \beta _{h_{2}} (g_{1} ,g_{2} )A^{g_{1} g_{2}}_{v} B^{h_{2}}_{p},
	\end{equation}
	where
	\begin{equation}
	\beta _{h_{2}} (g_{1} ,g_{2} ):=\frac{\alpha (g_{1} ,g_{2} ,h_{2} )\alpha \left( g_{1} g_{2} h_{2}\bar{g_2}\bar{g_1} ,g_{1} ,g_{2}\right)}{\alpha \left( g_{1} ,g_{2} h_{2} \bar{g_2} ,g_{2}\right)}.
	\end{equation}
	
	One immediate consequence is that $A_v^gB_p^h=B^{gh\bar g}_pA_v^g$, which yields an alternative formulation of the multiplication rule
	\begin{equation}
	B^{h_{1}} A^{g_{1}} B^{h_{2}} A^{g_{2}} =\delta _{h_{1} ,g_{1} h_{2} \bar{g_1}} \beta '_{h_{1}} (g_{1} ,g_{2} )B^{h_{1}} A^{g_{1} g_{2}},
	\end{equation}
	where
	\begin{equation}
	\beta '_{h_{1}} (g_{1} ,g_{2} ):=\beta_{\bar g_2\bar g_1 h_1 g_1 g_2}(g_1,g_2)=\frac{\alpha (h_{1} ,g_{1} ,g_{2} )\alpha \left( g_{1} ,g_{2} ,\bar{g_2}\bar{g_1} h_{1} g_{1} g_{2}\right)}{\alpha \left( g_{1} ,\bar{g}_{1} h_{1} g_{1} ,g_{2}\right)}.
	\end{equation}
This TQD algebra is the algebra of local operators that commute with Hamiltonian \eqref{eq:QDHam}.

An alternative way is to choose the base point of every triangle by the vertex with the greatest ordering number. The construction of operators $B_p^h$ and $A_v^g$,  depends on the choice, but the physics is independent of the choice, and hence we will stick to the current choice throughout the paper.
	
	\subsection{PEM duality between QD models and TQD models}
	\label{sec:Z4toZ2Z2}
	
	We first study the PEM duality in QD models with a finite group $G$. Let $N\subseteq G$ be a normal Abelian subgroup and $K=N\backslash G$ the corresponding quotient group. The $G$ can be written as a semidirect product group $N\rtimes_F K$, which is specified by a 2-cocycle $F\in H^2(K,N)$, i.e., a map $F:K\times K\rightarrow N$, such that
	\begin{equation}
	\delta _{K} F( k_{1} ,k_{2} ,k_{3}) = {^{k_{1}}F( k_{2} ,k_{3})} F( k_{1} k_{2} ,k_{3})^{-1} F( k_{1} ,k_{2} k_{3}) F( k_{1} ,k_{2})^{-1} =1,
	\end{equation}
	where ${^{k_1}F(k_2,k_3)}$ is a left action of $k_1$ on $F(k_2,k_3)$ by conjugation (see Appendix \ref{sec:semidirectStructure} for details).
    The multiplication rule in $G$ is given by
	\begin{equation}
    \label{eq:akakF}
	( a_{1} ,k_{1})( a_{2} ,k_{2}) =( a_{1} (^{k_1}a_{2})F( k_{1} ,k_{2}) ,k_{1} k_{2}),
	\end{equation}
	where $a_1,a_2\in N$ and $k_1,k_2\in K$.
		
	For a local basis vector $\ket{a,k}$ on one edge in the Hilbert space, we define the \textbf{partial Fourier transform} by
	\begin{equation}
	\label{eq:FTtranform}
	|k,\rho \rangle =\frac{1}{\sqrt{|N|}}\sum _{a} \overline{\rho ( a)} |a,k\rangle, 
	\end{equation}
	where $\rho\in \hat{N}$ and the bar means complex conjugation.
	
\begin{figure}[htb]
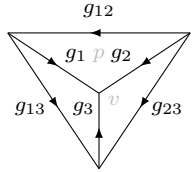

		\begin{center}
			\unitCellAA
		\end{center}
		\caption{A vertex $v$ and a neighbouring triangle $p$.}
		\label{fig:unitCellAA}
	\end{figure}

To define the dual operators, consider a vertex $v$ and a neighboring plaquette $p$ as shown in Fig. \ref{fig:unitCellAA}. (The remaining part of the graph is neglected. In general, the number of triangles neighboring to the vertex could be arbitrary.)
	
The operator $A_v^g$ acts as
	\begin{equation}
	\label{eq:Abasis}
	A^{g}_{v} \bket{\unitCellAB} =\bket{\unitCellAL}, 
	\end{equation}
	and the operator $B_p^h$ acts as
	\begin{equation}
	\label{eq:Bbasis}
	B^{h}_{p} \bket{\unitCellAM} =\delta_{g_{1}g_{12}\bar{g_2},h} \bket{\unitCellAM}, 
	\end{equation}
	where ${g_{1}{g_{12}}\bar{g_2}}$ is the holonomy around the triangle $p$.
	
It is convenient to factorize the operators into two parts due to the semidirect product structure. Let $A^a=A^{(a,1_K)}$ , $A^x=A^{(1_N,x)}$, $B_p^b=\sum_{y\in K}B^{(b,y)}$, and  $B_p^y=\sum_{b\in N}B^{(b,y)}$. It is straightforward to show $A_p^aA_p^x=A_p^{(a,x)}$ and $B_p^bB_p^y=B_p^{(b,y)}$.
	We define the dual operators by the partial Fourier transform
	\begin{equation}\label{eq:Akrho}
	\tilde{B}_v^{(\rho )} =\frac{1}{|N|}\sum _{a\in N}\overline{\rho(a)} A^{( a)}_{v},
	\end{equation}
	\begin{equation}\label{eq:Bkrho}
	\tilde{A}^{\eta}_{p} =\frac{1}{|N|}\sum _{b\in N} \eta ( b) B^{ b}_{p}.
	\end{equation}
The operators $A^x$ and $B^y$ remain the same, and we denote the corresponding dual operators by
\begin{equation}
\tilde{A}_v^x=A_v^x, \quad \tilde{B}_p^y=B_p^y.
\end{equation}
The matrix form of the dual operators in the basis $\ket{k,\rho}$ can be obtained as follows:	
	\begin{equation}
	\begin{aligned}
	& \tilde{A}^x_v \bket{\vertexDiagramAA}\\
	= &{\rho^{\bar{x}}_{1}( F( x,x_{1})) \rho ^{\bar{x}}_{2}( F( x,x_{2})) \rho^{\bar{x}} _{3}( F( x,x_{12})) }\bket{\vertexDiagramAB},
	\end{aligned}
	\end{equation}
	where $\rho^{\bar{x}}$ is defined by $\rho^{\bar{x}}(a)=\rho(^{\bar{x}}a)$,
	\begin{equation}
	 \tilde{B}^{\rho}_{v} \bket{\vertexDiagramAA}
	= \delta _{\rho _{1} \rho _{2} \rho _{3} ,\rho } \bket{\vertexDiagramAA}.
	\end{equation}
	\begin{equation}
	\tilde{A}_p^\eta\bket{\triangleDiagram}
	=  \overline{\eta (F(x_{1} ,x_{12} )F( y,x_{2})^{-1} )} \bket{\triangleDiagramAA}.
	\end{equation}
where $y=x_1x_{12}\bar{x_2}$, and
    \begin{equation}
	\tilde{B}_p^y \bket{\triangleDiagram}
	= \delta _{x_{1}x_{12}\bar{x_2} ,y} \bket{\triangleDiagram}.
	\end{equation}
The action of $\tilde{A}_p^\eta$ is derived by the partial Fourier transform as follows.
	\begin{equation}
	\begin{aligned}
	& \tilde{A}_p^\eta\bket{\triangleDiagram} \\
	= & \frac{1}{|N|}\sum _{b\in N}{ \eta (b)}\frac{1}{|N|^{3}}\sum _{a_{1} a_{2} a_{12}} \overline{\rho _{1} (a_{1} )\rho _{2} (a_{2} )\rho _{12} (a_{12} )}\frac{1}{|N|^{3}}\sum _{\rho '_{1} \rho '_{2} \rho '_{12}}{\rho '_{1} (a_{1} )\rho'_{2} (a_{2} )\rho '_{12} (a_{12} )}\\
	& \ \  \times \frac{1}{|N|}\sum _{\tilde{\eta }}\tilde{\eta }\left( a_{1} (^{x_1}a_{12}) F(x_{1} ,x_{12} )\bar{b} (^y\bar{a_2}) F( y,x_{2})^{-1}\right)\bket{\triangleDiagramAB} \\
	= & \overline{\eta (F(x_{1} ,x_{12} )F( y,x_{2})^{-1} )}\bket{\triangleDiagramAA}.
	\end{aligned}
	\end{equation}
The action of the other operators are derived in a similar way. 
	
\begin{figure}[!ht]
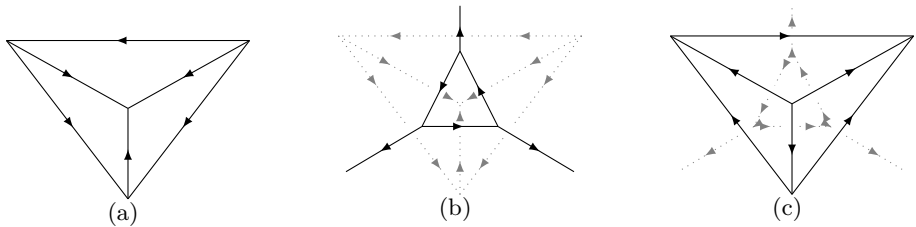

		\begin{center}
			\subfigure[]{\dualGraphAA}
			\qquad
			\subfigure[]{\dualGraphAB}
            \qquad
			\subfigure[]{\dualGraphAC}
		\end{center}
		\caption{For the triangulation $\Gamma$ in (a), we define the dual graph $\tilde \Gamma$ in (b) in which the direction of each dual edge is a $\pi/2$ clockwise rotation of the corresponding edge in $\Gamma$. The double dual graph $\tilde{\tilde{\Gamma}}$ in (c), obtained as the dual graph of $\tilde{\Gamma}$, is the same as $\Gamma$ but with all edge directions reversed. }
		\label{fig:dualGraph}
	\end{figure}

	First, we observe that the matrices of $\tilde{A}_p^\eta$ and $\tilde{B}_v^\rho$ in the dual basis of the Hilbert space define a QD$(\hat{N})$ model on the dual graph $\Tilde{\Gamma}$, where we identify the dual vertex $\tilde{v}$ with the original $p$, and the dual plaquette $\tilde{p}$ with $v$. See Fig. \ref{fig:dualGraph}. We thus rewrite the dual operators  $\tilde{A}_p^\eta$ and $\tilde{B}_v^\rho$  as $\tilde{A}_{\tilde v}^\eta$ and $\tilde{B}_{\tilde{p}}^\rho$, which form a QD Hamiltonian
        \begin{equation}
       \tilde{H}=-\sum_{\tilde{v}}\frac{1}{|\hat{N}|}\sum_{\eta}\tilde{A}_{\tilde{v}}^\eta-\sum_{\tilde{p}}\tilde{B}_{\tilde{p}}^{1_{\hat{N}}}
    \end{equation}
    on the dual graph $\tilde{\Gamma}$.	
	
	Now we recombine the tilde operators by setting $\tilde{A}_{v,\tilde{v}}^{(x,\eta)}=\tilde{A}_v^x\tilde{A}_{\tilde{v}}^\eta$ and $\tilde{B}_{p,\tilde{p}}^{(y,\rho)}=\tilde{B}_p^y\tilde{B}_{\tilde{p}}^\rho$, and observe that
	\begin{equation}
	\tilde{A}^{\left(x_{1}, \eta_1\right)} \tilde{A}^{\left(x_{2}, \eta_{2}\right)} \tilde{B}^{\left(1_{K}, 1_{\hat{N}}\right)}=\tilde{A}^{\left(x_1 x_{2}, \eta_{1}\eta_{2}\right)} \tilde{B}^{\left(1_{K}, 1_{\hat{N}}\right)},
	\end{equation}
where the pairs $(x,\eta)$ form a direct product group $G'=K\times \hat{N}$. As long as $G$ has a nontrivial semidirect product structure, i.e., $F$ is a nontrivial 2-cocycle, we have $G\neq G'$.
	
	More generically, the dual operators form a new algebra, with the multiplication rule being
	\begin{equation}
	\tilde{B}_{\tilde p}^{h'_{1}} \tilde{A}_{\tilde v}^{g'_{1}}\tilde{B}_{\tilde p}^{h'_{2}} \tilde{A}_{\tilde v}^{g'_{2}} =\delta _{h'_{1} ,g'_{1} h'_{2}\bar{g'_{1}}} \beta '_{h'_{1}} (g'_{1} ,g'_{2} )\tilde{B}_{\tilde p}^{h'_{1}} \tilde{A}_{\tilde v}^{g'_1g'_2},
	\end{equation}
	where\begin{equation}
	\beta '_{h'_{1}} (g'_{1} ,g'_{2} ):=\frac{\alpha '(h'_{1} ,g'_{1} ,g'_{2} )\alpha '\left( g'_{1} ,g'_{2} ,\bar{g'_2}\bar{g'_1} h'_{1} g'_{1} g'_{2}\right)}{\alpha '\left( g'_{1} ,\bar{g'_1} h'_{1} g'_{1} ,g'_{2}\right)},
	\end{equation}
	where the 3-cocycle $\alpha'\in H^3(G',U(1))$ is given by
	\begin{equation}
	\label{eq:omegaPrimeQD}
	\alpha '(( x_{1} ,\rho _{1}) ,( x_{2} ,\rho _{2}) ,( x_{3} ,\rho _{3})) =\rho _{1}( F( x_{2} ,x_{3})).
	\end{equation}
	Such an algebra is identified with the TQD algebra $D^{\alpha '}( G')$. See Appendix \ref{sec:semidirect} for the full definition of $D^{\alpha '}( G')$. Hence, under the partial Fourier transform, the algebra of local observables is mapped from $D( G)$ to $D^{\alpha '} G'$.
    
	We summarize the main features of the derived PEM duality as follows. First, the PEM duality always exists for any D$(G)$ model, with the dual group $G'=K\times \hat{N}$ and the  $3$-cocycle $\alpha'$ determined by the semidirect product structure $F$ in $G$.  Second, under the  partial Fourier transform in Eq. \eqref{eq:FTtranform}, we can construct the dual local operators $\tilde{A}^\eta$,$\tilde{A}^x $, $\tilde{B}^\rho$, and $\tilde{B}^y$, such that $\tilde{A}^\eta$ and $\tilde{B}^\rho$ form the QD algebra on the dual graph. 
	
	We denote by $\mathcal{T}$ the PEM duality transformation defined by Eqs. \eqref{eq:FTtranform}, \eqref{eq:Akrho}, and \eqref{eq:Bkrho}. As a self-consistency check, we perform the PEM transformation twice $\mathcal{T}^2$, which maps the graph $\Gamma$ to the double dual graph  $\tilde{\tilde{\Gamma}}$, as the dual graph of $\tilde{\Gamma}$. $\tilde{\tilde \Gamma}$ is the same as $\Gamma$ but with all edge arrows reversed (see Fig. \ref{fig:dualGraph}(c)). The transformation $\mathcal{T}^2$ on the operators is given by
\begin{equation}
\{A^a_v, B^b_p\}\text{ on }\Gamma
\overset{\mathcal{T}^2}{\longrightarrow} 
\{A^{\bar a}_{\tilde{\tilde{v}}}, B^{\bar b}_{\tilde{\tilde{p}}}\}\text{ on }\tilde{\tilde{\Gamma}}
\equiv
\{A^{ a}_{v}, B^{b}_{p}\}\text{ on }{\Gamma}
\end{equation}
The last equality is due to the identification of $a$ on an directed edge $e$ with $\bar{a}$ on an reversed edge.
    
\subsection{Examples}

We examine some examples of the PEM duality in the QD models.

\subsubsection{EM duality in $\mathbb{Z}_2$ QD model}
	
	In the QD model with $G=\mathbb{Z}_{2}$ (known as the toric code model), the Hilbert space is spanned by $\frac{1}{2}$-spins (to represent the group elements in $\mathbb{Z}_{2}$) on all edges, and the Hamiltonian terms can be expressed in terms of the Pauli matrices
	\begin{equation}
    \label{eq:ABsigma}
	A_{v} =\prod _{e\text{ into } v} \sigma ^{x}_{e} ,\ \ B_{p} =\prod _{e\text{ around } p} \sigma ^{z}_{e},
	\end{equation}
	where $e$ denotes the edges. Indeed, compared to Eq. \eqref{eq:AvTransform}, $\prod _{e\text{ into } v} \sigma ^{x}_{e}$ is a $\mathbb{Z}_{2}$ gauge transformation where $\mathbb{Z}_{2}=\left\{1,\sigma ^{x}\right\}$, and the delta function in Eq. \eqref{eq:BpOperator} becomes 
	\[\frac{1}{2}\left( 1+\prod _{e\text{ around } p} \sigma ^{z}_{e}\right).
	\]
	
	In this case, where the normal subgroup of $G$ is $G$ itself, the PEM duality becomes a full EM duality. This model is self-dual under the EM duality, where the vertices are mapped to triangles on the dual graph and vice versa: $v\mapsto\tilde{p},p\mapsto\tilde{v}$., and the local operators are transformed as
	\begin{equation}
	A_v\mapsto\tilde{B}_{\tilde{p}},B_p\mapsto\tilde{A}_{\tilde{v}}.
	\end{equation}
They can be written as in the same form as in Eq. \eqref{eq:ABsigma} where $v$ is replaced by $\tilde{v}$ (and $p$ by $\tilde{p}$). Such an EM self-duality can be generalized to all QD models with Abelian $G$.
	
	\subsubsection{Example $\mathbb{Z}_4\rightarrow \mathbb{Z}_{2} \times \mathbb{Z}_{2}$}
	\label{sec:Z4Z2Z2}
	
	The simplest example of a nontrivial PEM duality is in the QD model with $G=\mathbb{Z}_{4}$. Let $G=\mathbb{Z}_{4} =\{0,1,2,3\}$, and $N=\mathbb{Z}_2=\{0,2\}$ is a normal subgroup of $G$. The quotient group $K=N\backslash G$ consists of two elements $[ 0] =N+0$ and $[ 1] =N+1$. The $G$ is a semidirect product $N\rtimes K$, whose product structure is given by $F( k_{1} ,k_{2}) =k_{1} +k_{2} -\langle k_{1} +k_{2} \rangle _{2}$ where $\langle k\rangle _{2} =k\bmod 2$. We find that $G'=\mathbb{Z}_{2} \times \mathbb{Z}_{2} = \{(00),(01),(10),(11)\}$. By Eq. \eqref{eq:omegaPrimeQD}, we can derive a nontrivial 3-cocycle $\alpha'\in H^3(\mathbb{Z}_{2} \times \mathbb{Z}_{2},U(1))$:
	\begin{equation}
	\alpha '(( k_{1} ,n_{1}) ,( k_{2} ,n_{2}) ,( k_{3} ,n_{3})) =\exp[ \frac{\pi n_{3} \mathrm{i}}{2}( k_{1} +k_{2} -\langle k_{1} +k_{2} \rangle _{2})].
	\end{equation}
	This duality is summarized in Table \ref{tab:EMdualtiyZ4}.

	\begin{table}[!h]
		\begin{center}
			
			\begin{tabular}{lcc}
				\toprule 
				& QD$(G=\mathbb{Z}_{4})$ & dual model \\
				\midrule 
				gauge group & \makecell[c]{$G=\mathbb{Z}_{4} =\mathbb{Z}_{2} \ltimes \mathbb{Z}_{2}$} & \makecell[c]{$G'=\mathbb{Z}_{2} \times \mathbb{Z}_{2}$} \\
				\hline 
				algebra of observables & $D(\mathbb{Z}_{4})$ & $D^{\alpha'}(\mathbb{Z}_{2} \times \mathbb{Z}_{2})$ \\
				\toprule
			\end{tabular}
		\end{center}
		\caption{PEM duality in the QD model with $G=\mathbb{Z}_4$.}\label{tab:EMdualtiyZ4}
	\end{table}
	
    \subsubsection{Example $D_{m}\rightarrow D_m$}
	
	Let $G=D_{m}$, the $m$-th dihedral group. Denote the group elements by $r^{a} s^{k}$ for $k=0,1$ and $a=0,1,\dots ,m$, where $s$ and $r$ are the generators of reflections and $2\pi /3$-rotations, satisfying $s^{2} =1=r^{m}$ and $rs=s\bar{r}$. Let $N=\left\{r^{0} ,r^{1} ,\dotsc ,r^{m}\right\}$ be the normal subgroup consisting of all rotations, and the quotient group is $K=\left\{Ns^{0} ,Ns^{1}\right\}$. Denote $r^{a} s^{k}$ by $( a,k)$ and the semidirect product $G=N\rtimes K$ has the product structure
	\begin{equation}
	( a_{1} ,k_{1})( a_{2} ,k_{2}) :=\left( a_{1}\left(^{k_{1}} a_{2}\right) ,k_{1} k_{2}\right) \ 
	\end{equation}
	for $k=0,1$ and $a=0,1,\dotsc ,m$, where $^{k} a=( -1)^{k} a$ mod m.
	
	To compare with the generic formula, we set $F( k_{1} ,k_{2}) =0\in N$ and $\hat{F}( k_{1} ,k_{2}) =1$; hence we can choose $\epsilon ( k_{1} ,k_{2} ,k_{3}) =1$ for all $k_{1} ,k_{2} ,k_{3} \in K$. The corresponding equivalence is between $QD( G=D_{3})$ and $QD( G'=D_{3})$, where $G'=K\rtimes \hat{N}$ is again $D_{3}$. Since $F( a_{1} ,a_{2}) =0\in N$ and $\hat{F}( k_{1} ,k_{2}) =1$, we have $\alpha =1$ on $G$ and $\alpha '=1$ on $G'$.
	
    Both $G$ and $G'$ happen to be the same group $D_{m}$ in this case, but in general, they are distinct.
    
	\subsubsection{Example $D_{4}\rightarrow \mathbb{Z}_{2} \times \mathbb{Z}_{2} \times \mathbb{Z}_{2}$}
	
	It is known\cite{Hu2012a} that the QD$( D_{4})$ and the TQD$(\mathbb{Z}_{2} \times \mathbb{Z}_{2} \times \mathbb{Z}_{2} ,\alpha ')$ model with a particular type of $\alpha'$ are equivalent and thus describe the same topological order. We derive the PEM duality transformation as follows. Denote $r^{a} s^{k}$ by $(a,k)$. The semidirect product $G=N\rtimes K$ reads
	\begin{equation}
	(a_{1} ,k_{1} )(a_{2} ,k_{2} ):=\left( a_{1} a_{2} F(k_{1} ,k_{2} ),k_{1} k_{2}\right),
	\end{equation}
where
	\begin{equation}
	F\left( Nr^{x_{1}} s^{y_{1}} ,Nr^{x_{2}} s^{y_{2}}\right) =r^{x_{1} +x_{2} -\langle x_{1} +x_{2} \rangle +2y_{1} x_{2}}.
	\end{equation}

	Since $\alpha=1$, we can set $\hat{F} (a_{1} ,a_{2} )=1$, such that $\epsilon (k_{1} ,k_{2} ,k_{3} )=1$ for all $k_{1} ,k_{2} ,k_{3} \in K$. The PEM duality maps the the algebra of local observables from	$D(D_{4} =(\mathbb{Z}_{2} \times \mathbb{Z}_{2}) \ltimes \mathbb{Z}_{2})$ to $D^\alpha(\mathbb{Z}_{2} \times \mathbb{Z}_{2} \times \mathbb{Z}_{2} ,\alpha ')$, where $G'=K\rtimes \hat{N} =\mathbb{Z}_{2} \times \mathbb{Z}_{2} \times \mathbb{Z}_{2}$, and the 3-cocycle $\alpha '$ is given by
	\begin{equation}
	\alpha '((x_{1} ,\rho _{1} ),(x_{2} ,\rho _{2} ),(x_{3} ,\rho _{3} ))=\rho _{1} (F(x_{2} ,x_{3} )).
	\end{equation}
	According to the classification of the 3-cocycles on $\mathbb{Z}^{3}_{2}$ as listed in the Appendix \ref{sec:cocycleZZZ}, $\alpha '=\alpha _{III} \alpha _{II}$.

	\subsection{EM duality on the twisted quantum models}
	\label{sec:FTqtd}
	
	The previous subsections dealt with the PEM dualities in the QD models. In this subsection, we study the PEM duality in generic TQD models. 
	
	If there is an Abelian normal subgroup $N$ of $G$ such that $\alpha _{N}$ is trivial, we can define a partial Fourier transform
	\begin{equation}
	\label{eq:FTTQD}
	|x,\rho \rangle =\frac{1}{\sqrt{|N|}}\sum _{a\in N} \overline{\rho ( a)} |a,x\rangle. 
	\end{equation}
	The main result of this paper is that this partial Fourier transform maps the original TQD model to a dual model whose local operators form a TQD algebra $D^{\alpha '}( G')$.
	
	When the 3-cocycle $\alpha |_{N}$ restricted to $N$ is trivial, $G$ can be written as a semidirect product $G=N\rtimes_F K$ of the normal Abelian subgroup $N$ and the quotient group $K=N\backslash G$. The semidirect product structure 
	\begin{equation}
	(a_{1} ,k_{1} )(a_{2} ,k_{2} ):=\left( a_{1}\left(^{k_{1}} a_{2}\right) F(k_{1} ,k_{2} ),k_{1} k_{2}\right)
	\end{equation}
	is characterized by a 2-cocycle $F:K\times K\rightarrow N$ satisfying
	\begin{equation}
	\delta _{K} F( k_{1} ,k_{2} ,k_{3}) \equiv {^{k_{1}}F( k_{2} ,k_{3})} F( k_{1} k_{2} ,k_{3})^{-1} F( k_{1} ,k_{2} k_{3}) F( k_{1} ,k_{2})^{-1} =1,
	\end{equation}
	where $^ka$ is the conjugation of $a$ by $k$. see Appendix \ref{sec:semidirect} for detailed definition. The 3-cocycle $\alpha $ is cohomologically to
	\begin{equation}
	\label{eq:omegaTQD}
	\alpha (( a_{1} ,k_{1}) ,( a_{2} ,k_{2}) ,( a_{3} ,k_{3})) :=\hat{F}( k_{1} ,k_{2})( a_{3}) \epsilon ( k_{1} ,k_{2} ,k_{3}).
	\end{equation}
	We start with such a $G$ and $\alpha $. The local observables are characterized by the TQD algebra $D^{\alpha } G$, with the multiplication rule given by
	\begin{equation}
	\label{eq:ABABtqd}
	( A_v^{g_1}B_p^{h_1})( A_v^{g_2}B_p^{h_2}) :=\delta _{h_1,g_2h_2\bar{g_2}} \beta _{h_{2}} (g_{1} ,g_{2} ) A_v^{g_1g_2}B_p^{h_2},
	\end{equation}
	where $\beta _{h_{2}} (g_{1} ,g_{2} )$ is given by
	\begin{equation}
	\label{eq:betaTQD}
	\beta _{h_{2}} (g_{1} ,g_{2} ):=\frac{\alpha (g_{1} ,g_{2} ,h_{2} )\alpha \left( g_{1} g_{2} h_{2}\bar{g_2}\bar{g_1} ,g_{1} ,g_{2}\right)}{\alpha \left( g_{1} ,g_{2} h_{2} \bar{g_2} ,g_{2}\right)}.
	\end{equation}
	If we associate a 3-cocycle $\alpha$ to a tetrahedron, then $\beta _{h_{2}} (g_{1} ,g_{2} )$  is depicted in terms of tetrahedra as in Fig. \ref{fig:betaByTetrahedra}(a). See Appendix \ref{sec:TQDinTQD} for detailed explanation.

	\begin{figure}[htb]
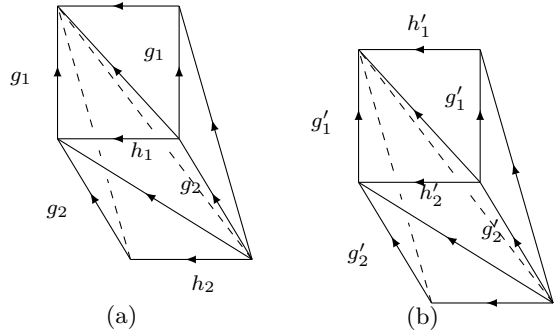

		\begin{center}
			\subfigure[]
			\betaByTetrahedraAA
			\quad
			\subfigure[]
			\betaByTetrahedraAB
		\end{center}
		\caption{(a) the multiplication coefficients $\beta _{h_{2}} (g_{1} ,g_{2} )$ is depicted as three tetrahedra glued together. (b) The Fourier transformed factor $\mathcal{F}[ \beta ]^{( \rho ,x)( \eta ,y)}_{( \rho _{1} ,x_{1})( \eta _{1} ,y_{1})( \rho _{2} ,x_{2})( \eta _{2} ,y_{2})}$ depicted.}
		\label{fig:betaByTetrahedra}
	\end{figure}

	Define the partial Fourier transform of the operator $( A_v^{g}B_p^{h})$ by
	\begin{equation}\label{eq:FTGamma}
	\begin{aligned}
	\Gamma^{( x,\eta )}_{( y,\rho )} & =\frac{1}{|N|}\sum _{a,b\in N} \overline{\rho ( a)} \eta ( b) A_v^{( a,x)}B_p^{( b,y)},
	\end{aligned}
	\end{equation}
	where we rewrite $g,h$ by the pairs: $g=(a,x),h=(b,y)$.
	
	We could derive the detailed matrix element of the new operators in a straightforward way, as we did in the previous section. In the following, we will not dwell on the detailed matrix form of the operators $\Gamma^{( x,\eta )}_{( y,\rho )}$ but follow a more convenient path by focusing on the algebra of the operators.
	
	Since the factor $\epsilon $ in $\alpha $ is not involved in the partial Fourier transform, we will ignore the parts involving $\epsilon $ during the calculation and restore it only in the final results.
	\begin{equation}
	\beta _{h_{2}} (g_{1} ,g_{2} )=\frac{\hat{F} (x_{1} ,x_{2})( b_{2} )\hat{F}\left( x_{1} x_{2} y_{2} \bar{x_2} \bar{x_1} ,x_{1}\right)( a_{2})}{\hat{F}\left( x_{1} ,x_{2} y_{2} \bar{x_2}\right)( a_{2})} \dotsc 
	\end{equation}
	
	The algebra of the new operators $\Gamma^{( x,\eta )}_{( y,\rho )}$ can be obtained by the partial Fourier transform as
	\begin{equation}
	\label{eq:GammaGammaGamma}
	\Gamma^{( x_{1} ,\eta _{1})}_{( y_{1} ,\rho _{1})}\Gamma^{( x_{2} ,\eta _{2})}_{( y_{2} ,\rho _{2})} =\sum _{\rho ,\eta }\mathcal{F}[ \beta ]^{( \rho ,x)( \eta ,y)}_{( \rho _{1} ,x_{1})( \eta _{1} ,y_{1})( \rho _{2} ,x_{2})( \eta _{2} ,y_{2})} \delta _{y_{1} ,x_{2} y_{2} \bar{x_2}}\Gamma^{( x,\eta )}_{( y,\rho )},
	\end{equation}
	where $x=x_{1} x_{2} ,y=y_{2}$, and the matrix elements of $\mathcal{F}[ \beta ]$ are given by the partial Fourier transform
	
	\begin{align*}
	& \mathcal{F} [\beta ]^{(\rho ,x)(\eta ,y)}_{(\rho _{1} ,x_{1} )(\eta _{1} ,y_{1} )(\rho _{2} ,x_{2} )(\eta _{2} ,y_{2} )}\\
	=\  & \frac{1}{|N|^{3}}\sum _{a_{1} a_{2} b_{1} b_{2}}\overline{\rho _{1} (a_{1} )\rho _{2} (a_{2} )} \eta _{1} (b_{1} )\eta _{2} (b_{2} )\rho \left( a_{1}^{x_{1}} a_{2} F(x_{1} ,x_{2} )\right)\overline{\eta (b_{2} )}\\
	& \ \ \ \ \ \ \ \delta _{b_{1}\left(^{x_{2} y_{2}\overline{x_{2}}} a_{2}\right) F\left( x_{2} y_{2}\overline{x_{2}} ,x_{2}\right) ,a_{2}^{x_{2}} b_{2} F(x_{2} ,y_{2} )}\frac{\hat{F} (x_{1} ,x_{2} )(b_{2} )\hat{F}\left( x_{1} x_{2} y_{2}\bar{x_{2}}\bar{x_{1}} ,x_{1}\right) (a_{2} )}{\hat{F}\left( x_{1} ,x_{2} y_{2}\bar{x_{2}}\right) (a_{2} )} \dotsc \\
	= & \frac{1}{|N|^{3}}\sum _{a_{1} a_{2} b_{2}}\overline{\rho _{1} (a_{1} )\rho _{2} (a_{2} )} \eta _{1}\left( a_{2} (^{x_{2}}b_{2}) \ ^{x_{2} y_{2}\bar{x_{2}}}\bar{a_{2}} F(x_{2} ,y_{2} )F\left( x_{2} y_{2}\overline{x_{2}} ,x_{2}\right)^{-1}\right) \eta _{2} (b_{2} )\\
	& \ \ \ \ \ \ \rho \left( a_{1} {^{x_{1}}a_{2}} F(x_{1} ,x_{2} )\right)\overline{\eta (b_{2} )}\\
	& \ \ \ \ \ \ \ \ \ \ \ \ \ \ \ \ \frac{\hat{F} (x_{1} ,x_{2} )(b_{2} )\hat{F}\left( x_{1} x_{2} y_{2}\bar{x_{2}}\bar{x_{1}} ,x_{1}\right) (a_{2} )}{\hat{F}\left( x_{1} ,x_{2} y_{2}\overline{x_{2}}\right) (a_{2} )} \dotsc \\
	= & \frac{1}{|N|^{3}}\sum _{a_{1} a_{2} b_{2}}\overline{\rho _{1} (a_{1} )\rho _{2} (a_{2} )} \eta _{1} (a_{2} )\eta ^{x_{2}}_{1} (b_{2} )\eta ^{x_{2} y_{2}\overline{x_{2}}}_{1}\left(\overline{a_{2}}\right) \eta _{1}\left( F(x_{2} ,y_{2} )F\left( x_{2} y_{2}\overline{x_{2}} ,x_{2}\right)^{-1}\right)\\
	& \ \ \ \ \ \ \ \ \eta _{2} (b_{2} )\rho (a_{1} )\rho ^{x_{1}} (a_{2} )\rho (F(x_{1} ,x_{2} ))\overline{\eta (b_{2} )}\\
	& \ \ \ \ \ \ \ \ \ \ \ \ \ \ \ \ \frac{\hat{F} (x_{1} ,x_{2} )(b_{2} )\hat{F}\left( x_{1} x_{2} y_{2}\bar{x_{2}}\bar{x_{1}} ,x_{1}\right) (a_{2} )}{\hat{F}\left( x_{1} ,x_{2} y_{2}\overline{x_{2}}\right) (a_{2} )} \dotsc, 
	\end{align*}

	where in the first equality the delta function is an expansion of $\delta_{h_1g_2,g_2h_2}$, and in the second equality $\bar{(^ab)}= {^a(\bar{b})}$ for all $a,b\in N$. According to Eq. \eqref{eq:FTGamma}, $\overline{\rho _{1}( a_{1}) \rho _{2}( a_{2})} \eta _{1}( b_{1}) \eta _{2}( b_{2})$ is the Fourier transform kernel that appears in $\Gamma^{( x_{1} ,\eta _{1})}_{( y_{1} ,\rho _{1})}\Gamma^{( x_{2} ,\eta _{2})}_{( y_{2} ,\rho _{2})} $, while ${\rho \left( a_{1}^{x_{1}} a_{2} F( x_{1} ,x_{2})\right) \overline{\eta ( b_{2})}}$ appears in $\Gamma^{( x,\eta )}_{( y,\rho )}$.
	The summation evaluates to
	\begin{equation}
	\begin{aligned}
	& \delta _{\rho ,\rho _{1}} \delta _{\rho ^{x_{1}} \eta _{1}\hat{F}\left( x_{1} x_{2} y_{2} \bar{x_2} \bar{x_1} ,x_{1}\right) ,\eta ^{x_{2} y_{2} \bar{x_2}}_{1} \rho _{2}\hat{F}\left( x_{1} ,x_{2} y_{2} \bar{x_2}\right)} \delta _{\eta ,\eta ^{x_{2}}_{1} \eta _{2}\hat{F} (x_{1} ,x_{2})}\\
	& \ \ \ \ \ \ \ \ \ \frac{\eta _{1}( F( x_{2} ,y_{2}))\rho ( F( x_{1} ,x_{2}))}{\eta _{1}\left( F\left( x_{2} y_{2} \bar{x_2} ,x_{2}\right)\right) } \dotsc 
	\end{aligned}
	\end{equation}
	The new algebra of $\Gamma^{( x,\eta )}_{( y,\rho )}$ becomes another TQD $D^{\alpha '} G'$, where $G'=\hat{K} \ltimes _{\hat{F}} N$, whose multiplication rule is
	\begin{equation}
	( x_{1} ,\eta _{1})( x_{2} ,\eta _{2}) :=\left( x_{1} x_{2} ,\eta ^{x_{2}}_{1} \eta _{2}\hat{F}( x_{1} ,x_{2})\right),
	\end{equation}
    and the 3-cocycle $\alpha '$ is given by (with $\epsilon$ factor restored back)
	\begin{equation}
	\label{eq:omegaDual}
	\alpha '(( x_{1} ,\rho _{1}) ,( x_{2} ,\rho _{2}) ,( x_{3} ,\rho _{3})) =\rho _{1}( F( x_{2} ,x_{3})) \epsilon ( x_{1} ,x_{2} ,x_{3}).
	\end{equation}
Let
	\begin{equation}
	\tilde{B}_{\tilde p}^{( \rho,xy\bar{x})}\tilde{A}_{\tilde v}^{(\eta,x)} \equiv\Gamma^{( \eta ,x)}_{( \rho ,y)},
	\end{equation}
	and relabel the group elements by $g'_1=(x_1,\eta_1)$, $h'_1=(x_1y_1\bar{x_1},\rho_1)$, $ g'_1=(x_1,\eta_1)$, $ h'_2=(x_2y_2\bar{x_2},\rho_2)$, $ g'_2=(x_2,\eta_2)$, then the algebra multiplication becomes
	\begin{equation}
	\tilde{B}_{\tilde p}^{h'_{1}} \tilde{A}_{\tilde v}^{g'_{1}}\tilde{B}_{\tilde p}^{h'_{2}} \tilde{A}_{\tilde v}^{g'_{2}} =\delta _{h'_{1} ,g'_{1} h'_{2}\bar{g'_{1}}} \beta '_{h'_{1}} (g'_{1} ,g'_{2} )\tilde{B}_{\tilde p}^{h'_{1}} \tilde{A}_{\tilde v}^{g'_1g'_2},
	\end{equation}
	where\begin{equation}
	\beta '_{h'_{1}} (g'_{1} ,g'_{2} ):=\frac{\alpha '(h'_{1} ,g'_{1} ,g'_{2} )\alpha '\left( g'_{1} ,g'_{2} ,\bar{g'_2}\bar{g'_1} h'_{1} g'_{1} g'_{2}\right)}{\alpha '\left( g'_{1} ,\bar{g'_1} h'_{1} g'_{1} ,g'_{2}\right)}
	\end{equation}
	is extracted from the partial Fourier transform above, and $g'_{1} ,h'_{1} ,g'_{2} ,h'_{2} \in G'$ are the dual group elements. The factor $\beta'$ is depicted in Fig. \ref{fig:betaByTetrahedra}.
	
Similar to Eqs. \eqref{eq:Akrho} and \eqref{eq:Bkrho}, each dual operator has an $\hat{N}$ part. Since $\alpha|_N=1=\alpha'|_{\hat{N}}$, the operators $A_v^a$ and $B_p^b$ in the TQD$(G,\alpha)$ model have the same matrix forms as those in the QD$(G)$ model. Under the partial Fourier transform \eqref{eq:FTTQD}, $\tilde{A}_{\tilde{v}}^{\eta}$ and $\tilde{B}_{\tilde{p}}^{\rho}$ define a QD$(\hat{N})$ model on the dual graph, in the same way as described in subsection \ref{sec:Z4toZ2Z2}.

\section{PEM duality transformation of anyons}

The elementary excitations in the TQD$(G,\alpha)$ model are anyons (i.e., gauge charges, gauge fluxes, and dyons) as the local topological quantum numbers. In this section, we study how anyons are transformed under the PEM duality. The anyons of a particular type carries an irreducible representation of $D^\alpha(G)$. Anyons are transformed under the PEM duality according the isomorphism $D^\alpha(G)\cong D^{\alpha'}(G')$ given by
\begin{equation}
\label{eq:isoTQD}
	A^{( a,x)} B^{( b,y)} \mapsto \frac{1}{|N|}\sum _{\rho ,\eta \in \hat{N}}\rho ( a)\overline{ \eta ( b)} \ \tilde{B}^{\left( xy\bar{x} ,\rho \right)}\tilde{A}^{( x,\eta )}.
\end{equation}

Recall that the irreducible representations of $D^\alpha(G)$ are labeled by pairs $(A,\mu)$\cite{Dijkgraaf1991} described as follows. Denote by ${^AC}$ the conjugacy classes of $G$, and for each $^AC$, we pick up a representative element ${^Ah_1}\in {^AC}$, and denote the centralizer of $^Ah_1$ by $Z_A:=\{g\in G|g{^Ah_1}={^Ah_1}g\}$. Let $(\mu,V_\mu)$ be an irreducible projective representation of $Z_A$ with the cocycle $\beta_{^Ah_1}$, i.e., 
\begin{equation}
    \mu(g)\mu(h)=\beta_{^Ah_1}(g,h)\mu(gh),
\end{equation}
for all $g,h\in Z_A$.
For every element in ${^Ah_i}\in {^AC}$, we introduce a unique representative $x_i\in G$ such that $x_i({^Ah_1})\bar{x_i}={^Ah_i}$. We can conveniently set $x_1=1$.

For every such pair $(A,\mu)$, we define an irreducible representation $(\pi^A_{\mu},V^A_{\alpha})$ of $D^{\alpha}(G)$, where the representation space is $V^A_{\mu}=\mathbb{C}[^AC]\otimes V_{\mu}$, and the representation matrix is can be obtained from
\begin{equation}
\label{eq:repTQD}
\pi^A_{\alpha}(A^gB^h)\ket{x_i,v}=\delta_{h,{^Ah_i}}\frac{\beta_{\bar{x_i}hx_i}(g,x_i)}{\beta_{\bar{x_i}hx_i}(x_k,\tilde g)}\ket{x_k,\mu(\bar{x_k}gx_i)v},
\end{equation}
where $v\in V_{\mu}$ and $x_k$ is determined by $g({^Ah_i})\bar{g}={^Ah_k}$, and $\tilde g=\bar{x_k}gx_i$. Here $\beta_h(g,x)$ refers to $\beta_h(g,u(x))$ where $u(x)=(1_N,x)$, and $\beta_h(x,g)$ refers to $\beta_h(u(x),g)$.
In the following, we will present a more explicit formula of the representation \eqref{eq:repTQD} when $G$ can be written as a semidirect product.

Let $G=N\rtimes_F K$ as described in the previous sections. The charges are identified with the irreducible representations $(A_1,\mu)$, where $A_1$ is the trivial conjugacy class $\{1\}$, and $\mu$ runs over all the irreducible representations of $G$. We have a more explicit formula of $\mu$ for $G=N\rtimes_F K$.

Fix $\xi\in\hat{N}$. Let 
\begin{equation}
\label{eq:Kxi}
K_\xi=\{x\in K|\xi^x=\xi\}.
\end{equation}
Let $x_i$ be the left coset representatives of $K/K_\xi$, i.e., unique representatives such that $\xi^{x_i}\neq \xi^{x_j}$ if $x_i\neq x_j$. Denote by $r$ an irreducible projective representation of $K_\xi$ such that $r(k_1)r(k_2)=\xi(F(k_1,k_2))r(k_1,k_2)$ for $k_1,k_2\in K_\xi$. Then we can construct an irreducible representation $\xi\otimes r$ of $N\rtimes_F K_\xi$ with $(\xi\otimes r)(a,k):=\xi(a)r(k)$ for $(a,k)\in N\rtimes_F K_\xi$. It induces an irreducible representation $(R^\xi_r,V^\xi_r)$ of $N\rtimes_F K$ defined by
\begin{equation}
\begin{aligned}
R^\xi_r(a,x)\ket{x_i,v}
=&\frac{\xi(F(x,x_i))}{\xi(F(x_k,\bar{x_k}xx_i))}\ket{x_k,\xi^{\bar{x_k}}(a)r(\overline{x_k}xx_i)v}
\end{aligned}
\end{equation}
where $v\in V_r$, $x_k$ is determined by $\xi^{xx_i}=\xi^{x_k}$. One verifies that $R^\xi_r(a_1,x_1)R^\xi_r(a_2,x_2)=R^\xi_r((a_1,x_1)(a_2,x_2))$ by using the identity
\begin{equation}
\frac{\xi(F(x_1,x_j))}{\xi(F(x_k,\widetilde{x_1}))}\frac{\xi(F(x_2,x_i))}{\xi(F(x_j,\widetilde{x_2}))}
\xi(F(\widetilde{x_1},\widetilde{ x_2}))
=\frac{\xi(F(x_1x_2,x_i))}{\xi(F(x_k,\widetilde{x_1}\widetilde{x_2}))}
\xi^{\bar{x_k}}(F(x_1,x_2)),
\end{equation}
where $x_j$ and $x_k$ is determined by $\xi^{x_1x_j}=\xi^{x_k}$ and $\xi^{x_2x_i}=\xi^{x_j}$, $\widetilde{x_1}=\bar{x_k}x_1x_j$, and $\widetilde{x_2}=\bar{x_j} x_2 x_i$. The representations $R^\xi_r$ form a complete set of all (inequivalent) irreducible representations of $N\rtimes_F K$  (e.g., see Section 8.2 of Ref\cite{Serre1977LinearGroups}). Hence, the charges can be classified by the representations $(A_1,R^\xi_r)$.

The representation $\pi^{A_1}_{R^\xi_r}$ induces an irreducible representation  of $D^{\alpha'}(G')$ under the Fourier transform \eqref{eq:isoTQD}, defined by
\begin{equation}
\label{eq:piTildeTQD}
\widetilde{\pi^{A_1}_{R^\xi_r}}(\tilde{B}^{(y,\rho)}\tilde{A}^{(x,\eta)})
=\frac{1}{|N|}\sum_{ab}\overline{\rho(a)}{\eta(b)}\pi^{A_1}_{R^\xi_r}(A^{(a,x)}B^{(b,y)})
\end{equation}
Since the Fourier transform is an isomorphism $D^\alpha(G)\cong D^{\alpha'}(G')$ that preserves the product structure, the representation $\widetilde{\pi^{A_1}_{R^\xi_r}}$ must be an irreducible representation of $D^{\alpha'}(G')$, which should be identified with a pair $(\tilde{A},\tilde{\mu})$ for certain conjugacy class $\tilde{A}$ of $G'$ and irreducible projective representation $\tilde{\mu}$ of the centralizer $Z_{\tilde{A}}$ in $G'$.

To identify the pair $(\tilde{A},\tilde{\mu})$, we evaluate the matrix element of $\widetilde{\pi^{A_1}_{R^\xi_r}}(\tilde{B}^{(y,\rho)}\tilde{A}^{(x,\eta)})$ by
\begin{equation}
\label{eq:tildepi}
\begin{aligned}
&\frac{1}{|N|}\sum_{ab}\overline{\rho(a)}{\eta(b)}\left(\delta_{y,1_K}\delta_{b,1_N}\frac{\xi(F(x,x_i))}{\xi(F(x_k,x))}\xi^{\bar{x_k}}(a)r(\overline{x_k}xx_i)\right)
\\
=&\delta_{y,1_K}\delta_{\rho,\xi^{\bar{x_k}}}\frac{\xi(F(x,x_i))}{\xi(F(x_k,x))}r(\overline{x_k}xx_i),
\end{aligned}
\end{equation}
where $h=(y,\rho),g=(x,\eta)$. From the delta functions, we recognize that $\tilde{A}$ is generated by $(1_K,\xi)$. Let $^{\tilde{A}}h_1=(1_K,\xi)$. One immediately sees that $^{\tilde{A}}C=\{\xi^{\bar{x_k}}\}$ and that the centralizer is $Z_{\tilde{A}}=K_\xi\ltimes_{\hat{F}}\hat{N}$, where $K_\xi$ is the subgroup defined in Eq. \eqref{eq:Kxi}, and $x_k$ is the corresponding coset representatives of $K/K_\xi$. Therefore, $K/K_\xi\cong {^{\tilde{A}}C}$, and we choose $\tilde u(x_k)=(x_k,1_{\hat{N}})$ as the representative to label the element ${^{\tilde{A}}h_k}=\tilde u(x_k) {^{\tilde{A}}h_1} \overline{\tilde u({x_k})}=\rho^{\bar{x_k}}$ in ${^{\tilde{A}}C}$.

With the identification above, together with that $\tilde{A}^{(x,\eta)}\tilde{B}^{(y,\rho)}=\tilde{B}^{(xy\bar{x},\rho^{\bar{x}})}\tilde{A}^{(x,\eta)}$, the representation $\widetilde{\pi^{A_1}_{R^\xi_r}}$ can be written in the standard form as in Eq. \eqref{eq:repTQD}, i.e.,
\begin{equation}
\widetilde{\pi^{A_1}_{R^\xi_r}}(\tilde{A}^{(x,\eta)}\tilde{B}^{(y,\rho)})\ket{\tilde u(x_i),v}
=\delta_{(y,1_K),(\rho,\xi^{\bar{x_i}})}\frac{\beta_{^{\tilde A}h_1}(g,x_i)}{\beta_{^{\tilde A}h_1}(x_k,\tilde g)}\ket{\tilde u(x_k),\tilde{\mu}(\overline{\tilde u(x_k)}(x,\eta)\tilde u(x_i))v},
\end{equation}
where $v\in V_{\tilde{A}}=V_r$, $\tilde g=\overline{u(x_k)}gu(x_i)$, and $\tilde{\mu}$ is an irreducible projective representation of $Z_{\tilde{A}}=K_\xi\ltimes_{\hat{F}}\hat{N}$ with the 2-cocycle $\beta_{^{\tilde A}h_1}$, defined by
\begin{equation}
\tilde{\mu}(x,\eta)=r(x).
\end{equation}

Loosely speaking, a charge $(A_1,R^\xi_r)$ can be factorized into a trivial flux $A_1$, a $N$-charge $\xi$, and a $K_{\xi}$-charge $r$. Then, the PEM duality maps the charge to a dyon with a $\hat{N}$-flux $\tilde{A}=(1_K,\xi)$, a trivial $\hat{N}$-charge, and the $K_\xi$-charge $r$.

We have explicitly shown how the charges are transformed under the PEM duality. The derivation can be extended to all types of anyons.

\section{Dual model: Reciprocal bilayer model}\label{sec:bilayerModel}
The PEM duality in a TQD model naturally casts the model and its dual model in a bilayer fomulation, which conversely makes the PEM duality more comprehensible.

\subsection{Duality transformation on observables}
	
The PEM duality transformation on the algebra of the TQD algebra $D^\alpha(G)$ reads
	\begin{equation}
	\frac{1}{|N|}\sum _{a,b\in N}\overline{\rho ( a)} \eta ( b) A^{( a,x)} B^{( b,y)} =\tilde{B}^{\left( xy\bar{x} ,\rho \right)}\tilde{A}^{( x,\eta )},
	\end{equation}
	with its inverse transformation being
	\begin{equation}
	\label{eq:algebraIsomorphism}
	A^{( a,x)} B^{( b,y)} =\frac{1}{|N|}\sum _{\rho ,\eta \in \hat{N}}\rho ( a)\overline{ \eta ( b)} \ \tilde{B}^{\left( xy\bar{x} ,\rho \right)}\tilde{A}^{( x,\eta )}.
	\end{equation}
	The individual operators are transformed as
	\begin{equation}
	A^{( a,x)} =\left(\sum _{\rho \in \hat{N}}{\rho ( a)}\sum _{y\in K}\tilde{B}^{( y,\rho )}\right)\tilde{A}^{( x,1)},
	\end{equation}
	where $1$ in the pair $( x,1)$ on the RHS is the identity element of $\hat{N}$ (identity representation of $N$), and
	\begin{equation}
	B^{( b,y)} =\left(\sum _{\rho \in \hat{N}}\tilde{B}^{( y,\rho )}\right)\frac{1}{|N|}\sum _{\eta \in \hat{N}}\overline{\eta ( b)} \ \tilde{A}^{( 1,\eta )},
	\end{equation}
	where $1$ in the $( 1,\eta )$ the RHS is the identity element of $K$.

	\subsection{Reciprocal bilayer model}
	In this subsection, we will examine the algebra structure of the operators $A^{( a,x)} , B^{( b,y)}$ and $\tilde{A}^{( x,\eta )} , \tilde{B}^{( y,\rho )}$, and show that the TQD$( G,\alpha )$ model or the dual model can be understood as a bilayer system. The dual model, understood as a bilayer model, consists of an upper layer that is a QD$(\hat{N})$ model with Hilbert subspace spanned by the configurations of $\hat{N}$ elements on the edges of the dual graph, and a lower layer on which the Hilbert subspace is spanned by the $K$ elements on the edges of the dual graph. We will call such a bilayer model a \textbf{reciprocal bilayer model}.
	
	We start by examining the operators $A^{( a,x)}_v$, $B^{( b,y)}_p$ in the TQD$( G,\alpha )$ model and the operators $\tilde{A}^{( x,\eta )}_{\tilde{v}}$, $\tilde{B}^{( y,\rho )}_{\tilde{p}}$ in the dual model TQD$(G',\alpha')$. To reduce confusion, we will suppress the subscript $v,p$ in the sequel. These operators are parameterized by pairs of group elements, so we can factorize each such operator into an $N$ (or $\hat{N}$) part and a $K$ part. Define
	\begin{equation}
	\label{eq:AaAx} A^{a} =A^{(a,1)} ,A^{x} =A^{(1,x)} ,B^{b} =\sum _{y\in K} B^{(b,y)} ,B^{y} =\sum _{b\in N} B^{(b,y)},
	\end{equation}
	and define
	\begin{equation}
	\label{eq:tildeAaAx}
	\tilde{A}^{\eta } =\tilde{A}^{(1,\eta )} ,\tilde{A}^{x} =\tilde{A}^{(1,x)} ,\tilde{B}^{\rho } =\sum _{y\in K}\tilde{B}^{(y,\rho )} ,\tilde{B}^{y} =\sum _{\rho \in \tilde{N}}\tilde{B}^{(y,\rho )}.
	\end{equation}
	According to the formulas \eqref{eq:omegaTQD} and \eqref{eq:omegaDual}, we have
	\begin{equation}
	A^{(a,x)} =A^{a} A^{x} ,B^{(b,y)} =B^{b} B^{y},
	\end{equation}
	\begin{equation}
	\tilde{A}^{(x,\eta )} =\tilde{A}^{x}\tilde{A}^{\eta } ,\tilde{B}^{(y,\rho )} =\tilde{B}^{y}\tilde{B}^{\rho }.
	\end{equation}
	Note that $A^{x} =\tilde{A}^{x}$ and $B^{y} =\tilde{B}^{y}$. 
	
The factorized operators in Eqs. \eqref{eq:AaAx} and \eqref{eq:tildeAaAx} may be used to build a bilayer system. The upper layer accommodates a QD model, whose local operators $A^{a} ,B^{b}$ form a QD algebra $D(N)$. More specifically, for every fixed configuration $\{k\}$ of $K$ elements on the edges of the lower level, the concrete matrix forms of $A^{a}, B^{b}$ depend on the configuration $\{k\}$. This dependence reveals how the upper layer is coupled to the lower layer.
	
	On the other hand, the lower layer model is neither a QD nor a TQD model defined by $K$. The Hilbert subspace in the lower layer is spanned by the configurations of the $K$ elements. The local operators, $A^{x} =\tilde{A}^{x}$ and $B^{y} =\tilde{B}^{y}$, in general do not form a quantum double algebra or twisted quantum double. The set $\left\{A^{x} ,B^{x}\right\}_{x\in K}$ is even not closed under the multiplication: $A^{x} A^{y}$ yields an $A^{F(x,y)} A^{xy}$ term with $F(x,y)\in N$. Nevertheless, we can rearrange the labels $(a,x)$ in the operators to make the set closed. In the following, we shall introduce the rearranged operators $\mathbf{A}^{(a,x)}\mathbf{B}^{(b,y)}$, which form a subalgebra on each layer.
	
	First, we introduce a new degree of freedom $x_{v}\in K$ on every vertex $v$. The group elements $x_v$ depends on the configuration of $k$'s on the edges of the triangulation, such that $x_v$ can be viewed as a function $x_v(\{k\})$. Suppose $\{k\}$ is a configuration, and $x_v$ is the corresponding group element at a vertex $v$. If another configuration $\{k'\}$ is related to $\{k\}$ by a gauge transformation at $v$, then the corresponding element $x_v(\{k'\})$ is given by
	\begin{equation}
	v_x\left(\configurationKAB\right) = yx_v\left(\configurationKAA\right).
	\end{equation}
	
	In short, $x_v$ transforms in the regular representation of $K$ under the above mapping. This implies that $x_{v}(\{k\})$ is a nonlocal function. The above constraint on $x_v$ does not uniquely determine the function $x_v$. In the following, we choose an arbitrary solution of $x_v$ to the constraint.

	It is convenient to introduce the projection operator $P_v^{x_0}$ as
	\begin{equation}
	P_v^{x_0}|\{k\}\rangle = \delta_{x_v(\{k\}),x_0}|\{k\}\rangle.
	\end{equation}
	The above constraint on the function $x_v$ is expressed as
	\begin{equation}
	P_v^{(xx_0)}A_v^{(a,x)} = A_v^{(a,x)}P_v^{(x_0)},
	\quad
	P_v^{(x_0)}B_p^{(a,x)} = A_v^{(a,x)}B_p^{(x_0)}.
	\end{equation}
	
	
	We define the rearranged operators as
	\begin{equation}
	P^{x_0}\mathbf{A}^{(a,x)}\mathbf{B}^{(b,y)} =\overline{\hat{F}(x,x_0)(b)} P^{x_0}A^{(\lambda_a,x)}B^{(^{x_0}b,y)}
	\end{equation}
	and
	\begin{equation}
	\mathbf{\tilde B}^{(xy\bar{x},\rho)}\mathbf{\tilde A}^{(x,\eta)} P^{x_0} = \overline{\rho(F(\bar{x_0},x))}\tilde B^{(xy\bar{x},\rho^{\bar{x_0}})}\tilde A^{(x,\tilde\lambda_\eta)}P^{x_0},
	\end{equation}
	where
	\begin{equation}
	\lambda_a = {^{x_0}\left(\frac{a}{F(\bar{x_0},x)}\right)}
	\end{equation}
	and
	\begin{equation}
	\tilde\lambda_{\eta} = \left(\frac{\eta}{\hat{F}(x,x_0)}\right)^{\bar{x_0}}
	\end{equation}
	are defined by $(1_N,\bar{x_0})(\lambda_a,x) = (a,\bar{x_0}x)$ and $(x,\tilde\lambda_\eta)(x_0,1_{\hat{N}}) = (xx_0,\eta)$.
	
	The map $a\mapsto \lambda_a$ is an automorphism on $N$, with an inverse map $a\mapsto \kappa_{\bar{x_0},(a,x)}={^{\bar{x_0}}a}F(\bar{x_0},x)$ . The $\lambda$ has the property that $(1,\bar{x_0})(\lambda_a,x)(\lambda_b,y) = (1,\bar{x_0})(\lambda_{ab},xy)$. Similarly, $\tilde\lambda_\eta$ is an automorphism on $\hat{N}$ satisfying $(x,\tilde\lambda_\eta)(y,\tilde{\lambda}_{\rho})(x_0,1_{\hat{N}}) = (xy,\tilde{\lambda}_{\eta\rho})(x_0,1_{\hat{N}})$. 
	See Appendix \ref{sec:rearrangeAB} for the detailed discussion about the rearranged operators.
	
	Define
	\begin{equation}
	\mathbf A^x \mathbf B^y = \mathbf A^{(1,x)}\sum_b\mathbf{B}^{(b,y)},
	\end{equation}
	\begin{equation}
	\mathbf{\tilde B}^{xy\bar{x}}\mathbf{\tilde A}^x = \sum_\rho\mathbf{\tilde B}^{(xy\bar{x},\rho)}\mathbf{\tilde A}^{(x,1)},
	\end{equation}
	which are clearly identical operators: $\mathbf A^x \mathbf B^y = \mathbf{\tilde B}^{xy\bar{x}}\mathbf{\tilde A}^x$. These operators form a subalgebra:
	\begin{equation}
	\label{eq:algebraAXBY}
	\mathbf A^{x_1}\mathbf{B}^{y_1}\mathbf{A}^{x_2}\mathbf B^{y_2}P^{x_0} = \delta_{y_1,x_2y_2\bar{x_2}}\Phi_{y_2}(x_1,x_2)\beta^\epsilon_{y_2}(x_1,x_2)\mathbf A^{x_1x_2}\mathbf{B}^{y_2}P^{x_0},
	\end{equation}
	where 
	\begin{equation}
	\begin{aligned}
	& \Phi _{y_{2}} (x_{1} ,x_{2} )\beta ^{\epsilon }_{y_{2}} (x_{1} ,x_{2} )\\
	= & \frac{\epsilon \left( x_{2} x_{0} ,\overline{x_{0}}\overline{x_{2}} ,x_{2} y_{2}\overline{x_{2}}\right) \epsilon \left( x_{2} x_{0} ,\overline{x_{0}} y_{2} x_{0} ,\overline{x_{0}}\right) \epsilon \left( x_{1} x_{2} x_{0} ,\overline{x_{0}} ,y_{2}\right)}{\epsilon \left( x_{2} x_{0} ,\overline{x_{0}} ,y_{2}\right) \epsilon \left( x_{2} x_{0} ,\overline{x_{0}} y_{2} x_{0} ,\overline{x_{0}}\overline{x_{2}}\right) \epsilon \left( x_{1} x_{2} x_{0} ,\overline{x_{0}}\overline{x_{2}} ,x_{2} y_{2}\overline{x_{2}}\right)}\\
	& \quad\frac{\epsilon \left( x_{1} x_{2} x_{0} ,\overline{x_{0}} y_{2} x_{0} ,\overline{x_{0}}\overline{x_{2}}\right) \epsilon \left( x_{2} y_{2}\overline{x_{2}} ,x_{2} x_{0} ,\overline{x_{0}}\overline{x_{2}}\right) \epsilon \left( x_{1} x_{2} y_{2}\overline{x_{2}}\overline{x_{1}} ,x_{1} x_{2} x_{0} ,\overline{x_{0}}\right)}{\epsilon \left( x_{1} x_{2} x_{0} ,\overline{x_{0}} y_{2} x_{0} ,\overline{x_{0}}\right) \epsilon \left( x_{2} y_{2}\overline{x_{2}} ,x_{2} x_{0} ,\overline{x_{0}}\right) \epsilon \left( x_{1} x_{2} y_{2}\overline{x_{2}}\overline{x_{1}} ,x_{1} x_{2} x_{0} ,\overline{x_{0}}\overline{x_{2}}\right)}
	\end{aligned}
	\end{equation}
	See Appendix \ref{sec:rearrangeAB} for the derivation.

	Let $\mathbf A^a =\sum_{by}\mathbf A^{(a,1)}\mathbf{B}^{(b,y)}=A^a$, $\mathbf{B}^{b}=\sum_{by}\mathbf{A}^{(1,1)}\mathbf{B}^{(y)}$ and $\mathbf A^x = \sum_{by}\mathbf A^{(1,x)}\mathbf{B}^{(b,y)}$, $\mathbf B^y =\sum_b\mathbf{A}^{(1,1)} \mathbf B^{(b,y)}$. We have
	\begin{equation}
	\mathbf A^{a_1} \mathbf B^{b_1}\mathbf A^{a_2} \mathbf B^{b_2} = \delta_{b_1,b_2}\mathbf A^{a_1a_2} \mathbf B^{b_2}.
	\end{equation}
	Some immediate consequences are $\mathbf{A}^x=\mathbf{\tilde{A}}^x$, $\mathbf{B}^y=\mathbf{\tilde{B}}^y$, $\mathbf{B}^{xy\bar{x}}\mathbf{A}^x=\mathbf{A}^x\mathbf{B}^y$.

	Having derived the two subalgebras above, one consisting of $\mathbf{A}^a,\mathbf{B}^b$ over $N$, and the othe consisting of $\mathbf{A}^x,\mathbf{B}^y$ over $K$, we can rewrite the TQD Hamiltonian in terms of these operators.

	Since $(\sum_a\mathbf A^{a})(\sum_x \mathbf A^{x})=\sum_{ax} A^{(a,x)}$ and $\mathbf B^{1_N} \mathbf B^{1_K}= B^{(1,1)}$, we can rewrite the TQD Hamiltonian as
	\begin{equation}\label{eq:TQDHamKH}
	H^{\text{bilayer}}=H^N+H^K,
	\end{equation}
	where
	\begin{equation}
	\label{eq:AaBN}
	H^N=-\sum _{v}\frac{1}{|N|}\sum _{a\in N} \mathbf{A}^{a}_{v}-\sum _{p} \mathbf{B}^{1_{N}}_{p}
	\end{equation}
	and
	\begin{equation}
	\label{eq:AxBK}
	H^K=
	-\frac{1}{|K|}\sum _{x\in K} \mathbf{A}^{x}_{v}\sum_{p\text{ around }v}\mathbf{B}_p^{1_K}
	-\sum_{p} \mathbf{B}^{1_{K}}_{p}.
	\end{equation}
	
	The $H^N$ and $H^K$ describes two subsystems respectively. The first term $H^N$ defines a QD model with $N$ on the triangulation $\Gamma$, with the Hilbert subspace spanned by the configurations of $N$ elements on the edges of $\Gamma$. In this model, the local operators $\mathbf{A}^a,\mathbf{B}^b$ form the QD algebra $D(N)$. We call this subsystem the upper layer (or first layer). The term $H^K$ defines a model with the Hilbert subspace spanned by the configurations of the $K$ elements on the edges on a duplicate of $\Gamma$. In this model, the local operators $\mathbf{A}^x,\mathbf{B}^y$ form the algebra given in Eq. \eqref{eq:algebraAXBY}. We call this subsystem the lower layer (or second layer).
	
	The $H^{\text{bilayer}}$ has the identical spectrum of of eigenstates as that of the original Hamiltonian \eqref{eq:TQDHam}. We also observe that all the four terms in Eqs. \eqref{eq:AaBN} and \eqref{eq:AxBK} are commuting projection operators and that $\sum_{ax}\mathbf{A}^a\mathbf{A}^x\mathbf{B}^{1_N}\mathbf{B}^{1_K}=\sum_{ax}A^{(a,x)}B^{(1_N,1_K)}$.

\begin{figure}[!ht]
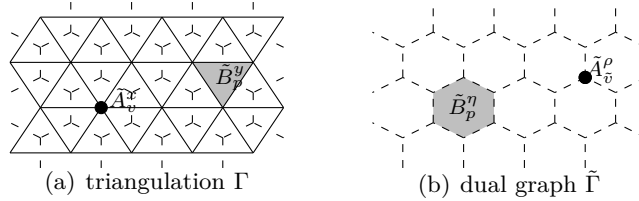

		\centering
		\subfigure[triangulation $\Gamma$]{\DoubleLayerGraph}
		\qquad
		\subfigure[dual graph $\tilde{\Gamma}$]{\DoubleLayerGraphAA}
		\caption{(a) The operators $\tilde{A}^{x}_{v},\tilde{B}^{y}_{p}$ defined on the original triangulation $\Gamma$. (b) The operators $\tilde{A}^{\rho}_{\tilde{v}},\tilde{B}^{\eta}_{\tilde{p}}$ defined on the dual graph $\tilde{\Gamma}$.}
		\label{fig:DoubleLayerGraph}
	\end{figure}

	Under the partial EM duality, $\{A^{a}_v \}$ are mapped to $\left\{\tilde{B}^{\rho }_v\right\}$, while $\left\{B^{b}_p\right\}$ to $\left\{\tilde{A}^{\rho }_p\right\}$. In terms of the rearranged operators, $\{\mathbf{A}^{a}_v \}$ are mapped to $\left\{\mathbf{\tilde{B}}^{\rho }_v\right\}$, and $\left\{\mathbf{B}^{b}_p\right\}$ to $\left\{\mathbf{\tilde{A}}^{\rho }_p\right\}$.  We introduce the dual trivalent graph $\tilde \Gamma$, and relabel the vertex $v$ and plaquette $p$ by the dual plaquette $\tilde{p}=v$ and the dual vertex $\tilde{v}=p$. The Hamiltonian of the dual model can be written as
	\begin{equation}
	\tilde{H} = \tilde{H}^K + \tilde{H}^{\hat{N}},
	\end{equation}
	where
	\begin{equation}\label{eq:HamK}
	\tilde{H}^K=-\sum _{v}\frac{1}{|K|}\sum _{x\in K}\mathbf{\tilde{A}}^{x}_{v} -\sum _{p}\frac{1}{|N|}\sum _{\eta \in \tilde{H}}\mathbf{\tilde{B}}^{1_{K}}_{p} 
	\end{equation}
	and
	\begin{equation}\label{eq:HamH}
	\tilde{H}^{\hat{N}}=-\sum _{\tilde{v}}\mathbf{\tilde{A}}^{\eta }_{\tilde{v}} - \sum _{\tilde{p}}\mathbf{\tilde{B}}^{1_{\tilde{N}}}_{\tilde{p}}.
	\end{equation}
	The $\tilde{H}^K$ defined on the original triangulation $\Gamma$ is the same as the first part of Eq. \eqref{eq:TQDHamKH} because $\mathbf A^{x}_{v} =\mathbf{\tilde{A}}^{x}_{v}$ and $\mathbf{B}^{y}_{p} =\mathbf{\tilde{B}}^{y}_{p}$. The operators are labeled by elements of $K$. The $\tilde{H}^{\hat{N}}$ defined on the dual graph $\tilde{\Gamma}$ (in dashed lines in Fig. \ref{fig:DoubleLayerGraph} ), with operators labeled by elements of $\hat{N}$, forms a QD model defined by the finite group $\hat{N}$. Similar to that in the original TQD model, the restricted Hilbert space for every fixed configuration $\{k_{1} ,k_{2} ,\dotsc \}$ of elements of $K$ on the triangulation give rise to an individual Hilbert space of this QD model.

\begin{figure}[htb]
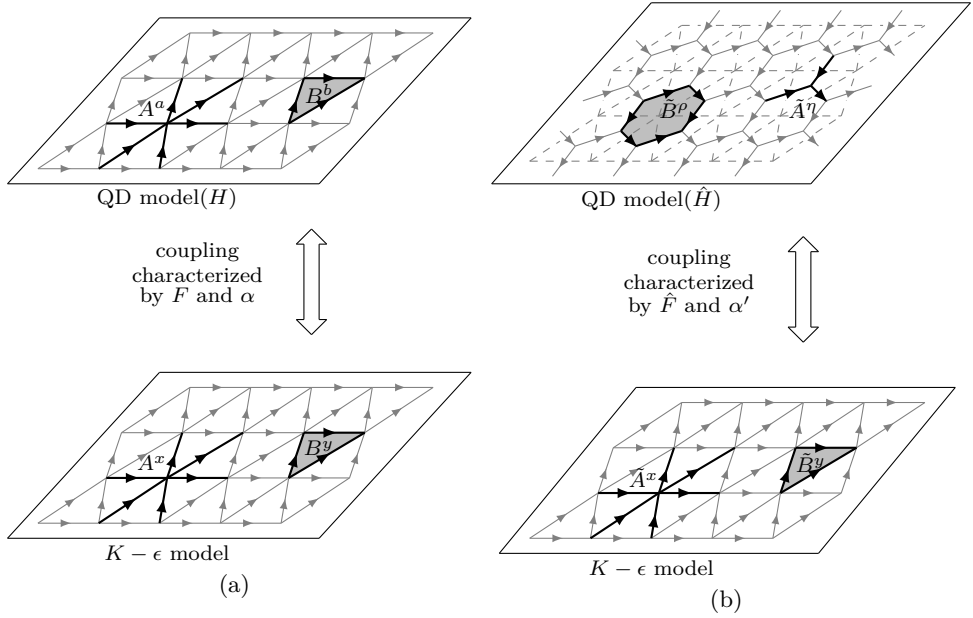

		\begin{center}
			\subfigure[]
			\bilayerModelAA
			\subfigure[]
			\bilayerModelAB
		\end{center}
		\caption{(a) The TQD$( G,\alpha )$ model can be understood as a bilayer system. The upper layer is a QD$( N)$ model and the lower layer is a $K$-$\epsilon$ model. Both layers are on the same graph. (b) Under the EM duality, the upper layer is mapped to a QD$(\hat{N})$ model and $K$-$\epsilon$ model on the lower layer remains unchanged. The graphs on the upper and the lower layer is dual to each other.}
		\label{fig:bilayerModel}
	\end{figure}

 	The partial EM duality maps the upper layer (a QD$(N)$ model) of the original model to a the upper layer (QD$(\hat{N})$ model) of the dual model, while preserves the lower layer ( $K$-$\epsilon$ model). The graphs on the upper and the lower layer are graph-dual to each other, we thus call the dual model \textit{the reciprocal bilayer model}. The coupling of the two layers is characterized by the semidirect product structure $F$ and the 3-cocycle $\alpha$ on $G$ ($\hat{F}$ and $\alpha '$ on $G'$ on the dual model respectively). See Fig. \ref{fig:bilayerModel}. Since $\alpha$ is further determined by $\hat{F}$ and $\alpha'$ by $F$, we see that in bot models, the coupling between two layers are characterized by $F$ and $\hat{F}$.
	
	As a special case, if we begin with a QD$(G)$ model, the lower layer model is a QD$(K)$ model since $\epsilon=1$. The QD$(G)$ is a QD$(N)$-QD$(K)$ bilayer system, with the coupling purely characterized by the semidirect product structure $F$. Under the PEM duality, the QD$(G)$ model is mapped to a TQD$(G',\alpha')$ model, where $G'=K\times \hat{N}$ is a direct product and $\alpha'$ is determined by the semidirect product structure $F$. The coupling is characterized by $\alpha'$ (and finally by $F$ which determines the form of $\alpha'$).

	\subsection{Invariant ground states under the PEM duality}
	
	Under the EM duality, the ground states of the original model are mapped to the ground states in the dual model. The ground states of the original TQD model are the simultaneous $+1$ eigenvectors of $\frac{1}{|G|}\sum_g A^gB^1$, while those of the dual model are the common $+1$ eigenvectors of $\frac{1}{|G'|}\sum_{g'} \tilde{A}^{g'}\tilde{B}^{1}=1$ in the dual model. Under the PEM duality, the constraint operator is transformed as
	\begin{equation}\label{eq:GroundStateOperator}
	\frac{1}{|G|}\sum_g A^gB^1\mapsto
	\frac{1}{|G'|}\sum_{g'} \tilde{B}^{1}\tilde{A}^{g'}.
	\end{equation}
	Hence the EM duality preserves the ground states.

	\section{Isomorphism between two twisted quantum double algebras}
	
	We have derived the PEM duality via the partial Fourier transform. The mathematics behind the derivation is that the Fourier transform over $N$ induces an isomorphism of the two TQD algebras dual to each other. 
	
	Given $(G,\alpha)$ and $(G',\alpha')$, where $G=N\rtimes _{F} K$ with $N$ an Abelian normal subgroup of $G$, $\alpha ((a_{1} ,k_{1} ),(a_{2} ,k_{2} ),(a_{3} ,k_{3} )):=\hat{F} (k_{1} ,k_{2} )(a_{3} )\epsilon (k_{1} ,k_{2} ,k_{3} )$, $G'=K\ltimes _{\hat{F}}\hat{N}$ with $\alpha '((x_{1} ,\rho _{1} ),(x_{2} ,\rho _{2} ),(x_{3} ,\rho _{3} ))=\rho _{1} (F(x_{2} ,x_{3} ))\epsilon (x_{1} ,x_{2} ,x_{3} )$, and $\epsilon\in C^3(K,U(1))$ satisfying $\delta_K(\epsilon)=\hat{F}\wedge F$,
	we have constructed an isomorphism via the partial Fourier transform over $N$:
	\begin{equation}
	\begin{aligned}
	\mathcal{F} :\  & D^{\alpha } G\rightarrow D^{\alpha '} G'\\
	& A^{(a,x)} B^{(b,y)} \mapsto \frac{1}{|N|}\sum _{\rho ,\eta \in \hat{N}}\rho (a)\overline{\eta (b)} \ \tilde{B}^{\left( xy\bar{x} ,\rho \right)}\tilde{A}^{(x,\eta )},
	\end{aligned}
	\end{equation}
	 As an immediate consequence, the individual operators $A^g\equiv \sum_h A^gB^h$ and $B^h\equiv A^{1}B^h$ are mapped as
	\begin{equation}
	A^{( a,x)} \mapsto \left(\sum _{\rho \in \hat{N}}{\rho ( a)}\sum _{y\in K}\tilde{B}^{( y,\rho )}\right)\tilde{A}^{( x,1)}
	\end{equation}
	and
	\begin{equation}
	B^{( b,y)} \mapsto \left(\sum _{\rho \in \hat{N}}\tilde{B}^{( y,\rho )}\right)\frac{1}{|N|}\sum _{\eta \in \hat{N}}\overline{\eta ( b)} \ \tilde{A}^{( 1,\eta )}.
	\end{equation}
	
	The isomorphism preserves the quasi-Hopf algebra structure. In Section \ref{sec:FTqtd}, we have concretely derived the partial Fourier transform of the product coefficient. The result can be summarized as a proof that $\mathcal{F}$ preserves the product structure:
	\begin{equation}
	\label{eq:preserveProduct}
	\mathcal{F}(A^{g_1}B^{h_1})\mathcal{F}(A^{g_2}B^{h_2})=\mathcal{F}(A^{g_1}B^{h_1}A^{g_2}B^{h_2}).
	\end{equation}
	The preservation of the coproduct structure is just dual to that of the product structure. The proof can be done in a similar way, which is beyond this work.
	
	The map of the key structures under the isomorphism is summarized in Table \ref{tab:KeyStructure}.

	\begin{table}[!h]
		\centering	
		\begin{tabular}{cc}
			\toprule
			TQD model
			&	Dual model
			\\
			\midrule
			$G=N\rtimes _{F} K$
			& 
			$G'=K\ltimes _{\hat{F}}\hat{N}$
			\\
			\hline 
			semidirect product structure $F$
			& 
			dual 3-cocycle $\alpha '$
			\\
			\hline 
			3-cocycle $\alpha $
			&
			semidirect product structure $\hat{F}$
			\\
			\bottomrule
		\end{tabular}
		\caption{The mapping of the key structures under the isomorphism.}
		\label{tab:KeyStructure}
		
	\end{table}

	\acknowledgments
	
	The authors thank Davide Gaiotto for inspiring discussions and Bernardo Uribe for encouraging communications after the first version of this paper. Y.H. thanks his supervisor Wenqing Zhang for his support. Y.W. is supported by NSF grant No. 11875109 and Shanghai Municipal Science and Technology Major Project (Grant No.2019SHZDZX01).
	
	
	\appendix

	\section{3-cocycles for $\mathbb{Z}^{3}_{m}$}
	\label{sec:cocycleZZZ}
	
The group elements are denoted by triples $a=( a_{1} ,a_{2} ,a_{3})$ with $a_{1} ,a_{2} ,a_{3} =0,1,\dotsc ,m-1.$ The cohomology group $H^{3}\left(\mathbb{Z}^{3}_{m} ,U(1)\right) =\mathbb{Z}^{7}_{m}$ has seven generators,
	\begin{equation}
	\begin{aligned}
	\alpha ^{(i)}_{I} (a,b,c) & =\exp\left\{\frac{2\pi \mathrm{i}}{m^{2}} a_{i}( b_{i} +c_{i} -\langle b_{j} +c_{j} \rangle )\right\}\\
	\alpha ^{(ij)}_{II} (a,b,c) & =\exp\left\{\frac{2\pi \mathrm{i}}{m^{2}} a_{i}( b_{j} +c_{j} -\langle b_{j} +c_{j} \rangle )\right\}\\
	\alpha _{III} (a,b,c) & =\exp\left\{\frac{2\pi \mathrm{i}}{m} a_{1} b_{2} c_{3}\right\}
	\end{aligned}
	\end{equation}
	where $1\leq i\leq 3$ and $1\leq i\leq j\leq 3$ are assumed respectively in the first two lines, and $\langle x\rangle = x \mod m$.
	
	\section{twisted quantum double $D^{\alpha } G$}
	\label{sec:semidirect}
	
	In what follows we briefly review the definition of the twisted Drinfeld's twisted double of a finite group\cite{Dijkgraaf1991} as a quasi-triangular quasi-Hopf algebra. Let $D^{\alpha } G$ be a finite-dimensional vector space with a basis $\{A^gB^x\}_{(g,x)\in G\times G}$. Define a product on $D^{\alpha } G$ by
	\begin{equation}
	( A^gB^x)( A^hB^y) :=\delta _{x,hy\bar{h}} \beta_y(g,h) A^{gh}B^y.
	\end{equation}
	This product admits a unit
	\begin{equation}
	1=\sum _{x\in G} A^1 B^x.
	\end{equation}
	Define a coproduct $\Delta:D^{\alpha } G\rightarrow D^{\alpha } G\otimes D^{\alpha } G$ and counit $\varepsilon:D^{\alpha } G\rightarrow \mathbb{C}$ by
	\begin{equation}
	\Delta ( A^gB^x) :=\sum _{a,b\in G:ab=x} \mu _{g} (a,b) (A^gB^a) \otimes (A^gB^b)
	\end{equation}
	and
	\begin{equation}
	\varepsilon ( A^gB^x) :=\delta _{x,1}.
	\end{equation}
	Besides, set the Drinfeld associator by
	\begin{equation}
	\Phi :=\sum _{x,y,z\in G} \alpha (x,y,z)^{-1} (A^1B^x)\otimes (A^1B^y)\otimes (A^1B^z)
	\end{equation}
	and the $R$ matrix by
	\begin{equation}
	R:=\sum _{x,y\in G} (A^1B^x)\otimes (A^xB^y).
	\end{equation}
	Finally, define the antipode by a linear map $S:D^{\alpha } G\rightarrow D^{\alpha } G$
	\begin{equation}
	S( A^gB^x) :=\frac{1}{\beta_{\bar{x}}\left(\bar{g}, g\right) \mu _{g}\left( x,\bar{x}\right)} A^{\bar{g}}B^{\bar{g}\bar{x}g},
	\end{equation}
	where $\beta_x(g,h)$ and $\mu _{g} (x,y)$ are expressed in terms of $\alpha $ by
	\begin{equation}
	\beta_y(g,h):=\frac{\alpha (g,h,x)\alpha \left( ghx\bar{h} \bar{g} ,g,h\right)}{\alpha \left( g,hx\bar{h} ,h\right)}
	\end{equation}
	and
	\begin{equation}
	\mu _{g} (x,y):=\frac{\alpha \left( gx\bar{g} ,gy\bar{g} ,g\right) \alpha (g,x,y)}{\alpha \left( gx\bar{g} ,g,y\right)}
	\end{equation}
	for all $g,h,x,y\in G$. As a special case, if $G$ is Abelian and $\alpha ( x,g,h) =\alpha ( x,h,g)$, then $\beta_x(g,h)=\alpha ( x,g,h)$ and $\mu _{g} (x,y)=\alpha ( g,x,y)$.

	\section{Semidirect structure in gauge group $G$ and its dual group $G'$}
	\label{sec:semidirectStructure}

	Let $N$ be a normal Abelian subgroup of a $G$, and denote by \ $K=N\backslash G$ the quotient group. Let $p:G\rightarrow K$ be the usual surjection: $p(g):=Ng$, $\forall g\in G$ , with $p( 1_{G}) =1_{K}$. For each $ x\in K$, choose a representative $u(x)$ in $G$ (such that $pu(x)=x$), with $u( 1_{K}) =1_{G}$. The quotient group $K$ is a right $G$-set with $x\triangleleft g:=p(u(x)g)$, for $x\in K$ and $g\in G$. Moreover, the set $u( K) =\{u(x)|x\in K\}$ is a right $G$-set: $u(x)\triangleleft g=u(x\triangleleft g)$, for $x\in K$ and $g\in G$. The elements $u(x)g$ and $u(x\triangleleft g)$ differ by an element $\kappa _{x,g} \in N$, for $x\in K$ and $g\in G$:
	\begin{equation}
	u(x)g=\kappa _{x,g} u(x\triangleleft g)
	\end{equation}
	The relation
	\begin{equation}
	\kappa _{x,g_{1} g_{2}} =\kappa _{x,g_{1}} \kappa _{xg_{1} ,g_{2}}
	\end{equation}
	holds for any $x\in K$ and $g_{1} ,g_{2} \in G$.
	
	Since $N$ is an Abelian normal subgroup $G,$ there is an induced $K$-left action on $N$ by conjugation:
	\begin{equation}
	^{k} a:=u(k)a\bar{u(k)} \ \ \text{ for } k\in K\ \ \text{ and } \ \ a\in N.
	\end{equation}
	The extension $1\rightarrow N\rightarrow G\rightarrow K\rightarrow 1$ can be classified by (the cohomology classes of) 2-cocycles $F\in H^{2} (K,N)$. Explicitly, $F: K\times K\rightarrow N$ is a map such that
	\begin{equation}
	\delta _{K} F( k_{1} ,k_{2} ,k_{3}) = {^{k_{1}}F( k_{2} ,k_{3})} F( k_{1} k_{2} ,k_{3})^{-1} F( k_{1} ,k_{2} k_{3}) F( k_{1} ,k_{2})^{-1} =1.
	\end{equation}
	With an appropriate choice of $F$, we can assume
	\begin{equation}
	G:=N\rtimes _{F} K,
	\end{equation}
	where the product structure of $G$ is given by the formula
	\begin{equation}
	( a_{1} ,k_{1})( a_{2} ,k_{2}) :=\left( a_{1}\left(^{k_{1}} a_{2}\right) F( k_{1} ,k_{2}) ,k_{1} k_{2}\right).
	\end{equation}
	With this explicit choice of $G$, we choose the function $u:K\rightarrow G$ to be $u(k):=( 1_{N} ,k)$ and therefore we have
	\begin{equation}
	\kappa _{k_{1} ,( a,k_{2})} = {^{k_{1}}a}F( k_{1} ,k_{2}),
	\end{equation}
    leading to $F( k_{1} ,k_{2}) =\kappa _{k_{1} ,( 1,k_{2})}$. For $x\in K$ and $g=(a,k)\in G$, the action of $G$ on $K$ is
	\begin{equation}
	x\triangleleft g=x\triangleleft (a,k)=xk.
	\end{equation}
	Denote the dual group of $N$ by $\hat{N} :=\operatorname{Hom}( N,U(1))$, consisting of the unitary irreducible representations of $N$. It has an induced $K$ action on $\hat{N}$ defined as 
	\begin{equation}
	    \rho ^{k} (a):=\rho \left(^{k} a\right)
	\end{equation}
	for $\rho \in \hat{N}$ and $k\in K$. 
	
	Define a 2-cocycle $\hat{F}\in H^2(K,\hat{N})$ by a function $K^2\rightarrow \hat{N}$ satisfying
	\begin{equation}
	\delta _{K}\hat{F}( k_{1} ,k_{2} ,k_{3}) =\frac{\hat{F}( k_{2} ,k_{3})\hat{F}( k_{1} ,k_{2} k_{3}) }{\hat{F}( k_{1} k_{2} ,k_{3})\hat{F}( k_{1} ,k_{2})^{k_{3}}}=1,
	\end{equation}
	where
	\begin{equation}
	\hat{F}( k_{1} ,k_{2})^{k}( a) =\hat{F}( k_{1} ,k_{2})( ^ka).
	\end{equation}
	
	Every 2-cocycle $\hat{F}$ defines a semidirect product group $G'=K\ltimes_{\hat{F}} \hat{N}$, whose product structure is given by
	\begin{equation}
	( k_{1} ,\rho _{1})( k_{2} ,\rho _{2}) :=\left( k_{1} k_{2} ,\rho ^{k_{2}}_{1} \rho _{2}\hat{F}( k_{1} ,k_{2})\right).
	\end{equation}
	In our convention, the group elements of $G$ and $G'$ are denoted by pairs in the way that $(a,1_K)(1_N,k)=(a,k)$ and $(k,1_{\hat{N}})(1_K,\rho)=(k,\rho)$.
	
	Suppose $F$ and $\hat{F}$ are chosen such that $\hat{F} \land F$ defined by $(\hat{F} \land F)( k_{1} ,k_{2} ,k_{3} ,k_{4}) :=\hat{F}( k_{1} ,k_{2})( F( k_{3} ,k_{4}))$ is cohomologically trivial in $H^{4}( K,U( 1))$. Then there exists a 3-cochain $\epsilon \in C^{3}( K,U( 1))$ such that $\delta _{K} \epsilon =\hat{F} \land F$, i.e.,
		\begin{equation}
		\label{eq:aFFepsilonCond}
		\delta_K\epsilon ( k_{1} ,k_{2} ,k_{3} ,k_{4}) 
		=\frac{\epsilon ( k_{2} ,k_{3} ,k_{4}) \epsilon ( k_{1} ,k_{2} k_{3} ,k_{4}) \epsilon ( k_{1} ,k_{2} ,k_{3} ) }{\epsilon ( k_{1} k_{2} ,k_{3} ,k_{4}) \epsilon ( k_{1} ,k_{2} ,k_{3} k_{4}) }
		=\hat{F}( k_{1} ,k_{2})( F( k_{3} ,k_{4})).
		\end{equation}
	Then we have two explicit (representatives of) 3-cocycles defines by
	\begin{equation}
	\alpha ((a_{1} ,k_{1} ),(a_{2} ,k_{2} ),(a_{3} ,k_{3} ))=\hat{F} (k_{1} ,k_{2} )(a_{3} )\epsilon (k_{1} ,k_{2} ,k_{3} )
	\end{equation}
	and
	\begin{equation}
	\alpha '((x_{1} ,\rho _{1} ),(x_{2} ,\rho _{2} ),(x_{3} ,\rho _{3} ))=\rho _{1} (F(x_{2} ,x_{3} ))\epsilon (x_{1} ,x_{2} ,x_{3} ).
	\end{equation}
	The 3-cocycle conditions $\delta_G\alpha=1$ and $\delta_{G'}\alpha'=1$ can be verified by the 2-cocycle conditions of $F$ and $\hat{F}$, together with the condition \eqref{eq:aFFepsilonCond}. For example, we can check that
	\begin{equation}
	\begin{aligned}
	&\delta_G\alpha((a_1,x_1),(a_2,x_2),(a_3,x_3),(a_4,x_4))\\ = &\frac{\hat{F}(x_2,x_3)(a_4)\hat{F}(x_1,x_2,x_3)(a_4)\hat{F}(x_1,x_2)(a_3)}{\hat{F}(x_1x_2,x_3)(a_4)\hat{F}(x_1,x_2)(a_3\left(^{x_3}a_4\right)F(x_3,x_4))}\delta_K\epsilon(x_1,x_2,x_3,x_4)\\ = &\delta_K\hat{F}(x_1,x_2,x_3)(a_4)\frac{\delta_K\epsilon(x_1,x_2,x_3,x_4)}{\hat{F}(x_1,x_2)(F(x_3,x_4))}\\ = &1.
	\end{aligned}
	\end{equation}
	Note also that $\alpha |_{N} =1$ and $\alpha '|_{\hat{N}} =1$.

\section{The TQD algebra in TQD model}
\label{sec:TQDinTQD}
	
\begin{figure}[!ht]
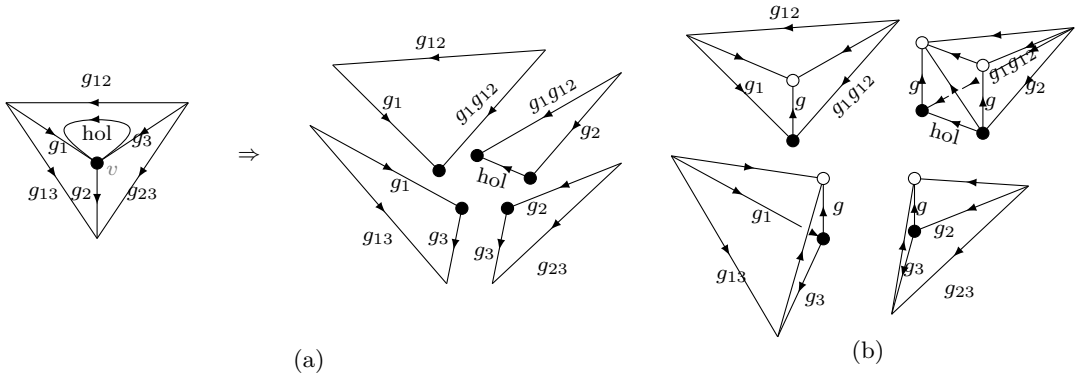

	    \centering
	    \subfigure[]{\unitCellAJ}\quad
	    \subfigure[]{\unitCellAK}
	    \caption{(a) We separate the triangles. Each solid dot denotes the vertex $v$. ()}
	    \label{fig:extendAvg}
	\end{figure}

Here We explain the definition of $A_v^g$ in Eq. \eqref{eq:extendAvgTQD}. We separate the three triangles near the vertex $v$ as in Fig. \ref{fig:extendAvg}(a). We also assign a triangle to the holonomy loop labeled by $\mathrm{hol}=g_1g_{12}\bar{g_2}$. We denote by a solid black dot the vertex $v$ of each triangle. For each solid black dot, we assign an $\alpha$ as follows. We create a new vertex $v'$ near each $v$ above the surface of the triangles (pointing out of the paper), and denote it by a small circle. Each edge from the solid black dot to the samll circle is labeled by $g$. Using the small circles, we extend the four triangles to five tetrahedra, from which the five $\alpha$'s in the action of $A_v^g$ are derived. To see the detailed rule how to assign a 3-cocycle to a tetrahedron, see Ref\cite{Hu2012a}.

\begin{figure}[!ht]
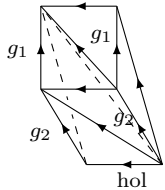

	    \centering
	    {\betaByTetrahedraAD}
	    \caption{Tetrahedral presentation of $\beta_{\mathrm{hol}}(g_1,g_2)$.}
	    \label{fig:betaHol}
	\end{figure}

Now fix the $\mathrm{hol}$. Using the tetrahedron presentation of 3-cocycles, we can prove that the matrix entries of $A_v^{g_1}A_v^{g_2}$ differ from those of $A_v^{g_1g_2}$ by $\beta_{\mathrm{hol}}(g_1,g_2)$, which is the product of three $\alpha$'s presented by the three tetrahedra in Fig. \ref{fig:betaHol}. 

Let $\mathrm{hol}=h_2\in G$, we derive the multiplication
	\begin{equation}
	A^{g_{1}}_{v} B^{h_{1}}_{p} A^{g_{2}}_{v} B^{h_{2}}_{p} =\delta _{h_{1} ,g_{1} h_{2} \bar{g}_{1}} \beta _{h_{2}} (g_{1} ,g_{2} )A^{g_{1} g_{2}}_{v} B^{h_{2}}_{p}.
	\end{equation}
By writing down the three $\alpha$'s presented by the three tetrahedra in Fig. \ref{fig:betaHol} explicitly, the $\beta$ reads
	\begin{equation}
	\beta _{h_{2}} (g_{1} ,g_{2} ):=\frac{\alpha (g_{1} ,g_{2} ,h_{2} )\alpha \left( g_{1} g_{2} h_{2}\bar{g_2}\bar{g_1} ,g_{1} ,g_{2}\right)}{\alpha \left( g_{1} ,g_{2} h_{2} \bar{g_2} ,g_{2}\right)}.
	\end{equation}	

	\section{Rearranged operators in the bilayer model}
	\label{sec:rearrangeAB}
	
	Since the partial Fourier transform only applies to the Hilbert subspace spanned by the configuration of the $N$ elements, the Hilbert subspace spanned by the $K$ elements are unaffected. On the other hand, the $K$-parts of the operators $\left\{A^{x},B^x\right\}_{x\in K}$ are not closed under the multiplication: $A^{x} A^{y}$ yields an $A^{F(x,y)} A^{xy}$ term with $F(x,y) \in N$. In the following, we will introduce the rearranged operators such that the $K$-parts of operators form a subalgebra.

	Let
	\begin{equation}
	\begin{aligned}
	P^{x_0}\mathbf{\tilde B}^{(xy\bar{x},\rho)}{\tilde A}^{(x,\eta)} = &\frac{1}{|N|}\sum_a\overline{\rho(a)}\eta(b)P^{x_0}\mathbf{A}^{(a,x)}B^{(b,y)}\\ = &\frac{1}{|N|}\sum_a\overline{\rho(a)}\eta(b)P^{x_0}A^{(\lambda_a,x)}B^{(b,y)}\\ = &\overline{\rho(F(\bar{x_0},x))}\frac{1}{|N|}\sum_a\overline{\rho^{\bar{x_0}}(a)}\eta(b)P^{x_0}A^{(a,x)}B^{(b,y)}\\ = &\overline{\rho(F(\bar{x_0},x))}P^{x_0}\tilde B^{(xy\bar{x},\rho^{\bar{x_0}})}\tilde A^{(x,\eta)},
	\end{aligned}
	\end{equation}
	where $\rho^{\bar{x_0}}(a) = \rho(^{\bar{x_0}}a)$.
	
	We define the rearranged operators by
	\begin{equation}
	\mathbf{\tilde B}^{(xy\bar{x},\rho)}\mathbf{\tilde A}^{(x,\eta)}P^{x_0} = \mathbf{\tilde B}^{(xy\bar{x},\rho)}{\tilde A}^{(x,\tilde\lambda_\eta)}P^{x_0}
	\end{equation}
	and
	\begin{equation}
	\mathbf A^{(a,x)}\mathbf{B}^{(b,y)}P^{x_0} = \frac{1}{|N|}
	\sum_{\rho\eta}\overline{\eta(b)}\rho(a)\mathbf{\tilde B}^{(xy\bar{x},\rho)}\mathbf{\tilde A}^{(x,\eta)}P^{x_0}.
	\end{equation}
	
	The relations between the rearranged operators and the original are:
	\begin{equation}
	\mathbf{A}^{(a,x)}\mathbf{B}^{(b,y)}P^{x_0} = \overline{\hat{F}(x,x_0)(b)}A^{(\lambda_a,x)}B^{(^{x_0}b,y)}P^{x_0}
	\end{equation}
	amd
	\begin{equation}
	P^{x_0}\mathbf{\tilde B}^{(xy\bar{x},\rho)}\mathbf{\tilde A}^{(x,\eta)} = \overline{\rho(F(\bar{x_0},x))}P^{x_0}\tilde B^{(xy\bar{x},\rho^{\bar{x_0}})}\tilde A^{(x,\tilde\lambda_\eta)}.
	\end{equation}
	The algebra mapping $\{A^{(a,x)} ,B^{(b,y)} \}\rightarrow \{\mathbf{A}^{(a ,x)} ,\mathbf{B}^{(b ,y)} \}$ is an isomorphism, which however does not preserve the algebra relations. These new operators depend on the configurations of $\{k\}$ nonlocally because $x_{v}(\{k\})$ is a nonlocal function.
	
	For the tilde operators, the algebra is
	\begin{equation}
	\begin{aligned}
	&P^{x_0}\mathbf{\tilde B}^{(y_1,\rho_1)}\mathbf{\tilde A}^{(x_1,\eta_1)}
	\mathbf{\tilde B}^{(y_2,\rho_2)}\mathbf{\tilde A}^{(x_2,\eta_2)}\\ = &P^{x_0}\mathbf{\tilde B}^{(y_1,\rho_1)}{\tilde A}^{(x_1,\tilde\lambda_{\eta_1})}
	\mathbf{\tilde B}^{(y_2,\rho_2)}{\tilde A}^{(x_2,\tilde\lambda_{\eta_2})}\\ = &\delta_{h_1,g_1h_2\bar{g_1}}
	\beta_{h_1}(g_1,g_2)\frac{\rho_1(F(\bar{x_0},x_1x_2))}{\rho_1(F(\bar{x_0},x_1))\rho_2(F(\bar{x_0}x_1,x_2))}
	P^{x_0}\mathbf{\tilde{B}}^{(y_1,\rho_1)}
	{\tilde A}^{(x_1,\tilde\lambda_{\eta_1})(x_2,\tilde\lambda_{\eta_2})}\\ = &\delta_{h_1,g_1h_2\bar{g_1}}
	\beta_{h_1}(g_1,g_2)\frac{\rho_1(F(\bar{x_0},x_1x_2))}{\rho_1(F(\bar{x_0},x_1))\rho_2(F(\bar{x_0}x_1,x_2))}
	P^{x_0}\mathbf{\tilde{B}}^{(y_1,\rho_1)}
	\mathbf{\tilde A}^{(x_1x_2,\eta_1\eta_2)},
	\end{aligned}
	\end{equation}
	where $g_1 = (x_1,\tilde\lambda_{\eta_1}),g_2 = (x_2,\tilde\lambda_{\eta_2}),h_{1} = \left(y_{1}, \rho_{1}^{\bar{x_0}}\right), h_{2} = \left(y_{2}, \rho_{2}^{\bar{x_0}x_1}\right)$, and
	\begin{equation}
	\begin{aligned}
	&\beta_{h_1}(g_1,g_2)
	\\ = &
	\frac{\alpha\left(g_{1}, g_{2},\bar{g_2}\bar{g_1} h_1\left(g_{1} g_{2}\right)\right) \alpha\left(h_1, g_{1}, g_{2}\right)}{\alpha\left(g_{1},\bar{g_1} h_1 g_{1}, g_{2}\right)}
	\\ = &
	\frac{\tilde\lambda_{\eta_1}(F(x_2,\bar{x_2}\bar{x_1}y_1(x_1x_2)))\rho_1^{\bar{x_0}}(F(x_1,x_2))}
	{\tilde\lambda_{\eta_1}(F(\bar{x_1}y_1x_1,x_2))}\beta^{\epsilon}_{y_1}(x_1,x_2).
	\end{aligned}
	\end{equation}
	
	We expand the delta function by 
	\begin{equation}
	\delta_{h_1,g_1h_2\bar{g_1}} = \delta_{y_1,x_1y_2\bar{x_1}}
	\delta_{\rho_2,\rho_1\left(\frac{\hat{F}(y_1,x_1)\tilde\lambda_{\eta_1}}{\hat{F}(x_1,y_2)\tilde\lambda_{\eta_1}^{y_2}}\right)^{\bar{x_1}x_0}}
	\end{equation}
	and substitute $\rho_2^{\bar{x_0}}$ by 
	\begin{equation}
	\rho_2^{\bar{x_0}} = \rho_1^{\bar{x_0}}\left(\frac{\hat{F}(y_1,x_1)\tilde\lambda_{\eta_1}}{\hat{F}(x_1,y_2)\tilde\lambda_{\eta_1}^{y_2}}\right)^{\bar{x_1}}.
	\end{equation}
	Using
	\begin{equation}
	\rho_1(\delta_KF(\bar{x_0},x_1,x_2)) = \frac{\rho_1(F(\bar{x_0},x_1x_2))\rho_1^{x_0}(F(x_1,x_2))}{\rho_1(F(\bar{x_0},x_1))\rho_1(F(\bar{x_0}x_1,x_2))} = 1,
	\end{equation}
	We can cancel all the factors involving $\rho_1$ except the delta functions in the coefficient. 
We obtain the algbera
	\begin{equation}
	\begin{aligned}
	& \mathbf{A}^{(a_{1} ,x_{1} )}\mathbf{B}^{(b_{1} ,y_{1} )}\mathbf{A}^{(a_{2} ,x_{2} )}\mathbf{B}^{(b_{2} ,y_{2} )} P^{x_{0}}\\ = & \delta_{y_{1} ,x_{2} y_{2} \bar{x_2}} \delta_{(^{x_{2} x_{0}}b_{2}) \lambda_{a_2}F(x_{2} ,y_{2}) , (^{x_{2} x_{0}} b_{1})({^{y_1}\lambda_{a_2}}) F(y_{1} ,x_{2})} 
	\beta ^{\epsilon }_{y_{2}}(x_{1} ,x_{2})
	\\
	&\quad
	\frac{\hat{F}\left(x_{1} y_{1} \bar{x_1} ,x_{1}\right)(\lambda _{a_{2}})}{\hat{F}(x_{1} ,y_{1})(\lambda _{a_{2}})}
	\hat{F}(x_{1} ,x_{2}x_0)^{\bar{x_0}\bar{x_2}}
	\left(\frac{F(y_{1} ,x_{2})(^{y_1}\lambda_{a_2})}{F(x_{2} ,y_{2})\lambda_{a_2}}\right)
	\mathbf{A}^{(a_{1} a_{2} ,x_{1} x_{2})}\mathbf{B}^{(b_{2} ,y_{2})} P^{x_{0}}.
	\end{aligned}
	\end{equation}
Similarly, we can compute the algebra of the dual operators:
	\begin{equation}
	\begin{aligned}
	& \mathbf{\tilde{B}}^{(x_{1} y_{1} \bar{x_1} ,\rho _{1} )}\mathbf{\tilde{A}}^{(x_{1} ,\eta _{1} )}\mathbf{\tilde{B}}^{(x_{2} y_{2} \bar{x_2} ,\rho _{2} )}\mathbf{\tilde{A}}^{(x_{2} ,\eta _{2} )} P^{x_{0}}\\ = & \delta _{y_{1} ,x_{2} y_{2} \bar{x_2}} \delta _{\rho _{2} ,\rho _{1}\left(\frac{\hat{F} (x_{1} y_{1} \bar{x_1} ,x_{1} )\tilde{\lambda }_{\eta _{1}}}{\hat{F} (x_{1} ,y_{1} )\tilde{\lambda }^{y_{1}}_{\eta _{1}}}\right)^{\bar{x_1} x_{0}}} \beta ^{\epsilon }_{y_{2}} (x_{1} ,x_{2} )\\
	& \frac{\tilde{\lambda }_{\eta _{1}} (F(x_{2} ,y_{2} ))}{\tilde{\lambda }_{\eta _{1}} (F(y_{1} ,x_{2} ))}\left(\frac{\hat{F} (x_{1} ,y_{1} )\tilde{\lambda }^{y_{1}}_{\eta _{1}}}{\hat{F} (x_{1} y_{1} \bar{x_1} ,x_{1} )\tilde{\lambda }_{\eta _{1}}}\right)
	\left[^{x_2x_0} (F(\bar{x}_{0} \bar{x_2} ,x_{2} ))\right] \ \mathbf{\tilde{B}}^{(x_{1} y_{1} \bar{x_1} ,\rho _{1} )}\mathbf{\tilde{A}}^{(x_{1} x_{2} ,\eta _{1} \eta _{2} )} P^{x_{0}},
	\end{aligned}
	\end{equation}
	where
	\begin{equation}
	\lambda_{a_2} = \frac{^{x_2x_0}a_2}{^{x_2x_0}F(\bar{x_0}\bar{x_2},x_2)},
	\end{equation}
	\begin{equation}
	\tilde\lambda_{\eta_1} = \frac{\eta^{\bar{x_0}\bar{x_2}}}{\hat{F}(x_1,x_2x_0)^{\bar{x_0}\bar{x_2}}}.
	\end{equation}
	
	Define
	\begin{equation}
	\mathbf A^x \mathbf B^y = \mathbf A^{(x,1)}\sum_b\mathbf{B}^{(y,b)}
	\end{equation}
	\begin{equation}
	\mathbf{\tilde B}^{xy\bar{x}}\mathbf{\tilde A}^x = \sum_\rho\mathbf{\tilde B}^{(xy\bar{x},\rho)}\mathbf{\tilde A}^{(x,1)}.
	\end{equation}
	We immediately observe that they are identical: $\mathbf B^y = \mathbf{\tilde B}^{xy\bar{x}}\mathbf{\tilde A}^x$. These operators form a subalgebra:
	\begin{equation}
	\mathbf A^{x_1}\mathbf{B}^{y_1}\mathbf{A}^{x_2}\mathbf B^{y_2}P^{x_0} = \delta_{y_1,x_2y_2\bar{x_2}}\Phi_{y_2}(x_1,x_2)\beta^\epsilon_{y_2}(x_1,x_2)\mathbf A^{x_1x_2}\mathbf{B}^{y_2}P^{x_0},
	\end{equation}where
	\begin{equation}
	\begin{aligned}
	& \Phi _{y_{2}} (x_{1} ,x_{2} )\\
	= & \frac{\hat{F}( x_{1} ,x_{2} x_{0})\left( F\left(\overline{x_{0}}\overline{x_{2}} ,x_{2}\right)\right)\hat{F}( x_{1} ,x_{2} x_{0})\left(\text{ }^{\overline{x_{0}}\overline{x_{2}}} F\left( x_{2} y_{2}\overline{x_{2}} ,x_{2}\right)\right)}{\hat{F}( x_{1} ,x_{2} x_{0})\left(\text{ }^{\overline{x_{0}}\overline{x_{2}}} F( x_{2} ,y_{2})\right)\hat{F}( x_{1} ,x_{2} x_{0})\left(\text{ }^{\overline{x_{0}} y_{2} x_{0}} F\left(\overline{x_{0}}\overline{x_{2}} ,x_{2}\right)\right)}\\
	&\quad \frac{\hat{F}\left( x_{1} ,x_{2} y_{2}\overline{x_{2}}\right)\left(\text{ }^{x_{2} x_{0}} F\left(\overline{x_{0}}\overline{x_{2}} ,x_{2}\right)\right)}{\hat{F}\left( x_{1} x_{2} y_{2}\overline{x_{2}}\overline{x_{1}} ,x_{1}\right)\left(\text{ }^{x_{2} x_{0}} F\left(\overline{x_{0}}\overline{x_{2}} ,x_{2}\right)\right)}.
	\end{aligned}
	\end{equation}

	By expanding the coefficient in term of $\epsilon $, we have
	\begin{equation}
	\begin{aligned}
	& \Phi _{y_{2}} (x_{1} ,x_{2} )\beta ^{\epsilon }_{y_{2}} (x_{1} ,x_{2} )\\
	= & \frac{\epsilon \left( x_{2} x_{0} ,\overline{x_{0}}\overline{x_{2}} ,x_{2} y_{2}\overline{x_{2}}\right) \epsilon \left( x_{2} x_{0} ,\overline{x_{0}} y_{2} x_{0} ,\overline{x_{0}}\right) \epsilon \left( x_{1} x_{2} x_{0} ,\overline{x_{0}} ,y_{2}\right)}{\epsilon \left( x_{2} x_{0} ,\overline{x_{0}} ,y_{2}\right) \epsilon \left( x_{2} x_{0} ,\overline{x_{0}} y_{2} x_{0} ,\overline{x_{0}}\overline{x_{2}}\right) \epsilon \left( x_{1} x_{2} x_{0} ,\overline{x_{0}}\overline{x_{2}} ,x_{2} y_{2}\overline{x_{2}}\right)}\\
	& \quad\frac{\epsilon \left( x_{1} x_{2} x_{0} ,\overline{x_{0}} y_{2} x_{0} ,\overline{x_{0}}\overline{x_{2}}\right) \epsilon \left( x_{2} y_{2}\overline{x_{2}} ,x_{2} x_{0} ,\overline{x_{0}}\overline{x_{2}}\right) \epsilon \left( x_{1} x_{2} y_{2}\overline{x_{2}}\overline{x_{1}} ,x_{1} x_{2} x_{0} ,\overline{x_{0}}\right)}{\epsilon \left( x_{1} x_{2} x_{0} ,\overline{x_{0}} y_{2} x_{0} ,\overline{x_{0}}\right) \epsilon \left( x_{2} y_{2}\overline{x_{2}} ,x_{2} x_{0} ,\overline{x_{0}}\right) \epsilon \left( x_{1} x_{2} y_{2}\overline{x_{2}}\overline{x_{1}} ,x_{1} x_{2} x_{0} ,\overline{x_{0}}\overline{x_{2}}\right)}.
	\end{aligned}
	\end{equation}

	Let $\mathbf A^a = \mathbf A^{(a,1)}$ and $\mathbf A^x = \mathbf A^{(1,x)},\mathbf B^b = \sum_y\mathbf B^{(b,y)}$ and $\mathbf B^y = \sum_b\mathbf B^{(b,y)}$. We have
	\begin{equation}
	\mathbf A^{a_1} \mathbf B^{b_1}\mathbf A^{a_2} \mathbf B^{b_2} = \delta_{b_1,b_2}\mathbf A^{a_1a_2} \mathbf B^{b_2},
	\end{equation}
	which form the quantum double $D(N)$, and
	\begin{equation}
	\mathbf A^{x_1} \mathbf B^{y_1}\mathbf A^{x_2} \mathbf B^{y_2} = \Phi _{y_{2}} (x_{1} ,x_{2} )\beta^\epsilon_{y_2}(x_1,x_2)
	\delta_{y_1,x_2y_2\bar{x_2}}\mathbf A^{x_1x_2} \mathbf B^{y_2}.
	\end{equation}


	\bibliographystyle{apsrev4-1}
	\bibliography{StringNet}
\end{document}